\documentclass[a4paper,11pt]{article}

\usepackage{jheppub}

\usepackage[T1]{fontenc}
\usepackage[utf8]{inputenc}
\usepackage{bm}
\usepackage{xcolor}
\usepackage{braket}
\usepackage{slashed}
\usepackage[normalem]{ulem}
\usepackage{makecell}
\usepackage{comment}
\usepackage{tabularx}
\usepackage{multirow}
\usepackage{booktabs}
\usepackage{latexsym}
\usepackage{relsize}
\usepackage{adjustbox}

\usepackage{orcidlink}

\usepackage{subcaption}

\definecolor{softgray}{rgb}{0.5, 0.5, 0.5}
\definecolor{deeporange}{rgb}{0.85, 0.4, 0.0}
\definecolor{deepcyan}{rgb}{0.0, 0.5, 0.7}

\def\Babar{{\mbox{\slshape B\kern-0.1em{\smaller A}\kern-0.1em B\kern-0.1em{\smaller A\kern-0.2em R}}}}
\def\rd#1{\textcolor{red}{#1}}
\def\bl#1{\textcolor{blue}{#1}}
\def\gr#1{\textcolor{softgray}{#1}}
\def\ora#1{\textcolor{deeporange}{#1}}
\def\cya#1{\textcolor{deepcyan}{#1}}

\allowdisplaybreaks[4]

\newcommand{\Dst}{{D^*}}
\newcommand{\Dgen}{{D^{(*)}}}
\newcommand{\rDst}{r_{D^*}}
\newcommand{\rD}{r_D}

\newcommand{\dhh}{{\delta \hat h}}

\title{\bf \boldmath $|V_{cb}|$ determinations from $\bar{B} \to D^{(*)} \ell \bar\nu$ decays within the SM and beyond}

\author[a]{Wen-Sheng Fang,}
\author[b,c]{Syuhei Iguro,}
\author[a,1]{Xin-Qiang Li,\note{Corresponding author.}\orcidlink{0000-0002-3962-3577}}
\author[a]{Ria Sain,}
\author[a,2]{Ryoutaro Watanabe,\note{Corresponding author.}\orcidlink{0000-0001-9317-5843}}
\author[a]{and Ben-Liang Zhang}

\affiliation[a]{Institute of Particle Physics and Key Laboratory of Quark and Lepton Physics (MOE), Central China Normal University, Wuhan, Hubei 430079, China}
\affiliation[b]{Institute for Advanced Research (IAR), Nagoya University, Nagoya 464--8601, Japan}
\affiliation[c]{Kobayashi-Maskawa Institute (KMI) for the Origin of Particles and the Universe, Nagoya University, Nagoya 464--8602, Japan}

\emailAdd{xqli@mail.ccnu.edu.cn}
\emailAdd{watanabe@ccnu.edu.cn}

\abstract{We investigate the $|V_{cb}|$ determinations from exclusive semi-leptonic $\bar{B} \to D^{(*)}\ell\bar\nu$ decays, together with comprehensive Bayesian fit analyses of the $\bar{B} \to D^{(*)}$ transition form-factors, by taking into account recent updates of experimental distribution data and theoretical evaluations. Several commonly adopted form-factor parameterizations, including BSZ, BGL and HQET, have been considered under different fit scenarios. We compare the fitted values and study how the $|V_{cb}|$ determinations depend on the form-factor parameterizations and on the treatment of the experimental inputs. 
Following the fit setup adopted by the PDG, we find that the BSZ, BGL, and HQET results are mutually consistent at specific truncation orders of the form-factor parameterizations. Our representative determinations are $|V_{cb}|=(39.3\pm0.3)\times10^{-3}$ for BGL $(N_F=2)$ and $(39.5\pm0.3)\times10^{-3}$ for HQET $(3/2/1)$, compatible with the official PDG average.
We also consider new-physics effects that can contribute to $\bar{B} \to D^{(*)}\ell\bar\nu$.
The allowed regions remain compatible with vanishing new-physics Wilson coefficients at the $95\%$ confidence level, and the improvements in agreement with the data are modest. The exclusive-inclusive $|V_{cb}|$ puzzle remains unresolved in the present analysis.}

\keywords{Bottom Quarks, Semi-Leptonic Decays, CKM Parameters}
\arxivnumber{2606.17138}

\begin{document}
\maketitle
\flushbottom

%\clearpage
%%%%%%%%%%%%%%%%%%%%%%%%%%%%%%%%%%%%%%%%%%%%%%%%%%%
\section{Introduction} 
\label{sec:Intro}
%%%%%%%%%%%%%%%%%%%%%%%%%%%%%%%%%%%%%%%%%%%%%%%%%%%

In recent years, several developments have been made in both the theoretical evaluations of $\bar{B} \to D^{(*)}$ transition form-factors and the experimental measurements of $\bar{B} \to D^{(*)} \ell\bar\nu$ decays for $\ell = e, \mu$. 
These exclusive semi-leptonic processes are significant for precise determinations of the Cabibbo–Kobayashi–Maskawa (CKM) matrix element $|V_{cb}|$~\cite{ParticleDataGroup:2024cfk,HFLAV:2024ctg,FLAG:2024oxs,Bernlochner:2024sfg}.
Together with $\bar{B} \to D^{(*)} \tau\bar\nu$ decays, they also play a central role in the testing of the lepton-flavor universality (LFU) through the ratios 
$R_D=\Gamma(\bar{B} \to D\tau\bar\nu)/\Gamma(\bar{B} \to D\ell\bar\nu)$ and $R_{D^*} = \Gamma(\bar{B} \to D^*\tau\bar\nu)/\Gamma(\bar{B} \to D^*\ell\bar\nu)$~\cite{Bernlochner:2021vlv,Gambino:2020jvv,Klaver:2024xdy,Albrecht:2026mdp}.  

The unquenched lattice studies~\cite{Bernard:2008dn,FermilabLattice:2014ysv,Harrison:2017fmw} first focused on $\bar{B} \to D^{(*)}$ transition form-factors at the zero-recoil point, 
where the momentum transfer squared $q^2=(p_B-p_{D^{(*)}})^2$ reaches its maximum $q^2_\text{max} = (m_B - m_{D^{(*)}})^2$, 
or equivalently, the velocity transfer $w = v \cdot v^\prime = p_B \cdot p_{D^{(*)}}/(m_B m_{D^{(*)}})$ reaches its minimum $w_\text{min}=1$. 
Near zero recoil, however, the $\bar{B} \to D^{(*)}\ell\bar\nu$ decay rates are suppressed by the available phase space, and extrapolating the experimental data to this point leads to significant uncertainty in the $|V_{cb}|$ determination. 
This has motivated lattice evaluations away from zero recoil, which are now available for $\bar{B} \to D$ form-factors in Refs.~\cite{MILC:2015uhg,Na:2015kha} and for $\bar{B} \to D^*$ form-factors in Refs.~\cite{FermilabLattice:2021cdg,Harrison:2023dzh,Aoki:2023qpa}. 
On the other hand, the $\bar{B} \to D^{(*)}$ transition form-factors in the large-recoil (small-$q^2$) region can be most reliably calculated using the light-cone sum rule (LCSR) approach~\cite{Khodjamirian:2023wol,Colangelo:2000dp,Khodjamirian:2020btr}. 
Latest LCSR results, including higher-order perturbative and power corrections, can be found in Refs.~\cite{Wang:2017jow,Gubernari:2018wyi,Gao:2021sav,Cui:2023jiw}, which extend the earlier leading-order calculation~\cite{Faller:2008tr}.
Since the lattice and LCSR calculations typically constrain complementary $q^2$ regions, their combination is useful for reducing the uncertainties of the form-factors, especially in the large-recoil region~\cite{Cui:2023jiw}. 
The lattice and LCSR inputs used in this study are summarized in Table~\ref{Tab:THsummary}. 

\begin{table}[t]
\renewcommand{\arraystretch}{1.5}
\begin{adjustbox}{width=0.98\textwidth,center,keepaspectratio}
\begin{tabular}{llccc}
\toprule
\textbf{Name} & \textbf{Process}  & \textbf{Method} & \textbf{Data points} & \textbf{Tensor form-factors} \\[0.5em] \hline
MILC15~\cite{MILC:2015uhg}  & $\bar{B} \to D$ & Lattice  &  $w=1.00,\,1.08,\,1.16$ & No   \\[0.2em] \hline
HPQCD15~\cite{Na:2015kha} & $\bar{B} \to D$ & Lattice  & (not explicitly indicated) & No   \\[0.2em] \hline
MILC21~\cite{FermilabLattice:2021cdg} & $\bar{B} \to D^*$ & Lattice & $w=1.03,\,1.10,\,1.17$  &  No   \\[0.2em] \hline
HPQCD23~\cite{Harrison:2023dzh}  & $\bar{B} \to D^*$ & Lattice & $w=1.504,\,1.378,\,1.252,\,1.126,\,1.000$ &  Yes   \\[0.2em] \hline
JLQCD23~\cite{Aoki:2023qpa}  & $\bar{B} \to D^*$ & Lattice  & $w=1.025,\,1.060,\,1.100$ &  No   \\[0.2em] \hline
LCSR18~\cite{Gubernari:2018wyi}  & $\bar{B} \to D^{(*)}$ & LCSR  & $q^2/\text{GeV}^2 = -15,\,-10,\,-5,\,0$ & Yes  \\[0.2em] \hline
LCSR23~\cite{Cui:2023jiw}  & $\bar{B} \to D^{(*)}$ & LCSR  & $q^2/\text{GeV}^2 =-3,\,-2,\,-1,\,0,\,1,\,2,\,3$ &  No   \\
\bottomrule
\end{tabular}
\end{adjustbox}
\caption{
 Available data points of the lattice and LCSR evaluations of the $\bar{B} \to D^{(*)}$ transition form-factors. 
 \label{Tab:THsummary}}
\end{table}

The kinematic distributions of $\bar{B} \to D^{(*)} \ell\bar\nu$ decays measured by the BaBar, Belle, and Belle~II collaborations are also valuable for constraining the $\bar{B} \to D^{(*)}$ transition form-factors. 
These include the binned differential decay rates $\Delta\Gamma_i^D/\Delta w$ for $\bar{B} \to D\ell\bar\nu$, together with $\Delta\Gamma_i^{x}$ and event numbers $\Delta N_i^x$ for $\bar{B} \to D^*\ell\bar\nu$.
Here, $x = w$, $\cos\theta_\ell$, $\cos\theta_V$, and $\chi$ denote the recoil and angular variables to be introduced later.
Currently, two statistically independent measurements~\cite{Belle:2015pkj,Belle-II:2025rna} are available for $\bar{B} \to D \ell\bar\nu$,  
while three~\cite{Belle:2018ezy,Belle:2023bwv,Belle-II:2023okj} are available for $\bar{B} \to D^* \ell\bar\nu$, some of which have been released recently.\footnote{
The unpublished studies of Refs.~\cite{Belle:2017rcc,Belle-II:2022ffa} (Belle17 and BelleII22) have been superseded by Belle23 ($\bar{B} \to D^{*} \ell\bar\nu$) and BelleII25 ($\bar{B} \to D \ell\bar\nu$), respectively.
Two additional studies of angular distributions are also available in Refs.~\cite{Belle:2023xgj,Belle-II:2023svm}.
The former~\cite{Belle:2023xgj} uses the same dataset as Belle23 to construct angular coefficients.
The latter~\cite{Belle-II:2023svm} provides angular observables in only two recoil bins and has relatively large uncertainties.
We therefore do not include these studies in our analysis.
}
Together with the branching ratio measurements, these observables provide key inputs for $|V_{cb}|$ fit study~\cite{ParticleDataGroup:2024cfk,HFLAV:2024ctg,FLAG:2024oxs}. 
The experimental inputs used throughout this work are summarized in Table~\ref{Tab:EXsummary}.

\begin{table}[t]
\renewcommand{\arraystretch}{1.5}
\begin{adjustbox}{width=0.98\textwidth,center,keepaspectratio}
\begin{tabular}{lcccc}
\toprule
\textbf{Name} & \textbf{Process}  & \textbf{Data points}  & \textbf{Branching ratio} & \textbf{$\boldsymbol{|V_{cb}| \times 10^{3}}$ (*)} \\[0.5em] \hline
Belle15~\cite{Belle:2015pkj}   & \makecell{$B^0 \to D^{-} \ell^+ \nu~(\ell=e,\mu)$ \\[0.5em]  $B^+ \to \bar D^{0} \ell^+\nu~(\ell=e,\mu)$}  & $\displaystyle {\frac{\Delta\Gamma_i^D}{\Delta w}}$ (10 bins) 
& \makecell{$(2.39 \pm 0.04 \pm 0.11) \%$ \\[0.5em] $(2.54 \pm 0.04 \pm 0.13) \%$} & $40.12 \pm 1.34$ \\[0.4em] \hline
BelleII25~\cite{Belle-II:2025rna}   & \makecell{ $B^0 \to D^{-} \ell^+ \nu~(\ell=e,\mu)$ \\[0.5em]  $B^+ \to \bar D^{0} \ell^+\nu~(\ell=e,\mu)$} & $\displaystyle{\frac{\Delta \Gamma_i^D}{\Delta w}}$ (10 bins) 
& \makecell{$(2.06 \pm 0.05 \pm 0.10) \%$ \\[0.5em] $(2.31 \pm 0.04 \pm 0.09) \%$} & $39.2 \pm 0.9$ \\[0.4em]  \hline
%
%Belle17~\cite{Belle:2017rcc}   & $\bar B^0 \to D^{*+} \ell^-\bar\nu ~(\ell=e,\mu)$ & $\displaystyle {\Delta\Gamma_i^x}$ (40 bins) & $(4.95 \pm 0.11 \pm 0.22) \%$ & $37.4 \pm 1.3$ \\[0.4em] \hline
%
Belle18~\cite{Belle:2018ezy}   & $\bar B^0 \to D^{*+} \ell^-\bar\nu ~(\ell=e,\mu)$ & $\displaystyle {\Delta N_i^x}$ (40 bins)  & $(4.90 \pm 0.02 \pm 0.16) \%$ & $38.3 \pm 1.0$ \\ \hline
Belle23~\cite{Belle:2023bwv}   & \makecell{ $\bar B^0 \to D^{*+} \ell^-\bar\nu ~(\ell=e,\mu)$ \\[0.5em]  $B^- \to D^{*0} \ell^-\bar\nu ~(\ell=e,\mu)$} & $\displaystyle{\frac{\Delta\Gamma_i^x}{\Gamma^\Dst}}$ (40 bins)  & Not available & $40.6 \pm 0.9$ \\[0.4em] \hline
BelleII23~\cite{Belle-II:2023okj} & $\bar B^0 \to D^{*+} \ell^-\bar\nu ~(\ell=e,\mu)$ & $\displaystyle {\Delta\Gamma_i^x}$ (38 bins) & $(4.922 \pm 0.023 \pm 0.220)\%$ & $40.57 \pm 1.16$ \\
\bottomrule
\end{tabular}
\end{adjustbox}
\caption{Available data points of the kinematic distributions of $\bar{B} \to D^{(*)} \ell\bar\nu$ decays with respect to the variables $x = w$, $\cos\theta_\ell$, $\cos\theta_V$, and $\chi$, where the decay distributions $\Delta\Gamma_i^{D(x)}$ and the event numbers $\Delta N_i^x$ will be described later. 
As the BaBar measurements~\cite{BaBar:2019vpl,BaBar:2023kug} do not provide explicit values of their measurements, it is not possible for us to use their results for our fit analysis. 
The LHCb analysis of Ref.~\cite{LHCb:2026amx} has been reported recently, but one-dimensional distributions or binned measurements are not currently available.
(*) The $|V_{cb}|$ values shown here are the ones reported in each reference. \label{Tab:EXsummary} }
\end{table}

Since the lattice and LCSR approaches provide results only at a finite set of kinematic points, 
a parameterization of the $\bar{B} \to D^{(*)}$ transition form-factors is required to interpolate and extrapolate these theoretical inputs across the full semi-leptonic region, $0\leq q^2 \leq (m_B-m_{D^{(*)}})^2$. 
Two types of parameterizations had been widely used for this purpose: model-independent parameterization based on fundamental requirements (analyticity, unitarity, crossing symmetry), and heavy-quark-symmetry-based parameterization involving the unitarity bounds. They are generally referred to as the Boyd-Grinstein-Lebed (BGL)~\cite{Boyd:1997kz,Boyd:1995sq,Boyd:1995cf,Boyd:1994tt}\footnote{
There exists the Bourrely-Caprini-Lellouch (BCL) parameterization~\cite{Bourrely:2008za}, which can avoid issues arising from a truncated BGL expansion at the threshold $q^2=(m_B+m_{D^{(*)}})^2$.
See Ref.~\cite{Bourrely:2008za} for details. 
However, since the BCL parameterization assumes a more complicated form and these artifacts occur only beyond the semi-leptonic region in $\bar{B}\to D^{(*)} \ell\bar\nu$, we do not consider this parameterization in this study.
} and Caprini-Lellouch-Neubert (CLN)~\cite{Caprini:1997mu} parameterizations, respectively. Since the approximations made in Ref.~\cite{Caprini:1997mu} are no longer sufficiently justified in view of the current precision, the latter has been extended by including higher-order radiative and power corrections~\cite{Bernlochner:2017jka,Bordone:2019vic,Bernlochner:2022ywh,Bordone:2025jur}. 
In this paper, we refer to this framework as the Heavy-Quark-Effective-Theory (HQET) parameterization~\cite{Bernlochner:2017jka,Bigi:2017jbd,Jaiswal:2017rve,Jung:2018lfu,Bordone:2019vic,Bordone:2019guc,Bernlochner:2022ywh,Li:2024weu,Bordone:2025jur}. Another useful choice is the Bharucha-Straub-Zwicky (BSZ) parameterization~\cite{Bharucha:2015bzk}, which employs a ``simplified series expansion (SSE)'' about $q^2=0$. This form makes it straightforward to impose exact kinematic relations among the form-factors at $q^2=0$. Motivated by these observations, we consider the BSZ, BGL and HQET parameterizations of the $\bar{B} \to D^{(*)}$ transition form-factors throughout this paper.

The BGL and BSZ parameterizations have the advantage that the form-factors are treated independently. 
Thus, the relevant data points constrain separate sets of parameters for the corresponding form-factors. 
%For example, the binned kinematic distributions of $\bar{B} \to D \ell\bar\nu$ and $\bar{B} \to D^{*} \ell\bar\nu$ decays constrain individual parameters of the $\bar{B} \to D$ and $\bar{B} \to D^*$ transition form-factors, respectively. 
By contrast, the HQET parameterization describes all these form-factors in terms of some common unknown functions, the so-called Isgur-Wise (IW) functions. Consequently, all data relevant to $\bar{B} \to D^{(*)} \ell\bar\nu$ decays provide correlated constraints on the IW functions. 

In light of recent updates from lattice and experimental studies summarized in Tables~\ref{Tab:THsummary} and \ref{Tab:EXsummary}, the Particle Data Group (PDG)~\cite{ParticleDataGroup:2024cfk} and the Heavy Flavor Averaging Group (HFLAV)~\cite{HFLAV:2024ctg} employ the BGL parameterization and report their official values of $|V_{cb}|$ as 
\begin{align} \label{eq:VcbPDGofficial}
 \text{PDG:}~~ |V_{cb}| \times 10^{3} = 
 \begin{cases}  
    38.9 \pm 0.7 & (\bar{B} \to D \ell\bar\nu)  \\[0.5em] 
    40.0 \pm 0.7 & (\bar{B} \to D^* \ell\bar\nu) 
    \end{cases}
    ~= 39.5 \pm 0.5 \quad (\text{combined}) \,.
\end{align}
Their fitting strategy for the form-factor parameters and $|V_{cb}|$ is based on a $\chi^2$ fit including the available inputs listed in Tables~\ref{Tab:THsummary} and \ref{Tab:EXsummary}, except for the LCSR evaluations, but with tuned decay distributions. 
To be precise, the averaged decay distributions are constructed by combining the available measurements and are converted into the normalized form $\Delta\Gamma_i/\Gamma$, where $\Delta\Gamma_i$ denotes the averaged binned decay distribution and $\Gamma = \sum_i \Delta\Gamma_i$ is the total sum over bins~\cite{HFLAV:2024ctg}.
These normalized observables $\Delta\Gamma_i/\Gamma$ are crucial because they are independent of $|V_{cb}|$, and hence the values of $|V_{cb}|$ in Eq.~\eqref{eq:VcbPDGofficial} are determined only by the measured branching ratios.
Several minor points in their fit setup will be discussed later. 

In this paper, we will perform comprehensive Bayesian fit analyses with a Markov-Chain-Monte-Carlo (MCMC) method to determine the CKM matrix element $|V_{cb}|$. In particular, we will clarify the following points: 
\begin{itemize} 
\item 
 We examine whether the kinematic decay distributions can be incorporated directly, so that $|V_{cb}|$ is constrained by both the binned decay distributions $\Delta\Gamma_i$ and the branching ratios. 
 We also reproduce the PDG setup and compare it with our other fit scenarios. 
\item 
 We perform parameter fits using the BSZ, BGL and HQET parameterizations of the $\bar{B} \to D^{(*)}$ transition form-factors and compare their results quantitatively.
\item 
 We include possible new physics (NP) operators contributing to $\bar{B} \to D^{(*)} \ell\bar\nu$ decays and obtain the allowed ranges of the corresponding NP Wilson coefficients in the $b \to c\ell\bar\nu$ transition, together with the fitted values of $|V_{cb}|$ and the form-factor parameters. 
 We then discuss whether non-negligible NP contributions are still allowed by the current data on $\bar{B} \to D^{(*)} \ell\bar\nu$ decays. 
 Previous studies~\cite{Jung:2018lfu,Iguro:2020cpg,Crivellin:2009sd,Crivellin:2014zpa,Fedele:2023ewe} performed a similar analysis, while the present work includes several developments, such as the comparison of different form-factor parameterizations and updated measurements. 
\end{itemize}

This paper is organized as follows. A general description of the differential decay distributions of $\bar{B} \to D^{(*)} \ell\bar\nu$ decays in terms of the helicity amplitudes is given in section~\ref{sec:HA}, where we consider the most generic low-energy effective Hamiltonian relevant for the $b \to c\ell\bar\nu$ transition. The BSZ, BGL and HQET parameterizations of the $\bar{B} \to D^{(*)}$ transition form-factors, together with the relevant definitions are introduced in section~\ref{sec:FFdef}. Our fit procedure is then described in section~\ref{sec:setup}, along with a summary of the available experimental and theoretical datasets. Then, our fit results are presented in section~\ref{sec:result}. Finally, our conclusion is made in section~\ref{sec:summary}. For convenience, we collect in appendices~\ref{App:FFinput} and \ref{App:FFresult_all} the analytical formulae of the one-loop $\mathcal{O}(\alpha_s)$ corrections to the HQET parameterization and our fit details, respectively.

%%%%%%%%%%%%%%%%%%%%%%%%%%%%%%%%%%%%%%%%%%%%%%%%%%%
\section{Differential decay distributions} 
\label{sec:HA}
%%%%%%%%%%%%%%%%%%%%%%%%%%%%%%%%%%%%%%%%%%%%%%%%%%%
 
We start with the most generic low-energy effective Hamiltonian relevant for the $b \to c\ell\bar\nu$ transition,  
\begin{align} \label{eq:Hamiltonian}
 \mathcal{H}_{\rm{eff}} 
 = 2\sqrt 2  G_F V_{cb} \biggl[ 
 & (\bar c \gamma^\mu P_Lb)(\bar\ell \gamma_\mu P_L \nu) + C_{V_R} (\bar c \gamma^\mu P_R b)(\bar\ell \gamma_\mu P_L \nu) + C_{S_L} (\bar c P_Lb)(\bar\ell P_L \nu) \notag\\[0.3em]
 & + C_{S_R} (\bar c P_Rb)(\bar\ell P_L \nu) + C_{T} (\bar c \sigma^{\mu\nu} P_Lb)(\bar\ell \sigma_{\mu\nu} P_L \nu)  \biggl] +\text{h.c.}\,, 
\end{align}
where $P_L=(1-\gamma_5)/2$, $P_R=(1+\gamma_5)/2$, and $\sigma^{\mu\nu}=i/2\,[\gamma^\mu,\gamma^\nu]$, with the sign convention $\epsilon^{0123}=-1$, which implies that $\sigma^{\mu\nu} \gamma_5 = - i/2 \,\epsilon^{\mu\nu\rho\sigma} \sigma_{\rho\sigma}$. Since an SM-like NP merely rescales the value of $V_{cb}$, we do not consider this case here, as its effects are better probed through indirect or combined analyses. We have also assumed that neutrinos in Eq.~\eqref{eq:Hamiltonian} are always left-handed. We set the NP Wilson coefficients $C_X$ as real and the SM case is recovered by setting $C_{V_R}=C_{S_L}=C_{S_R}=C_T=0$. The invariant amplitude for $\bar{B} \to D\ell\bar\nu$ is expressed in terms of the helicity amplitudes as 
\begin{equation} \label{eq:BDLnu}
 \mathcal{M}(q^2,\theta_\ell) 
  = \frac{G_F}{\sqrt{2}} V_{cb} 
 \left[(1+C_{V_R}) \sum_{\lambda} \eta_\lambda  H_{\lambda} L_{\lambda}^{\ell}
 - \left(C_{S_R}+C_{S_L}\right) H L^\ell
 - C_T \sum_{\lambda,\lambda'} \eta_\lambda \eta_{\lambda'} H_{\lambda,\lambda'} L^\ell_{\lambda,\lambda'}
 \right], 
\end{equation}
while that for the sequential decay $\bar{B} \to \Dst (\to D\pi) \ell \bar\nu$ is given by
\begin{align}
 \mathcal{M}(q^2,\theta_\ell, \theta_V, \chi) 
 = \frac{G_F}{\sqrt{2}} V_{cb} \sum_{\lambda_{\Dst},\lambda'_{\Dst}} \mathcal{S}^{\lambda_{\Dst}} (q^2,\theta_\ell)\, \mathcal{D}^1_{\lambda_{\Dst}\,\lambda'_{\Dst}}(\chi)\, \mathcal{T}^{\lambda'_{\Dst}}(\theta_V)  \,, 
 \label{eq:BDstDpiLnu}
\end{align}
with $q^\mu=p_B^\mu-p_{D^{(*)}}^\mu=p_\ell^\mu+p_\nu^\mu$ and $q^2=(p_B-p_{D^{(*)}})^2$, 
where $\mathcal{S}^{\lambda_{\Dst}}(q^2,\theta_\ell)$ denotes the amplitude for $\bar{B}\to \Dst \ell \bar\nu$ and is described by 
\begin{align}
 \mathcal{S}^{\lambda_{\Dst}} (q^2,\theta_\ell) 
 = 
 & \sum_{\lambda} \eta_\lambda \left( H^{\lambda_{\Dst}}_{\lambda} - C_{V_R} H^{-\lambda_{\Dst}}_{-\lambda} \right) L_{\lambda}^{\ell} - \left(C_{S_R}-C_{S_L}\right) H^{\lambda_{\Dst}} L^\ell \notag \\[0.3em]
 & - C_T \sum_{\lambda,\lambda'} \eta_\lambda \eta_{\lambda'} H^{\lambda_{\Dst}}_{\lambda,\lambda'} L^\ell_{\lambda,\lambda'} \,, 
\end{align}
with $\eta_\lambda = +1\,(-1)$ for $\lambda = \pm1, 0\,(s)$. $\mathcal{T}^{\lambda'_{\Dst}}(\theta_V)$ represents the amplitude for $\Dst \to D \pi$, and $\mathcal{D}^1_{\lambda_{\Dst}\,\lambda'_{\Dst}}(\chi)$ is the Wigner rotation that connects the two decay planes defined by the helicity angles $\theta_\ell$ ($\ell$--$\nu$ plane in the $W^*$ rest frame) and $\theta_V$ ($D$--$\pi$ plane in the $\Dst$ rest frame) in the $B$-meson rest frame, with $\chi$ being the azimuthal angle between the two planes~\cite{Gratrex:2015hna,Jacob:1959at,Haber:1994pe,Korner:1989qb}. Explicit forms of these quantities will be given later. 

\subsection{Helicity amplitudes}
\label{sec:StandardForm}
%%%%%%%%%%%%%%%%%%%%%%%%%%%%%

Explicit expressions of the leptonic amplitudes $L_{\lambda}^{\ell}$, 
$L^\ell$ and $L^\ell_{\lambda,\lambda'}$ present in Eqs.~\eqref{eq:BDLnu} and \eqref{eq:BDstDpiLnu} could be found, \textit{e.g.}, in Ref.~\cite{Tanaka:2012nw}. For the hadronic amplitudes, we have to specify the matrix elements of all the SM and NP quark currents present in Eq.~\eqref{eq:Hamiltonian}. These matrix elements can be decomposed into a finite set of Lorentz structures, multiplied by some scalar functions of the momentum transfer $q^2$, with the latter defined as the $\bar{B} \to \Dgen$ transition form-factors. To this end, let us begin with the standard forms defined by 
\begin{align} 
\braket{D|\bar c \gamma^\nu b| \bar{B}} 
& = \left[ (p_B+p_D)^\nu - \frac{m_B^2-m_D^2}{q^2} q^\nu \right] {\color{red}f_+(q^2)} + \frac{m_B^2-m_D^2}{q^2} q^\nu \,{\color{red}f_0(q^2)} \,, \label{eq:fplus-fzero-standard} \\[1em]
 \braket{D|\bar c b| \bar{B}} 
 & = (m_B+m_D) \gr{f_S(q^2)} \,, \\[1em]
 \braket{D|\bar c \sigma^{\mu\nu} b| \bar{B}} 
 & = \frac{-2i}{m_B+m_D} \left( p_B^\mu p_D^\nu - p_B^\nu p_D^\mu \right) \,{\color{red}f_T(q^2)} \,, 
\end{align}
for the $\bar{B} \to D$, and 
\begin{align}
 \label{eq:DstBv}
 \braket{ D^*_{\lambda_\Dst} | \bar c \gamma^\mu b | \bar{B} } 
 & = i \frac{2\rd{V(q^2)}}{m_B+m_\Dst} \varepsilon^{\mu\nu\rho\sigma} (\epsilon^*_{\lambda_\Dst})_\nu \,(p_\Dst\!)_\rho \,(p_B)_\sigma \,, \\[1em]
 \braket{ D^*_{\lambda_\Dst} | \bar c \gamma^\mu \gamma^5 b | \bar{B} } 
 & = (m_B+m_\Dst) \epsilon^{*\mu}_{\lambda_\Dst} \rd{A_1(q^2)} - \frac{(\epsilon^*_{\lambda_\Dst} \cdot q)}{m_B+m_\Dst} (p_B+p_\Dst)^\mu \gr{A_2(q^2)} \notag \\[0.5em]
 & \quad - \frac{2m_\Dst}{q^2} (\epsilon^*_{\lambda_\Dst} \cdot q) q^\mu A_3(q^2) \,, \notag \\[1em]
\braket{D^*_{\lambda_\Dst} |\bar c \gamma^5 b| \bar{B}}
& =\left(\epsilon^*_{\lambda_\Dst} \cdot q\right) \gr{P(q^2)}, \\[1em]
\braket{D^*_{\lambda_\Dst} |\bar c \sigma^{\mu\nu} b| \bar{B}}
& = \epsilon^{\mu \nu \rho \sigma }
\Biggl\{ 
(\epsilon^*_{\lambda_\Dst})_{\sigma} 
\left[\left(p_B+p_\Dst\right)_\rho {\color{red}T_1(q^2)} - q_\rho \frac{m_B^2-m_{D^*}^2}{q^2}\left( {\color{red}T_1(q^2)}- {\color{red}T_2(q^2)} \right)\right] \notag \\[0.5em]
& \hspace{-0.5cm} - 2 \left(\epsilon^*_{\lambda_\Dst} \cdot q\right) \frac{ (p_B)_\sigma (p_\Dst)_\rho}{q^2}\left[ {\color{red}T_1(q^2)}- {\color{red}T_2(q^2)} -\frac{q^2}{m_B^2-m_{D^*}^2} \gr{T_3(q^2)}\right]\Biggr\} \,, \label{eq:tensor-standard}
\end{align}
with 
\begin{align}
 A_3(q^2) = \frac{m_B+m_\Dst}{2m_\Dst} \rd{A_1(q^2)} - \frac{m_B-m_\Dst}{2m_\Dst} \gr{A_2(q^2)} - \rd{A_0(q^2)} \,, 
\end{align}
for the $\bar{B} \to D^*$ transition, where $\epsilon^{\mu}_{\lambda_\Dst}$ is the polarization vector of the $D^*$ meson. The form-factors marked in gray, $\{ \gr{f_S}, \gr{P}, \gr{A_2}, \gr{T_3} \}$, can be expressed in terms of the other relevant ones. To be specific, $\gr{f_S}$ and $\gr{P}$ are related to $f_0$ and $A_0$ through 
\begin{align}
 \gr{f_S(q^2)} = \frac{m_B-m_D}{m_b-m_c} {\color{red}f_0(q^2)} \,, 
 \qquad
 \gr{P(q^2)} = -\frac{2m_\Dst}{m_b+m_c}  {\color{red}A_0(q^2)} \,. 
 \label{eq:FFscalarEOM}
\end{align}
For $\gr{A_2}$ and $\gr{T_3}$, on the other hand, we will describe their explicit forms when introducing the specific form-factor parameterization methods. 

Taking the above standard forms, the hadronic amplitudes are obtained as~\cite{Duan:2024ayo} 
\begin{align}
 H_V & \equiv H_0 = \sqrt{\frac{Q_+Q_-}{q^2}} \rd{f_+(q^2)} \,, \label{eq:HVD}  \\[1em]
 H_{V_t} & \equiv H_s = \frac{m_B^2 - m_D^2}{\sqrt{q^2}} \rd{f_0(q^2)} \,, \label{eq:HVtD} \\[1em]
 H_S & \equiv H = (m_B+m_D) \gr{f_S(q^2)} = \frac{m_B^2-m_D^2}{m_b-m_c} \rd{f_0(q^2)}\,, 
 \\[1em]
 H_T & \equiv H_{+,-} = H_{0,s} = - H_{-,+} = - H_{s,0} = - \frac{\sqrt{Q_+Q_-}}{m_B+m_D} \rd{f_T(q^2)} \,, \label{eq:HTD} 
\end{align}
for $\bar{B} \to D$, and 
\begin{align}
 H_{V_+} & \equiv H^{\lambda_\Dst=+}_+ = (m_B+m_\Dst) \rd{A_1(q^2)} - \frac{\sqrt{Q_+Q_-}}{m_B+m_\Dst} \rd{V(q^2)} \,, \\[1em]
 H_{V_-} & \equiv H^{\lambda_\Dst=-}_- = (m_B+m_\Dst) \rd{A_1(q^2)} + \frac{\sqrt{Q_+Q_-}}{m_B+m_\Dst} \rd{V(q^2)} \,, \\[1em]  
 H_{V_0} & \equiv H^{\lambda_\Dst=0}_0 = \frac{m_B+m_\Dst}{2m_\Dst \sqrt{q^2}} \left[ \frac{Q_+Q_-}{(m_B+m_\Dst)^2} \gr{A_2(q^2)} - (m_B^2-m_\Dst^2-q^2) \rd{A_1(q^2)} \right] \,, \\[1em]
 H_{V_s} & \equiv H^{\lambda_\Dst=0}_s = - \frac{\sqrt{Q_+Q_-}}{\sqrt{q^2}} \rd{A_0(q^2)} \,, \\[1em]
 H_P & \equiv H^{\lambda_\Dst=0} = \frac{\sqrt{Q_+Q_-}}{2m_\Dst} \gr{P(q^2)} = - \frac{\sqrt{Q_+Q_-}}{m_b+m_c} \rd{A_0(q^2)} \,, \\[1em]
 H_{T_+} & \equiv +H^{\lambda_\Dst=+}_{+,0} = +H^{\lambda_\Dst=+}_{+,s} = -H^{\lambda_\Dst=+}_{0,+} = -H^{\lambda_\Dst=+}_{s,+} \\[0.5em]
 &= \frac{1}{\sqrt{q^2}} \Big[\sqrt{Q_+Q_-}\, \rd{T_1(q^2)} + (m_B^2-m_\Dst^2) \rd{T_2(q^2)} \Big] \,, \\[1em]
 H_{T_-} & \equiv + H^{\lambda_\Dst=-}_{-,0} = -H^{\lambda_\Dst=-}_{-,s} = -H^{\lambda_\Dst=-}_{0,-} = +H^{\lambda_\Dst=-}_{-,s} \\[0.5em]
 &= \frac{1}{\sqrt{q^2}} \Big[\sqrt{Q_+Q_-}\, \rd{T_1(q^2)} - (m_B^2-m_\Dst^2) \rd{T_2(q^2)} \Big] \,, \\[1em]
 H_{T_0} & \equiv H^{\lambda_\Dst=0}_{+,-} = H^{\lambda_\Dst=0}_{0,s} = -H^{\lambda_\Dst=0}_{-,+} = -H^{\lambda_\Dst=0}_{s,0}  \\[0.5em] 
 &= \frac{1}{2m_\Dst} \left[\frac{Q_+Q_-}{m_B^2-m_\Dst^2} \gr{T_3(q^2)} -(m_B^2+3m_\Dst^2-q^2) \rd{T_2(q^2)} \right] \,, 
\end{align}
for $\bar{B} \to D^*$, with the hadronic amplitudes $H^{\lambda_\Dst}_{\lambda,\lambda'}=0$ for all other helicity components. Here, for convenience, we have introduced the abbreviations
\begin{align}
 Q_\pm = (m_B \pm m_\Dgen)^2 - q^2 \,.
\end{align}
%with $q^\mu=p_B^\mu-p_{D^{(*)}}^\mu=p_\ell+p_\nu$ and $q^2=(p_B-p_{D^{(*)}})^2$. 

\subsection{Decay distributions}
\label{sec:DecayDist}
%%%%%%%%%%%%%%%%%%%%%%%%%%%%%

The differential decay rate for $\bar{B} \to D \ell \bar\nu$ can be written as
\begin{align} \label{eq:BDlnu}
  \frac{d\Gamma(\bar{B}\to D \ell\bar\nu)}{dq^2}  &=  \frac{G_F^2 \left|V_{cb}\right|^2}{192\pi^3m_B^3} q^2 \sqrt{Q_+Q_-} |\eta_\mathrm{EW}|^2 \notag \\[0.5em]
 & \quad \times \left[(1+ C_{V_R})^2 \, (H_{V})^2 + \frac{3}{2} (C_{S_L} + C_{S_R} )^2 (H_S)^2 + 8 C_T^2 (H_T)^2 \right] \,, 
\end{align}
where the helicity amplitudes $H_X$ are given in Eqs.~\eqref{eq:HVD}--\eqref{eq:HTD}, and $\eta_\mathrm{EW}=1.00662$ accounts for the leading electroweak corrections~\cite{Sirlin:1981ie,FermilabLattice:2014ysv}.\footnote{Although the electroweak corrections generally depend on the operator structures, such differences can be absorbed into the Wilson coefficients of the operators. Therefore, we use a common Sirlin factor $\eta_\mathrm{EW}$.} Coulomb corrections for final-state charged particles in neutral $B$-meson decays~\cite{Atwood:1989em,Ginsberg:1968pz} are neglected.

The $\Dst \to D \pi$ amplitude $\mathcal T^{\lambda'_{\Dst}}(\theta_V)$ and the Wigner rotation $\mathcal D^1_{\lambda_{\Dst}\,\lambda'_{\Dst}}(\chi)$ present in the decay amplitude $\mathcal{M}(q^2,\theta_\ell, \theta_V, \chi) $ of $\bar{B} \to \Dst(\to D \pi) \ell\bar\nu$ are obtained, respectively, as  
\begin{align}
 \mathcal T^{0} &= \frac{N_D}{2} \sqrt{\frac{3}{\pi}} \cos\theta_V \,, & \mathcal T^{\pm1}  &= \mp \frac{N_D}{2} \sqrt{\frac{3}{2 \pi}} \sin\theta_V \,,  & &\\[1em]
 \mathcal D^1_{00} &= 1 \,, & \mathcal D^1_{\pm1\pm1} &= e^{\pm i \chi} \,, & \text{others} &= 0 \,, 
\end{align}
where the normalization factor $N_D$ is determined so that
\begin{align} \label{eq:fullprocess}
 \int_{-1}^1 d\cos\theta_V\int_{-\pi}^{\pi} d\chi \frac{d\Gamma_\text{full}}{dq^2\,d\cos\theta_\ell\,d\cos\theta_V\,d\chi} = \frac{d\Gamma(\bar{B} \to \Dst \ell\bar\nu)}{dq^2\,d\cos\theta_\ell} \mathcal B ({\Dst \to D \pi}) \,, 
\end{align}
is satisfied. In this way, we determine the full angular distribution as
\begin{align}
 \frac{d \Gamma_\text{full}}{dq^2 d\cos\theta_\ell d\cos\theta_V d\chi} =& \frac{3 G_F^2|V_{cb}|^2}{4096\pi^4 m_B^3} q^2 \sqrt{Q_+Q_-}\,|\eta_\mathrm{EW}|^2\, \mathcal B(D^{*} \to \bar D \pi) \notag \\[0.5em]
 & \times \sum_{i=1}^{10} \mathcal J_i (\theta_\ell, \theta_V, \chi)\, \mathcal H_i (q^2) \,, \label{eq:BDstlnu}
\end{align}
with
\begin{align}
 \mathcal H_1 (q^2) = & \left( H_{V_+} - C_{V_R} H_{V_-} \right)^2 \,, \\[0.5em]
 \mathcal H_2 (q^2) = & \left( H_{V_-} - C_{V_R} H_{V_+} \right)^2 \,, \\[0.5em] 
 \mathcal H_3 (q^2) = & \left( 1-C_{V_R} \right)^2 H_{V_0}^2 \,, \\[0.5em]
 \mathcal H_4 (q^2) = & \left( C_{S_R} - C_{S_L} \right)^2 H_P^2 \,, \\[0.5em]
 \mathcal H_5 (q^2) = & C_T^2 \left(H_{T_+}^2 +H_{T_-}^2  \right) \,, \\[0.5em]
 \mathcal H_6 (q^2) = & C_T^2\, H_{T_0}^2 \,, \\[0.5em] 
 \mathcal H_7 (q^2) = & \left( H_{V_+} - C_{V_R} H_{V_-} \right) \left( H_{V_-} - C_{V_R} H_{V_+} \right) + 16\, C_T^2\, H_{T_+} H_{T_-} \,, \\[0.5em]
 \mathcal H_8 (q^2) = & (1-C_{V_R}) H_{V_0} \left( H_{V_+} - C_{V_R} H_{V_-} \right) - 8\, C_T^2\, H_{T_0} \left( H_{T_+} - H_{T_-} \right) \notag \\[0.2em]
 & + 2\, C_T \left(C_{S_R}-C_{S_L} \right) H_P \left(H_{T_+} - H_{T_-} \right)
  \,, \\[0.5em]
 \mathcal H_9 (q^2) = & (1-C_{V_R}) H_{V_0} \left( H_{V_-} - C_{V_R} H_{V_+} \right) - 8\, C_T^2\, H_{T_0} \left( H_{T_+} - H_{T_-} \right) \notag \\[0.2em]
 & - 2\, C_T \left(C_{S_R}-C_{S_L} \right) H_P \left(H_{T_+} - H_{T_-} \right)
 \,, \\[0.5em]
 \mathcal H_{10} (q^2) = & \left(C_{S_R} - C_{S_L}\right) C_T \, H_P\, H_{T_0} \,, 
\end{align}
and 
\begin{align}
 \mathcal J_1 & = (1 - \cos\theta_\ell)^2 \sin^2\theta_V \,, \\[0.5em]
 \mathcal J_2 & = (1 + \cos\theta_\ell)^2 \sin^2\theta_V \,, \\[0.5em]
 \mathcal J_3 & = +4\sin^2\theta_\ell \cos^2\theta_V \,, \\[0.5em]
 \mathcal J_4 & = +4\cos^2\theta_V \,, \\[0.5em]
 \mathcal J_5 & = +16\sin^2\theta_\ell \sin^2\theta_V \,, \\[0.5em]
 \mathcal J_6 & = +64\cos^2\theta_\ell \cos^2\theta_V \,, \\[0.5em] 
 \mathcal J_7 & = -2 \sin^2\theta_\ell \sin^2\theta_V \cos2\chi \,, \\[0.5em] 
 \mathcal J_8 & = +4 (1-\cos\theta_\ell) \sin\theta_\ell \cos\theta_V\sin\theta_V\cos\chi \,, \\[0.5em] 
 \mathcal J_9 & = -4 (1+\cos\theta_\ell) \sin\theta_\ell \cos\theta_V\sin\theta_V\cos\chi \,, \\[0.5em]
 \mathcal J_{10} & = +32 \cos\theta_\ell \cos^2\theta_V \,.
 %
 %\mathcal J_{11} & = +16\sin\theta_\ell \cos\theta_V\sin\theta_V\cos\chi \,, 
 %
\end{align}
Note that we are now considering the NP contributions with real Wilson coefficients $C_X$, and the interferences occur only between the left- and right-handed currents, and between pseudo-scalar and tensor contributions in the limit of massless leptons~\cite{Duan:2024ayo}.

%%%%%%%%%%%%%%%%%%%%%%%%%%%%%%%%%%%%%%%%%%%%%%%%%%%
\section{Form-factor parameterizations} 
\label{sec:FFdef}
%%%%%%%%%%%%%%%%%%%%%%%%%%%%%%%%%%%%%%%%%%%%%%%%%%%

In addition to the standard forms of the $\bar{B} \to \Dgen$ transition form-factors,
\begin{align}
 \big\{ \rd{f_+}, ~\rd{f_0},~ \gr{f_S}, ~\rd{f_T}, ~\rd{V}, ~\rd{A_1}, ~\gr{A_2}, ~\rd{A_0}, ~\gr{P},  ~\rd{T_1},  ~\rd{T_2},  ~\gr{T_3} \big\} \,,
\end{align}
there are several other form-factor representations depending on the specific parameterization method used. In particular, the gray parts $\{\gr{f_S}, \gr{A_2}, \gr{P}, \gr{T_3} \}$ are not used directly in any method and should be replaced by other form-factors, as will be seen later. In the following, for convenience, we summarize the relations of such representations with the standard forms defined by Eqs.~\eqref{eq:fplus-fzero-standard}~--~\eqref{eq:tensor-standard}. 

\subsection{BSZ parameterization (\rd{red})}
%%%%%%%%%%%%%%%%%%%%%%%%%%%%%

The so-called BSZ parameterization~\cite{Bharucha:2015bzk} simply introduces fit-parameters for the following set of form-factors:
\begin{align}
 \big\{ \rd{f_+}, ~\rd{f_0}, ~\rd{f_T}, ~\rd{V}, ~\rd{A_1}, ~\rd{A_{12}}, ~\rd{A_0},  ~\rd{T_1},  ~\rd{T_2},  ~\rd{T_{23}} \big\} \,, \label{eq:BSZfunc}
\end{align}
where $\rd{A_{12}}$ and $\rd{T_{23}}$ are introduced instead of $\gr{A_2}$ and $\gr{T_3}$ themselves through
\begin{align}
 \gr{A_2(q^2)} & = \frac{\Big[ \left(m_B+m_\Dst\right)^2 \left(m_B^2-m_\Dst^2-q^2\right) \rd{A_1(q^2)} - 16 m_B m_\Dst^2 \left(m_B+m_\Dst\right) \rd{A_{12}(q^2)} \Big]}{Q_+Q_-} \,, \label{eq:A12def} \\[0.5em]
 \gr{T_3(q^2)} & = \frac{\Big[ \left(m_B^2-m_\Dst^2\right) \left(m_B^2+3m_\Dst^2-q^2\right)  \rd{T_2(q^2)} - 8 m_B m_\Dst^2 \left(m_B-m_\Dst\right) \rd{T_{23}(q^2)} \Big]}{Q_+Q_-} \,. \label{eq:T23def}
\end{align}
In terms of these two derivative form-factors, the hadronic amplitudes $H_{V_0}$ and $H_{T_0}$ are simplified, respectively, as 
\begin{align}
 H_{V_0}  = - 8 \frac{m_B\,m_\Dst}{\sqrt{q^2}} \rd{A_{12}(q^2)} \,, \qquad
 H_{T_0}  = -4 \frac{m_B\,m_\Dst}{m_B+m_\Dst} \rd{T_{23}(q^2)} \,. 
\end{align}
For the scalar and pseudo-scalar form-factors, on the other hand, we just follow the relations given by Eq.~\eqref{eq:FFscalarEOM}. Then, each form-factor of Eq.~\eqref{eq:BSZfunc} is parameterized in terms of the unknown parameters $a_n^F$ as~\cite{Bharucha:2015bzk,Gubernari:2018wyi}  
\begin{align} \label{eq:BSZ-parameterization}
 \rd{F(q^2)} \equiv \frac{1}{1- q^2/M_F^2} \sum_{n=0}^{\rd{N_F}} \rd{a_n^F} 
 \left[z(q^2) - z(0)\right]^n \,, 
\end{align}
with 
\begin{align}
 & z(q^2) = \frac{\sqrt{t_+ - q^2} - \sqrt{t_+ - t_0}}{\sqrt{t_+ - q^2} + \sqrt{t_+ - t_0}} \,, \\[1em]
 & t_+ = \left(m_B+m_{D^{(*)}}\right)^2 \,, \quad\quad  t_0 =  \left(m_B+m_{D^{(*)}}\right) \left(\sqrt{m_B} - \sqrt{m_{D^{(*)}}}\right)^2 \,.
\end{align}
The mass $M_F$ of the lowest-lying resonance compatible with the quantum
numbers of the form-factor $F$ is listed as~\cite{Gubernari:2018wyi}
\begin{align}
 6.275\,\text{GeV} & = M_{A_0} \,, \\[0.5em]
 6.330\,\text{GeV} & = M_{f_+} = M_{f_T} = M_{V} = M_{T_1} \,, \\[0.5em]
 6.420\,\text{GeV} & = M_{f_0} \,, \\[0.5em] 
 6.767\,\text{GeV} & = M_{A_1} = M_{A_{12}} = M_{T_2} = M_{T_{23}} \,. 
\end{align}
The prefactor in Eq.~\eqref{eq:BSZ-parameterization} factors out the contribution of the lowest-lying pole, which improves the analytic properties of the remaining function and leads to a faster convergence of the series. The BSZ parameterization is also characterized by imposing the exact kinematic relations of the form-factors at $q^2=0$ at the level of the SSE coefficients~\cite{Bharucha:2015bzk}. In this work, we will take this simplest parameterization as a reference model. 

\subsection{BGL parameterization (\ora{orange})}
%%%%%%%%%%%%%%%%%%%%%%%%%%%%%

The BGL parameterization~\cite{Boyd:1997kz,Boyd:1995sq,Boyd:1995cf,Boyd:1994tt}, on the other hand, considers the following set of form-factors for the SM part: 
\begin{align}
 \big\{ \ora{f_+}, ~\ora{f_0}, ~\ora{g}, ~\ora{f}, ~\ora{F_1}, ~\ora{F_2} \big\} \,. \label{eq:BGLfunc}
\end{align}
They are related to the standard forms, respectively, through 
\begin{align} \label{eq:BGL1}
  \rd{f_+(q^2)} &= \ora{f_+(q^2)} \,, &
  \rd{f_0(q^2)} &= \ora{f_0(q^2)}\,, \\[1em]
  \rd{V(q^2)} &= \frac{m_B+m_{D^*}}{2} \ora{g(q^2)} \,, &
  \rd{A_1(q^2)} &= \frac{1}{m_B+m_{D^*}} \ora{f(q^2)} \,, \\[0.5em]
  \rd{A_{12}(q^2)} &= \frac{1}{8m_B m_\Dst} \ora{F_1(q^2)} \,, &
  \rd{A_0(q^2)} &= \frac{1}{2} \ora{F_2(q^2)} \,,
  \label{eq:BGL2}
\end{align}
where $\ora{f_+(q^2)}$ and $\ora{f_0(q^2)}$ are the same as in the BSZ parameterization~\cite{Bharucha:2015bzk} but with a different parameterization. Specifically, each form-factor of Eq.~\eqref{eq:BGLfunc} is now parameterized in terms of the unknown parameters $\ora{b_n^F}$ as~\cite{Boyd:1997kz,Boyd:1995sq,Boyd:1995cf,Boyd:1994tt}
\begin{align} \label{eq:BGL-parameterization-1}
 \ora{F(q^2)} \equiv  \frac{1}{P_F(z(q^2)) \,\phi_F(z(q^2))} \sum_{n=0}^{\ora{N_F}} \ora{b_n^F} z(q^2)^n \,, 
\end{align}
with 
\begin{align}
 & z(q^2) = \frac{\sqrt{t_+ - q^2} - \sqrt{t_+ - t_0}}{\sqrt{t_+ - q^2} + \sqrt{t_+ - t_0}} \,, \\[1em]
 & t_+ = (m_B+m_{D^{(*)}})^2 \,, \quad\quad  t_0 =  (m_B-m_{D^{(*)}})^2 \,.
\end{align}
The outer functions $\phi_F(z(q^2))$ in Eq.~\eqref{eq:BGL-parameterization-1} are given, respectively, as~\cite{Bigi:2016mdz,Bigi:2017njr,Bigi:2017jbd,Duan:2024ayo} 
\begin{align}
 \phi_{f_+}(z) & = 12.43 \times (1+z)^2(1-z)^{1/2} \left[(1+r_D)(1-z)+2 \sqrt{r_D}(1+z)\right]^{-5} \,,  \\[0.5em]
 \phi_{f_0}(z) & = 10.11 \times (1+z)(1-z)^{3/2} \left[(1+r_D)(1-z)+2 \sqrt{r_D}(1+z)\right]^{-4} \,, \\[0.5em]
 \phi_g(z) & = 53.79 \times (1+z)^2(1-z)^{-1/2} \left[(1+r_\Dst)(1-z)+2 \sqrt{r_\Dst}(1+z)\right]^{-4}\,\mathrm{GeV} \,, \\[0.5em]
 \phi_f(z) & = 1.454 \times (1+z)(1-z)^{{3/2}} \left[(1+r_\Dst)(1-z)+2 \sqrt{r_\Dst}(1+z)\right]^{-4}\,\mathrm{GeV}^{-1} \,, \\[0.5em]
 \phi_{F_1}(z) & = 0.195 \times (1+z)(1-z)^{{5/2}} \left[(1+r_\Dst)(1-z)+2 \sqrt{r_\Dst}(1+z)\right]^{-5}\,\mathrm{GeV}^{-1} \,, \\[0.5em]
 \phi_{F_2}(z) & = 10.71 \times (1+z)^2 (1-z)^{-1/2} \left[(1+r_\Dst)(1-z)+2 \sqrt{r_\Dst}(1+z)\right]^{-4} \,, 
\end{align}
with $r_D \equiv m_{D}/m_B$ and $r_\Dst \equiv m_{\Dst}/m_B$, while the Blaschke products $P_F(z(q^2))$ are defined by~\cite{Boyd:1997kz,Boyd:1995sq,Boyd:1995cf,Boyd:1994tt}
\begin{align}
 P_F(z) = \prod_n \frac{z - z_n^F}{1 -z\,z_n^F} \,, 
\end{align} 
with 
\begin{align}
 &z_1^{f_0} = -0.433 \,,& &z_2^{f_0} = -0.819 \,,  \\[0.5em]
 & z_1^{f_+} = -0.308 \,,& &z_2^{f_+} = -0.555 \,,& &z_3^{f_+} = -0.646 \,,  \\[0.5em]
 & z_1^{g} = -0.286 \,,&  &z_2^{g} = -0.479 \,,&  &z_3^{g} = -0.537 \,,&  &z_4^{g} = -0.890 \,, \\[0.5em]
 & z_1^{f,F_1} = -0.402 \,,&  &z_2^{f,F_1} = -0.406 \,,& &z_3^{f,F_1} = -0.637 \,,&  &z_4^{f,F_1} = -0.642 \,, \\[0.5em]
 & z_1^{F_2} = -0.274 \,,& &z_2^{F_2} = -0.443 \,,&  &z_3^{F_2} = -0.791 \,.
\end{align} 
For further detailed descriptions of the prefactors $P_F(z)$ and $\phi_F(z)$, we refer the readers to Ref.~\cite{Duan:2024ayo} and the references therein. 

In this paper, we simply consider the following two cases for the BGL parameterization:
\begin{itemize}
 \item SM--BSZ \& Tensor--BSZ, 
 \item SM--BGL \& Tensor--BSZ.
\end{itemize}
We call the latter the ``BGL'' for simplicity. 
The BGL parameterization has the practical advantage of a compact expansion in the conformal variable $z(q^2)$ and provides rigorous theoretical control over the expansion coefficients $b_n^F$ within its domain of validity, as will be explained in section~\ref{sec:setup}. Nevertheless, one important limitation is that it implicitly neglects sub-threshold branch cuts~\cite{Gubernari:2026sqc,Simula:2025lpc,Simula:2025fft,Gopal:2024mgb}.

\subsection{HQET parameterization (\bl{blue})}
%%%%%%%%%%%%%%%%%%%%%%%%%%%%%

We have another way to describe the $\bar{B} \to D^{(*)}$ transition form-factors based on HQET~\cite{Manohar:2000dt,Neubert:1993mb}, with which all the form-factors can be related to each other via heavy-quark symmetry, augmented by higher-order perturbative and power corrections~\cite{Bernlochner:2017jka,Bigi:2017jbd,Jaiswal:2017rve,Jung:2018lfu,Bordone:2019vic,Bordone:2019guc,Bernlochner:2022ywh,Bordone:2025jur}.

\subsubsection{HQET forms}
%%%%%%%%%%%%%%%%%%%%%%%%%%%%%

In HQET, the $\bar{B} \to D^{(*)}$ transition form-factors are defined, respectively, by~\cite{Bernlochner:2022ywh,Sakaki:2013bfa}\footnote{The form-factor $h_{T_3}$ in Eq.~\eqref{eq:hT3def} is introduced in Ref.~\cite{Bernlochner:2022ywh}, in which the higher-order power and perturbative corrections $\delta \hat{h}_{X}^{(Y)}$ are evaluated. It is different from $h'_{T_3}$ defined in Ref.~\cite{Sakaki:2013bfa}, in which only the leading IW is taken and $h'_{T_3}\to 0$ in this limit. To be precise, we can see from the identity $\varepsilon^{\mu\nu\rho\sigma} (v+v')_\rho (v-v')_\sigma h'_{T_3}(w) = \varepsilon^{\mu\nu\rho\sigma} v_\rho v'_\sigma h_{T_3}(w)$ that $h'_{T_3}(w) = - \frac{1}{2} h_{T_3}$.} 
\begin{align}
 \langle D | \bar c \gamma^\mu b | \bar{B} \rangle_\text{HQET} 
 & = \sqrt{m_Bm_D} \big[ (v+v^\prime)^\mu \cya{h_+(w)} + (v-v^\prime)^\mu \cya{h_-(w)} \big] \,, \\[1em]
 \langle D |\bar c b| \bar{B} \rangle_\text{HQET}
 & = \sqrt{m_Bm_D} (w+1) \cya{h_S(w)} \,, \\[1em]
 \langle D |\bar c \sigma^{\mu\nu} b| \bar{B} \rangle_\text{HQET}
 & = -i \sqrt{m_Bm_D} \big[ v^\mu v^{\prime\nu} - v^{\prime\mu} v^\nu \big] \cya{h_T(w)} \,, \\[1em]
 \langle D^* | \bar c \gamma^\mu b | \bar{B} \rangle_\text{HQET} 
 & = i \sqrt{m_Bm_\Dst} \varepsilon^{\mu\nu\rho\sigma} \epsilon^*_\nu v'_\rho v_\sigma \cya{h_V(w)} \,, \\[1em]
 \langle D^* | \bar c \gamma^\mu \gamma^5 b | \bar{B} \rangle_\text{HQET} 
 & = \sqrt{m_Bm_\Dst} \big[ (w+1) \epsilon^{*\mu} \cya{h_{A_1} (w)} - (\epsilon^* \cdot v) \left( v^\mu \cya{h_{A_2}(w)} + v^{\prime\mu} \cya{h_{A_3}(w)} \right) \big] \,, \notag \\[1em]
  \langle D^* | \bar c \gamma^5 b | \bar{B} \rangle_\text{HQET} 
 & = -\sqrt{m_Bm_\Dst} (\epsilon^* \cdot v) \cya{h_P(w)} \,, \\[1em]
 \langle D^* | \bar c \sigma^{\mu\nu} b | \bar{B} \rangle_\text{HQET} 
 & = -\sqrt{m_Bm_\Dst} \varepsilon^{\mu\nu\rho\sigma} \big[ \epsilon^{*}_\rho (v+v')_\sigma \cya{h_{T_1}(w)} + \epsilon^{*}_\rho (v-v')_\sigma \cya{h_{T_2}(w)} \notag \\[0.2em]
 & \hspace{9em} + (\epsilon^* \cdot v) v_\rho v'_\sigma \cya{h_{T_3}(w)} \big] \label{eq:hT3def}
 \,, 
\end{align} 
with
\begin{align} \label{eq:w-q2-relation}
 v^\mu = \frac{p_B^\mu}{m_B} \,, 
 \qquad 
 v^{\prime\mu} = \frac{p_{\Dgen}^\mu}{m_{\Dgen}} \,, 
 \qquad 
 w =v \cdot v' = \frac{m_B^2+m_{\Dgen}^2-q^2}{2m_Bm_{\Dgen}} \,. 
\end{align} 
The form-factors $\cya{h_X}$ are not directly used for parameterization. In HQET, they are instead described as~\cite{Bernlochner:2022ywh,Bernlochner:2017jka}
\begin{align}
 \cya{h_X(w)} = \bl{\xi(w)} \hat h_X(w) \,, 
\end{align} 
with 
\begin{align} \label{eq:HQET-hat-hX}
 \hat{h}_X
 =
 \hat{h}_{X}^{(0)}+\frac{\alpha_s}{\pi} \delta \hat{h}_{X}^{(\alpha_s)} 
 +\frac{\bar{\Lambda}}{2 m_b} \cya{\delta \hat{h}_{X}^{(m_b)}}
 +\frac{\bar{\Lambda}}{2 m_c} \cya{\delta \hat{h}_{X}^{(m_c)}}
 + \frac{\alpha_s}{\pi} \cdot \frac{\bar{\Lambda}}{2 m_c} \cya{\delta \hat{h}_{X}^{(\alpha_sm_c)}} 
 +\left(\frac{\bar{\Lambda}}{2 m_c}\right)^2 \cya{\delta \hat{h}_{X}^{(m_c^2)}} \,, 
\end{align}
and
\begin{align}
 \hat h_{X}^{(0)} 
 & = \begin{cases} 1 & \text{for}~X = +,V,A_1,A_3,S,P,T,T_1 \\[0.5em]
 0 & \text{for}~X = -,A_2,T_2,T_3 \end{cases} \,, 
\end{align} 
where $\bl{\xi(w)}$ is referred to as the leading IW function~\cite{Isgur:1989vq}, whereas $\delta \hat{h}_{X}^{(Y)}$ account for higher-order corrections in $\alpha_s$ and $1/m_{b,c}$ expansions. This form indicates that all the $\bar{B} \to D^{(*)}$ transition form-factors are represented by one unknown IW function $\bl{\xi(w)}$ in the exact heavy-quark limit (corresponding to $1/m_{b,c} \to 0$).

The power correction terms of $\cya{\delta \hat{h}_{X}^{(m_b)}}$, $\cya{\delta \hat{h}_{X}^{(m_c)}}$, and $\cya{\delta \hat{h}_{X}^{(m_c^2)}}$ include additional unknown functions, the so-called subleading and subsubleading IW functions, described as~\cite{Falk:1992wt,Bernlochner:2017jka,Iguro:2020cpg}
\begin{align}
 \cya{\dhh_{+}^{(m_b)}} = 
 \cya{\dhh_{+}^{(m_c)}} = 
 \cya{\dhh_{T_1}^{(m_b)}} = 
 -4(w-1) \bl{\hat\chi_2(w)} + 12 \bl{\hat\chi_3(w)} \,, 
\end{align}
\begin{align}
 \cya{\dhh_{-}^{(m_b)}} = 
 -\cya{\dhh_{-}^{(m_c)}} =
 \cya{\dhh_{T_2}^{(m_b)}} =
 1 - 2\bl{\eta(w)} \,, 
\end{align}
\begin{align}
 \cya{\dhh_{V}^{(m_b)}} 
  = &~ \cya{\dhh_{A_3}^{(m_b)}} = \cya{\dhh_{P}^{(m_b)}} = \cya{\dhh_{T}^{(m_b)}} = \cya{\dhh_{T}^{(m_c)}}  
  =  1 - 2 \bl{\eta(w)} -4(w-1) \bl{\hat\chi_2(w)} + 12 \bl{\hat\chi_3(w)} \,, \notag
\end{align}
\begin{align}
 \cya{\dhh_{V}^{(m_c)}} = 
 & 1 - 4 \bl{\hat\chi_3(w)} \,, 
\end{align}
\begin{align}
 \cya{\dhh_{A_1}^{(m_b)}} 
 = &~ \cya{\dhh_{S}^{(m_b)}} = \cya{\dhh_{S}^{(m_c)}} 
 = (w-1) \left[ (w+1)^{-1} \big( 1-2\bl{\eta(w)} \big) -4 \bl{\hat\chi_2(w)} \right] +12 \bl{\hat\chi_3(w)}  \,, 
\end{align}
\begin{align}
 \cya{\dhh_{A_1}^{(m_c)}} 
 = \, (w-1) (w+1)^{-1} -4\bl{\hat\chi_3(w)}  \,, \qquad 
 \cya{\dhh_{A_2}^{(m_b)}} = \, \cya{\dhh_{T_3}^{(m_b)}} = 0 \,, 
\end{align}
\begin{align}
 \cya{\dhh_{A_2}^{(m_c)}} = 
 & -2(w+1)^{-1} \big( 1 + \bl{\eta(w)} \big) + 4 \bl{\hat\chi_2(w)}  \,, 
\end{align}
\begin{align}
 \cya{\dhh_{A_3}^{(m_c)}} = 
 &\, 1 - 2 (w+1)^{-1} \big( 1 + \bl{\eta(w)} \big) -4 \bl{\hat\chi_2(w)} -4\bl{\hat\chi_3(w)} \,, 
\end{align}
\begin{align}
 \cya{\dhh_{P}^{(m_c)}} = 
 &\, - 1 + 2 \big( 1+\bl{\eta(w)} \big) +4(w-1) \bl{\hat\chi_2(w)} - 4 \bl{\hat\chi_3(w)} \,, 
\end{align}
\begin{align}
 \cya{\dhh_{T_1}^{(m_c)}} = -4 \bl{\hat\chi_3(w)} \,, \quad
 \cya{\dhh_{T_2}^{(m_c)}} = -1 \,, \quad
 \cya{\dhh_{T_3}^{(m_c)}} = (w+1)^{-1} \big( 1 + \bl{\eta(w)} \big) + 2 \bl{\hat\chi_2(w)} \,, 
\end{align}
\begin{align} \label{eq:subsubIW1}
 \cya{\dhh_{+}^{(m_c^2)}} & = \bl{\hat\ell_1 (w)} \,, &
 \cya{\dhh_{-}^{(m_c^2)}} & = \bl{\hat\ell_4 (w)} \,, \\[1em]
 \cya{\dhh_{V}^{(m_c^2)}} & = \bl{\hat\ell_2 (w)} - \bl{\hat\ell_5 (w)} \,, &
 \cya{\dhh_{A_1}^{(m_c^2)}} & = \bl{\hat\ell_2 (w)} - \frac{w-1}{w+1} \bl{\hat\ell_5 (w)} \,, \\[1em]
 \cya{\dhh_{A_2}^{(m_c^2)}} & = \bl{\hat\ell_3 (w)} + \bl{\hat\ell_6 (w)} \,, &
 \cya{\dhh_{A_3}^{(m_c^2)}} & = \bl{\hat\ell_2 (w)} - \bl{\hat\ell_3 (w)} - \bl{\hat\ell_5 (w)} + \bl{\hat\ell_6 (w)}  \,, 
\end{align}
\begin{align}
 \cya{\dhh_{S}^{(m_c^2)}} & = \bl{\hat\ell_1 (w)} - \frac{w-1}{w+1} \bl{\hat\ell_4 (w)} \,, \\[1em]
 \cya{\dhh_{P}^{(m_c^2)}} & = \bl{\hat\ell_2 (w)} + (w-1) \bl{\hat\ell_3 (w)} + \bl{\hat\ell_5 (w)} - (w+1) \bl{\hat\ell_6 (w)} \,, \label{eq:mod4}
\end{align}
\begin{align}
 \cya{\dhh_{T}^{(m_c^2)}} & = \bl{\hat\ell_1 (w)} - \bl{\hat\ell_4 (w)} \,, &
 \cya{\dhh_{T_1}^{(m_c^2)}} & = \bl{\hat\ell_2 (w)} \,, \\[1em]
 \cya{\dhh_{T_2}^{(m_c^2)}} & = \bl{\hat\ell_5 (w)} \,, &
 \cya{\dhh_{T_3}^{(m_c^2)}} & = \frac{1}{2} \big( \bl{\hat\ell_3 (w)} - \bl{\hat\ell_6 (w)} \big) \,,
 \label{eq:subsubIW2}
\end{align}
where $\hat\chi_i(w) = \chi_i(w)/\xi(w)$ and $\hat\ell_i(w) = \ell_i(w)/\xi(w)$. 
The one-loop $\mathcal{O}(\alpha_s)$ corrections $\delta \hat{h}_{X}^{(\alpha_s)}$ result from matching QCD onto HQET for the heavy-quark currents~\cite{Falk:1990yz,Falk:1990cz,Neubert:1992qq,Bernlochner:2017jka}, which do not include additional IW functions. 
Finally, we also implement the $\alpha_s/m_c$ corrections $\delta \hat{h}_{X}^{(\alpha_sm_c)}$ obtained recently in Ref.~\cite{Bernlochner:2022ywh}, which are functions of $\eta(w)$, $\hat\chi_2(w)$, and $\hat\chi_3(w)$. 
For convenience, we collect in appendix~\ref{App:FFinput} the explicit expressions of $\delta \hat{h}_{X}^{(\alpha_s)}$ and $\delta \hat{h}_{X}^{(\alpha_sm_c)}$. 
Other corrections are expected to be sufficiently suppressed due to the following power counting within HQET: 
\begin{align}
 \left(\frac{\bar{\Lambda}}{2m_b}\right)^2 
 ~~\ll~~ \frac{\alpha_s}{\pi} \cdot \frac{\bar{\Lambda}}{2m_c} 
 ~~<~~ \left(\frac{\bar{\Lambda}}{2m_c} \right)^2
  ~~\lesssim~~ \frac{\bar{\Lambda}}{2m_b} 
  ~~\lesssim~~ \frac{\alpha_s}{\pi} < \frac{\bar{\Lambda}}{2m_c} \,. 
\end{align}
Thus, we have ten unknown IW functions in total, and the HQET form-factors $\cya{h_X}$ are not independent of each other but controlled by these IW functions. 

\subsubsection{Relations to the standard forms}
%%%%%%%%%%%%%%%%%%%%%%%%%%%%%

The HQET form-factors $\cya{h_X}$ are related to the standard forms defined by Eqs.~\eqref{eq:fplus-fzero-standard}~--~\eqref{eq:tensor-standard}, respectively, as 
\begin{align}
 \label{eq:HQETfp}
 \rd{f_+(q^2)} & = \frac{1}{2 \sqrt{\rD}} \Big[ \left(1+r_D\right) \cya{h_+ (w)} - \left(1-r_D\right) \cya{h_-(w)} \Big] \,, \\[0.5em]
 \rd{f_0(q^2)} & = \sqrt{\rD} \left[ \frac{w + 1}{1+\rD} \cya{h_+ (w)} - \frac{w - 1}{1-\rD} \cya{h_- (w)} \right] \,,  \\[0.5em] 
 \gr{f_S(q^2)} & = \frac{\sqrt{\rD}}{1+\rD} (w+1) \cya{h_S(w)} \,, \\[0.5em]
 \rd{f_T(q^2)} & = \frac{1+\rD}{2 \sqrt{\rD}} \cya{h_T (w)} \,, \\[0.5em]
 \rd{A_1(q^2)} & = \frac{\sqrt{\rDst}}{1+\rDst} (w+1) \cya{h_{A_1} (w)} \,, \label{eq:A1-hA1} \\[0.5em] 
 \gr{A_2(q^2)} &= \frac{1+\rDst}{2 \sqrt{\rDst}} \Big[\rDst \cya{h_{A_2}(w)} + \cya{h_{A_3}(w)} \Big] \,, \\[0.5em]
 \rd{A_{12}(q^2)} & = - \frac{w+1}{8 \sqrt{\rDst}} \Big[ (\rDst - w) \cya{h_{A_1}(w)} +  (w-1) \left( \rDst\cya{h_{A_2}(w)} +  \cya{h_{A_3}(w)} \right) \Big] \,, \label{eq:A12-hA12} \\[0.5em]
 \rd{A_0(q^2)} & = \frac{1}{2 \sqrt{\rDst}} \Big[ (w+1) \cya{h_{A_1} (w)} + (w\, \rDst - 1) \cya{h_{A_2} (w)} +(\rDst -w) \cya{h_{A_3} (w)} \Big] \,, \\[0.5em]
 \rd{V(q^2)} & = \frac{1+r_\Dst}{2\sqrt{r_\Dst}} \cya{h_V (w)} \,, \\[0.5em]
 \gr{P(q^2)} & = - \sqrt{r_\Dst} \cya{h_P(w)} \,,  \\[0.5em]
 \rd{T_1(q^2)} & = \frac{1}{2\sqrt{r_\Dst}} \Big[ (1+r_\Dst) \cya{h_{T_1} (w)} - (1-r_\Dst) \cya{h_{T_2} (w)} \Big] \,, \\[0.5em]
 \rd{T_2(q^2)} & = \sqrt{\rDst} \left[ \frac{w+1}{1+\rDst} \cya{h_{T_1} (w)} - \frac{w-1}{1-\rDst} \cya{h_{T_2} (w)} \right] \,, \label{eq:HQETT2}\\[0.5em]
 \gr{T_3(q^2)} & = \frac{1}{2\sqrt{r_\Dst}} \Big[ (1-r_\Dst) \cya{h_{T_1} (w)} - (1+r_\Dst) \cya{h_{T_2} (w)} + (1-r_\Dst^2) \cya{h_{T_3} (w)} \Big] \,, \\[0.5em]
 \rd{T_{23}(w)} & = \frac{1+r_\Dst}{4 \sqrt{r_\Dst}} \Big[ (w+1) \cya{h_{T_1} (w)} + (w-1) \cya{h_{T_2} (w)} - (w^2-1) \cya{h_{T_3} (w)} \Big] \,. \label{eq:HQETT23}
\end{align} 
With the aid of these relations, we can rewrite the hadronic amplitudes $H_X$ in terms of the HQET form-factors $\cya{h_Y}$. 

\subsubsection{Parameterization of IW functions}
%%%%%%%%%%%%%%%%%%%%%%%%%%%%%

The HQET parameterization has been proposed so that the ten IW functions 
\begin{align}
 \big\{ \bl{\xi(w)},\, \bl{\eta(w)},\, \bl{\hat\chi_{2,3}(w)},\, \bl{\hat\ell_{1,2,3,4,5,6}(w)} \big\} \,, \label{eq:HQETfunc}
\end{align}
are directly parameterized. Here we follow the parameterization first proposed in Ref.~\cite{Bordone:2019vic} and later used in the literature (see, \textit{e.g.}, Refs.~\cite{Iguro:2020cpg,Bordone:2025jur}). This parameterization is based on the observation that, in the complex half plane of $\mathrm{Re}(w) \geq 1$, the HQET form-factors $h_X$, the matching coefficients $C_X$, and the ten IW functions given by Eq.~\eqref{eq:HQETfunc} are free of singularities due to QCD dynamics, while singularities of kinematic origin can always be removed by redefining the form-factors (c.f. section~\ref{Sec:kinematic} for details). Consequently, we can expand the ten IW functions in Eq.~\eqref{eq:HQETfunc} around $w = 1$ as~\cite{Bordone:2019vic,Caprini:1997mu}
\begin{align}
 \xi(w) &= \bl{c_0^{\xi} } + \bl{c_1^{\xi} } (w-1) + \frac{1}{2} \bl{c_2^{\xi} } (w-1)^2 + \frac{1}{6} \bl{c_3^{\xi} } (w-1)^3 \,, \label{eq:LO-IW-parameterization} \\[0.5em]
f(w) &= \bl{c_0^{f} } + \bl{c_1^{f} } (w-1) + \frac{1}{2} \bl{c_2^{f} } (w-1)^2 \, \quad ( f = \eta,\, \hat\chi_2,\, \hat\chi_3 )\,, \label{eq:NLO-IW-parameterization}\\[0.5em]
\hat\ell_n(w) &= \bl{c_0^{\ell_n} } + \bl{c_1^{\ell_n} } (w-1) \, \quad (n = 1, \cdots, 6) \,, \label{eq:NNLO-IW-parameterization}
\end{align}
for the leading, subleading, and subsubleading IW functions, respectively. Furthermore, the HQET property indicates that $\bl{c_0^{\xi}}=1$ and $\bl{c_0^{\chi_3}}=0$~\cite{Neubert:1993mb,Luke:1990eg}. We name the parameterization given by Eqs.~\eqref{eq:LO-IW-parameterization}~--~\eqref{eq:NNLO-IW-parameterization} as HQET $(3/2/1)$, where $3/2/1$ refers to the order up to which the leading/subleading/subsubleading IW functions are expanded in $w$ around $w = 1$. For comparison, we also consider the case of HQET $(2/1/0)$ obtained by taking $\bl{c_3^{\xi}}=\bl{c_2^{\eta}}=\bl{c_2^{\chi_2}}=\bl{c_2^{\chi_3}}=\bl{c_1^{\ell_n}}=0$. Note that these two form-factor modelings contain $23$ and $13$ free parameters to be fitted, respectively. 

Regarding the numerical input for the heavy-quark expansion, we take the matching scale as $\mu_b = 4.2\,\text{GeV}$ and fix~\cite{Iguro:2020cpg}  
\begin{align}
 \frac{\alpha_s}{\pi} = 0.0716 \,,\qquad 
 \frac{\bar{\Lambda}}{2m_b} = 0.0522 \,,\qquad 
 \frac{\bar{\Lambda}}{2m_c} = 0.1807 \,,\qquad 
 z_{cb} = \frac{m_c}{m_b} = 0.2890 \,,
\end{align}
which means that we keep these values during our fit analysis and do not consider uncertainties arising from these inputs.

%%%%%%%%%%%%%%%%%%%%%%%%%%%%%
\subsection{Kinematic constraints}
\label{Sec:kinematic}
%%%%%%%%%%%%%%%%%%%%%%%%%%%%%

Here we summarize possible constraints on the $\bar{B} \to \Dgen$ transition form-factors from the kinematic structures of the semi-leptonic $\bar{B} \to \Dgen \ell\bar\nu$ decays at the endpoints $q^2_\text{max} = (m_B-m_\Dgen)^2$ and $q^2 = 0$. 
Usually, $q^2_\text{max}$ ($q^2 = 0$) is called the zero (maximal) recoil point in the sense that the emitted lepton pair ($\ell,\bar\nu$) carries the largest (smallest) energy. 

Getting back to the form-factors $\rd{A_{12}(q^2)}$ and $\rd{T_{23}(q^2)}$ introduced in Eqs.~\eqref{eq:A12def} and \eqref{eq:T23def}, one can find the endpoint relations 
\begin{align}
 & \rd{A_1(q^2_\text{max})} = \frac{8 m_B m_\Dst}{m_B^2-m_\Dst^2} \rd{A_{12}(q^2_\text{max})} \,,  \\[1em]
 & \rd{T_2(q^2_\text{max})} = \frac{4 m_B m_\Dst}{(m_B+m_\Dst)^2} \rd{T_{23}(q^2_\text{max})} \,,
\end{align}
because of $Q_- = (m_B-m_\Dst)^2 - q^2_\text{max} = 0$ and the regularity of the form-factors $\gr{A_2(q^2)}$ and $\gr{T_3(q^2)}$ at $q^2 = q^2_\text{max}$. These two relations are also in accordance with the symmetries of the helicity amplitudes at the kinematic endpoint~\cite{Hiller:2013cza,Bharucha:2015bzk}. 

On the other hand, getting back to the definitions of the form-factors $\rd{f_{0,+}(q^2)}$, $\rd{T_{1,2}(q^2)}$, and $A_3(q^2)$, one finds that they must become zero at $q^2 \to 0$ to avoid the $1/q^2$ pole for the decay distributions. This requirement leads to the following three endpoint relations:
\begin{align}
 \rd{f_0(0)} =  \rd{f_+(0)} \,, \qquad \rd{T_1(0)} = \rd{T_2(0)} \,, \qquad \quad A_3(0) = 0 \,, 
\end{align}
where the last one can also be represented as 
\begin{align}
 \rd{A_0(0)} = \frac{8 m_B m_\Dst}{m_B^2-m_\Dst^2} \rd{A_{12}(0)} \,.
\end{align}

The exact kinematic endpoint conditions for the form-factors derived above imply further relations among the form-factor parameters for each parameterization method. They are explicitly summarized below.

\subsubsection*{\uline{BSZ}}
%%%%%%%%%%%%%%%%%%%%%%%%%%%%%
Up to the expansion order $N_F=2$, we find that  
\begin{align}
 & \rd{f_0(0)} = \rd{f_+(0)} &   &\Rightarrow&  &\rd{a^{f_0}_0} = \rd{a^{f_+}_0}&  &\text{($N_F$ independent)} \,,& \\[1em]
 &\rd{T_1(0)} =  \rd{T_2(0)}& &\Rightarrow&  &\rd{a^{T_2}_0} = \rd{a^{T_1}_0}& &\text{($N_F$ independent)} \,,& \\[1em]
 & \rd{A_0(0)} = \frac{8m_Bm_\Dst}{m_B^2-m_\Dst^2} \rd{A_{12}(0)}& &\Rightarrow&  &\rd{a^{A_0}_0} = 3.562520 \rd{a^{A_{12}}_0}& &\text{($N_F$ independent)}\,,&
\end{align}
\begin{align}
 & \rd{A_1(q^2_\text{max})} = \frac{8m_Bm_\Dst}{m_B^2-m_\Dst^2} \rd{A_{12}(q^2_\text{max})}  \\[0.5em]
 & \Rightarrow~ \rd{a^{A_{12}}_0} =  0.2807 \rd{a^{A_1}_0} - 0.015755 \rd{a^{A_1}_1} + 0.000884 \rd{a^{A_1}_2} + 0.056127 \rd{a^{A_{12}}_1} -  0.003150 \rd{a^{A_{12}}_2} \,, \notag \\[1.5em]
 & \rd{T_2(q^2_\text{max})} = \frac{4m_Bm_\Dst}{(m_B+m_\Dst)^2} \rd{T_{23}(q^2_\text{max})} \\[0.5em]
 & \Rightarrow ~ \rd{a^{T_{23}}_0} = 1.25178 \rd{a^{T_2}_0} - 0.070258 \rd{a^{T_2}_1} + 0.003943 \rd{a^{T_2}_2} + 0.056127 \rd{a^{T_{23}}_1} -  0.003150 \rd{a^{T_{23}}_2} \,. \notag 
\end{align}
This means that the form-factor parameters $\{ \rd{a^{f_0}_0}, \rd{a^{A_{12}}_0}, \rd{a^{A_0}_0}, \rd{a^{T_2}_0}, \rd{a^{T_{23}}_0} \}$ are not taken as free in the fit but fixed through these relations. 

\subsubsection*{\uline{BGL} }
%%%%%%%%%%%%%%%%%%%%%%%%%%%%%
In this case and also up to the expansion order $N_F=2$, we find that 
\begin{align}
 & \ora{f_0(0)} = \ora{f_+(0)}  \\[0.5em]
 &\Rightarrow~~ \ora{b^{f_0}_0} = -0.064441 \ora{b^{f_0}_1} - 0.004153 \ora{b^{f_0}_2} + 4.93792 \ora{b^{f_+}_0} + 0.318205 \ora{b^{f_+}_1} + 0.0205054 \ora{b^{f_+}_2} \,, \notag \\[1em]
 & \rd{A_0(0)} = \frac{8m_Bm_\Dst}{m_B^2-m_\Dst^2} \rd{A_{12}(0)}  \\[0.5em]
 &\Rightarrow~~ \ora{b^{F_1}_0} = -0.056082 \ora{b^{F_1}_1} - 0.003145 \ora{b^{F_1}_2} + 0.047004 \ora{b^{F_2}_0} + 0.002636 \ora{b^{F_2}_1} + 0.000148 \ora{b^{F_2}_2} \,, \notag \\[1em]
 & \rd{A_1(q^2_\text{max})} = \frac{8m_Bm_\Dst}{m_B^2-m_\Dst^2} \rd{A_{12}(q^2_\text{max})} \\[0.5em]
 &\Rightarrow~~  \ora{b^{F_1}_0} = 0.167455 \ora{b^f_0} \qquad \text{($N_F$ independent)} \,. \notag 
\end{align}
Recall that the tensor form-factors take the same forms as in the BSZ parameterization even when we take the ``BGL'' case. This means that $\{ \ora{b^{f_0}_0}, \ora{b^{F_{1}}_0}, \ora{b^{F_2}_0}, \ora{b^{T_1}_0}, \ora{b^{T_{23}}_0} \}$ are already fixed by these relations and should not be treated as free parameters in the fit. 

\subsubsection*{\uline{HQET} }
%%%%%%%%%%%%%%%%%%%%%%%%%%%%%
From Eq.~\eqref{eq:w-q2-relation}, we can see that 
\begin{align}
 q^2_\text{max} = (m_B-m_\Dst)^2 ~~\Rightarrow~~ w_\text{min}=1 \,, 
 \qquad q^2 = 0 ~~\Rightarrow~~ w_\text{max}=1.50356 \,. 
\end{align}
At the zero-recoil point $w_\text{min}=1$, we can derive from Eqs.~\eqref{eq:A1-hA1} and \eqref{eq:A12-hA12} as well as \eqref{eq:HQETT2} and \eqref{eq:HQETT23} that
\begin{align}
 & \rd{A_1(w_\text{min})} = \frac{8m_Bm_\Dst}{m_B^2-m_\Dst^2} \rd{A_{12}(w_\text{min})} = 0.867151 + 0.0291845 \bl{c^{\ell_2}_0} \,, \\[1em]
 & \rd{T_2(w_\text{min})} = \frac{4m_Bm_\Dst}{(m_B+m_\Dst)^2} \rd{T_{23}(w_\text{min})} = 0.888682 + 0.0291845 \bl{c^{\ell_2}_0} \,, 
\end{align}
which means that these kinematic endpoint conditions are automatically satisfied in this parameterization. On the other hand, we find that the kinematic conditions at the maximal recoil point $w_\text{max}$ are not such automatic but lead to non-trivial combinations of the relations among the free parameters $\bl{c_n^F}$. Indeed, it is hard to perform a fit analysis under such a constraint. Although we do not take into account such constraints to reduce the number of the free parameters, we introduce   
\begin{align}
 \delta_f & = f_0(w_\text{max}) - f_+(w_\text{max}) \,, \\[1em]
 \delta_A &= A_0(w_\text{max}) - \frac{8m_Bm_\Dst}{m_B^2-m_\Dst^2} A_{12}(w_\text{max}) \,, \\[1em]
 \delta_T &= T_1(w_\text{max}) - T_2(w_\text{max}) \,, 
\end{align}
to parameterize the deviations from the exact kinematic endpoint conditions. We have checked that our results obtained in this work satisfy $\delta_X \approx 0$ at least at the $5\%$ level. 

\subsection{Summary of form-factor parameterizations}
%%%%%%%%%%%%%%%%%%%%%%%%%%%%%

As a summary, we will consider the following form-factors and their parameterizations:
\begin{itemize}
    \item \textbf{BSZ}: it is applied to the standard forms of the $\bar{B} \to \Dgen$ transition form-factors, $F \in \{\rd{f_+},\, \rd{f_0},\, \rd{f_T},\, \rd{A_1},\, \rd{A_2},\, \rd{A_0},\, \rd{V},\, \rd{T_1},\, \rd{T_2},\, \rd{T_{23}}\}$. The parameterization is defined as
    \begin{align}
        \rd{F(q^2)} \equiv \frac{1}{1- q^2/M_F^2} \sum_{n=0}^{\rd{N_F}} \rd{a_n^F} \left[z(q^2) - z(0) \right]^n \,. 
    \end{align}
    Concerning the kinematic range of $-0.03 \lesssim z(q^2) \lesssim 0.03$, we expect that the two cases with $N_F=1$ and $N_F=2$ for all $F$ are sufficient, 
    which include $9$ and $15$ free form-factor parameters for SM $+$ NP but without the tensor contribution, respectively (after taking into account the kinematic endpoint constraints). For SM $+$ Tensor, the form-factor model introduces additional $6$ and $10$ free parameters for the two cases, respectively. We treat this parameterization as a reference form-factor model to be compared. 
    
    \item \textbf{BGL}: it is applied to the BGL forms with $F \in \{\ora{f_+},\, \ora{f_0},\, \, \ora{f},\, \ora{F_1},\, \ora{F_2},\, \ora{g}\}$. The parameterization is now given as 
    \begin{align}
        \ora{F(q^2)} \equiv  \frac{1}{P_F(z(q^2)) \,\phi_F(z(q^2))} \sum_{n=0}^{\ora{N_F}} \ora{b_n^F} z(q^2)^n \,, 
    \end{align}
    while the tensor form-factors take the same forms as in the BSZ parameterization. 
    In this case, the kinematic range is $0 \lesssim z(q^2) \lesssim 0.06$.
    Hence, we also consider the two cases with $N_F=1$ and $N_F=2$ for all $F$, including the same number of free form-factor parameters as in the BSZ case. 
    It is noted that the BGL parameterization has been employed for the official determination of the CKM matrix element $|V_{cb}|$, where several sets of $N_F$ have been surveyed~\cite{HFLAV:2024ctg,ParticleDataGroup:2024cfk}. We will get back to this point later in our numerical discussions. 

    \item \textbf{HQET}: it is applied to the IW functions $F \in \{\bl{\xi},\, \bl{\eta},\, \bl{\hat\chi_2},\, \bl{\hat\chi_3},\, \bl{\hat\ell_1},\, \bl{\hat\ell_2},\, \bl{\hat\ell_3},\, \bl{\hat\ell_4},\, \bl{\hat\ell_5},\, \bl{\hat\ell_6}\}$. The parameterization is introduced such that 
    \begin{align}
        \bl{F(w)} = \sum_{n=0}^{\bl{N_F}} \frac{1}{n!} \bl{c_n^F} (w-1)^n \,,
    \end{align}
    for each $F$. We consider the two cases with $N_F$ specified as HQET $(3/2/1)$ and HQET $(2/1/0)$, see below Eq.~\eqref{eq:NNLO-IW-parameterization}. They include $23$ and $13$ fitting parameters, respectively.
    The form-factors are related with each other unlike BSZ and BGL.
\end{itemize} 

It should be noted that, although these parameterizations are widely used in phenomenological analyses, they do not all rely on the same amount of theory input nor provide the same level of control over truncation uncertainties, as reviewed in Refs.~\cite{Gambino:2020jvv,Gubernari:2026sqc}.

%%%%%%%%%%%%%%%%%%%%%%%%%%%%%%%%%%%%%%%%%%%%%%%%%%%
\section{Analysis setup} 
\label{sec:setup}
%%%%%%%%%%%%%%%%%%%%%%%%%%%%%%%%%%%%%%%%%%%%%%%%%%%

The setup for our fit analysis is shown in this section. First, we recapitulate all the available data points that will be considered in our fit analysis. Then, we describe our fit procedure. 

%%%%%%%%%%%%%%%%%%%%%%%%%%%%%%%%%%%%%%%%%%%%%%%%%%%
\subsection{Data points} 
\label{sec:data}
%%%%%%%%%%%%%%%%%%%%%%%%%%%%%%%%%%%%%%%%%%%%%%%%%%%

\subsubsection[\texorpdfstring{$\boldsymbol{\bar{B} \to D\ell\bar\nu}$}{B2Dlnu} distribution]{\texorpdfstring{$\boldsymbol{\bar{B} \to D\ell\bar\nu}$}{B2Dlnu} distribution}
%%%%%%%%%%%%%%%%%%%%%%%%

The differential decay widths of $\bar{B} \to D\ell\bar\nu$ have been measured in Refs.~\cite{Belle:2015pkj} and \cite{Belle-II:2025rna}, which we name as Belle15 and BelleII25, respectively. Their data forms are given as 
\begin{align}
  \frac{\Delta\Gamma_i^D}{\Delta w} \equiv \frac{1}{\Delta w_i} \int_{w_i}^{w_{i+1}} \frac{d\Gamma(\bar{B} \to D \ell \bar\nu)}{dw} \,dw \,, \label{eq:Belle15obs}
\end{align} 
for $\Delta w_i = w_{i+1} - w_i$ and the $w$-bin being $w_i = \{1.00$, $1.06$, $1.12$, $1.18$, $1.24$, $1.30$, $1.36$, $1.42$, $1.48$, $1.54$, $w_\text{max}\}$ with $w_\text{max}= 1.59055$ (Belle15) and $1.591$ (BelleII25). We will use the combined result of the four individual processes $B^0 \to D^- e^+\nu$, $B^0 \to D^-\mu^+\nu$, $B^+ \to \bar D^0 e^+\nu$, and $B^+ \to \bar D^0 \mu^+\nu$. 

To reproduce the PDG setup, we also construct the averaged results by combining Belle15 and BelleII25, which are denoted by $\Delta\Gamma_{i, \text{ave}}^D$ and $\Gamma^D_\text{ave}$, with $\Gamma^D_\text{ave} = \sum_{i} \Delta\Gamma_{i, \text{ave}}^D$.

\subsubsection[\texorpdfstring{$\boldsymbol{\bar{B} \to D^*\ell\bar\nu}$}{B2Dstarlnu} distribution]{\texorpdfstring{$\boldsymbol{\bar{B} \to D^*\ell\bar\nu}$}{B2Dstarlnu} distribution}
%%%%%%%%%%%%%%%%%%%%%%%%

In Ref.~\cite{Belle:2018ezy} and \cite{Belle-II:2023okj}, which are named as Belle18 and BelleII23 respectively, the following kinematic distribution observables are introduced: 
\begin{align}
 \Delta\Gamma_{i}^w & \equiv \int_{w_i}^{w_{i+1}} dw \int_{-1}^1 d\cos\theta_\ell \int_{-1}^1 d\cos\theta_V \int_{0}^{2\pi} d\chi \frac{d\Gamma_\text{full}}{dw \,d\cos\theta_\ell d\cos\theta_V d\chi} \,,  \\[0.5em]
 \Delta\Gamma_{i}^{\cos\theta_\ell} & \equiv \int_{1}^{w_\text{max}} dw \int_{c_\ell^i}^{c_\ell^{i+1}} d\cos\theta_\ell \int_{-1}^1 d\cos\theta_V \int_{0}^{2\pi} d\chi \frac{d\Gamma_\text{full}}{dw \,d\cos\theta_\ell d\cos\theta_V d\chi} \,, \\[0.5em] 
 \Delta\Gamma_{i}^{\cos\theta_V} & \equiv \int_{1}^{w_\text{max}} dw \int_{-1}^{1} d\cos\theta_\ell \int_{c_V^i}^{c_V^{i+1}} d\cos\theta_V \int_{0}^{2\pi} d\chi \frac{d\Gamma_\text{full}}{dw \,d\cos\theta_\ell d\cos\theta_V d\chi} \,, \\[0.5em] 
 \Delta\Gamma_{i}^\chi & \equiv \int_{1}^{w_\text{max}} dw \int_{-1}^{1} d\cos\theta_\ell \int_{-1}^{1} d\cos\theta_V \int_{\chi_i}^{\chi_{i+1}} d\chi \frac{d\Gamma_\text{full}}{dw \,d\cos\theta_\ell d\cos\theta_V d\chi} \,, 
\end{align}
where the bin lists are taken as 
\begin{align}
    w_i &= \{1.00, 1.05, 1.10, 1.15, 1.20, 1.25, 1.30, 1.35, 1.40, 1.45, w_\text{max}\}\,, \\[0.5em]
    c_{\ell}^i &= \begin{cases} 
                    \{-1.0, -0.8, -0.6, -0.4, -0.2, 0.0, 0.2, 0.4, 0.6, 0.8, 1.0\} & \text{for Belle18} \\[0.5em]
                    \{-1.0, -0.4, -0.2, 0.0, 0.2, 0.4, 0.6, 0.8, 1.0\} & \text{for BelleII23}
                  \end{cases} \,, \\[0.5em]
    c_{V}^i &= \{-1.0, -0.8, -0.6, -0.4, -0.2, 0.0, 0.2, 0.4, 0.6, 0.8, 1.0\}\,, \\[0.5em]
    \chi_i &= \{0, \pi/5, 2\pi/5, 3\pi/5, 4\pi/5, \pi, 6\pi/5, 7\pi/5, 8\pi/5, 9\pi/5, 2\pi\}\,.
\end{align}
Regarding Belle18, the distribution bins are folded and the data points are given as 
\begin{align}
    N_i^x = N_{B^0} \tau_{B^0} \mathcal{B}(\bar{D}^0 \rightarrow K^{-} \pi^{+}) \mathcal{R}_{i j} \varepsilon_j \Delta \Gamma_{j}^x \,, 
\end{align}
for $x = \{w, \cos\theta_\ell, \cos\theta_V, \chi\}$, where $N_{B^0} = 7.50384 \times 10^6$ is the number of $B^0$ mesons in the data sample, $\tau_{B^0}$ denotes the $B^0$ lifetime, and $\mathcal{B}(\bar{D}^0 \rightarrow K^{-} \pi^{+}) = 3.947\%$ is the branching ratio of $\bar{D}^0 \rightarrow K^{-} \pi^{+}$~\cite{ParticleDataGroup:2024cfk}; $\mathcal{R}_{ij}$ and $\varepsilon_j$ represent the detector response matrix (\textit{i.e.}, the probability that an event generated in bin $j$ is observed in bin $i$) and efficiency (\textit{i.e.}, the probability that an event generated in bin $j$ is reconstructed and passes the analysis selection criteria), respectively. All information about these factors can be found in Ref.~\cite{Belle:2018ezy}.

In addition, Ref.~\cite{Belle:2023bwv} (named as Belle23) provides the distribution data for $x = \{w, \cos\theta_\ell, \cos\theta_V, \chi\}$, but the observables are given in the normalized forms 
\begin{align}
  \frac{\Delta\Gamma_{i}^x}{\Gamma^\Dst} \,, 
\end{align}
for $\Gamma^\Dst =\sum_i \Delta\Gamma_{i}^x$, which implies that the Belle23 data points are irrelevant to the overall factor of $|V_{cb}|$. This is also the PDG form as explained in the introduction.
 
Again, we combine Belle18, Belle23, and BelleII23 to construct the averaged results, which are denoted by $\Delta\Gamma_{i, \text{ave}}^x$ and $\Gamma^\Dst_\text{ave} = \sum_i \Delta\Gamma_{i, \text{ave}}^x$. One can check that $\Gamma^\Dst$ is universal for any of the kinematic variables $x = \{w, \cos\theta_\ell, \cos\theta_V, \chi\}$. 

As shown above, the Belle and Belle II collaborations provide three angular distributions with respect to the kinematical variables $\cos\theta_\ell$, $\cos\theta_V$, and $\chi$, along with the $w$ distribution. However, it should be noted that these four distributions are not statistically independent but constructed from the common data samples for each variable. 

\subsubsection{\texorpdfstring{$\boldsymbol{\bar{B} \to \Dgen\ell\bar\nu}$}{B2Dstarlnu} branching ratios}
%%%%%%%%%%%%%%%%%%%%%%%%

For the present case, we need to concern the branching ratio measurements when including them in our fit analysis, because the PDG~\cite{ParticleDataGroup:2024cfk} and HFLAV~\cite{HFLAV:2024ctg} averages for $\mathcal{B}(B^0 \to D^{(*)-} \ell^+ \nu)$ include the Belle15~\cite{Belle:2015pkj} and Belle~II~22~\cite{Belle-II:2022ffa} (Belle18~\cite{Belle:2018ezy} and BelleII23~\cite{Belle:2023bwv}) measurements. Hence, a more conservative choice is to take only the PDG average for $\mathcal{B}(B^+ \to \bar D^{(*)0} \ell^+ \nu)$, with~\cite{ParticleDataGroup:2024cfk}
\begin{align}
 \mathcal{B}(B^+ \to \bar D^{0} \ell^+ \nu) = (2.26 \pm 0.07) \% \,, \qquad
 \mathcal{B}(B^+ \to \bar D^{*0} \ell^+ \nu) = (5.26 \pm 0.10) \% \,, 
\end{align}
which do not include the distribution measurements shown above, and therefore avoid a possible overlap of experimental inputs in our fit analysis.
A detailed study of the difference between charged and neutral $B$-meson decays has recently been performed in Ref.~\cite{Jung:2026ewj} in this context. 

\subsubsection[Lattice input of \texorpdfstring{$\boldsymbol{\bar{B} \to \Dgen}$}{B2Dstar} form-factors]{Lattice input of \texorpdfstring{$\boldsymbol{\bar{B} \to \Dgen}$}{B2Dstar} form-factors}
%%%%%%%%%%%%%%%%%%%%%%%%

There is an unquenched lattice study~\cite{MILC:2015uhg} of the $\bar{B} \to D$ transition form-factors, which provides the numerical results for $f_+(w)$ and $f_0(w)$ at $w=\{1.00, 1.08, 1.16\}$. It is named as MILC15 throughout this paper. 
Another study of Ref.~\cite{Na:2015kha} named as HPQCD15 does not provide detailed $q^2$ nor $w$  points but we follow the implementation from the \texttt{EOS} software~\cite{EOSAuthors:2021xpv}, in which two data points are given for $f_+(w)$ and $f_0(w)$.

For the $\bar{B} \to D^*$ transition form-factors, three independent studies from three different collaborations are now available and named as MILC21~\cite{FermilabLattice:2021cdg}, HPQCD23~\cite{Harrison:2023dzh}, and JLQCD23~\cite{Aoki:2023qpa}, respectively. The numerical results of MILC21 are given for both the BGL and HQET forms of $\{f(w)$, $F_1(w)$, $F_2(w)$, $g(w)\}$ and $\{h_{A_1}(w)$, $h_{A_2}(w)$, $h_{A_3}(w)$, $h_{V}(w)\}$ at $w = \{1.03$, $1.10$, $1.17\}$. The HPQCD23 provides the numerical results for the HQET forms (including also the three tensor form-factors) of $\{h_{A_1}(w)$, $h_{A_2}(w)$, $h_{A_3}(w)$, $h_{V}(w)$, $h_{T_1}(w)$, $h_{T_2}(w)$, $h_{T_3}(w)\}$ at $w=\{1.000$, $1.126$, $1.252$, $1.378$, $1.504\}$. Note that we transform them to the standard forms when performing the fit analysis for the BSZ/BGL parameterizations. The JLQCD23 provides the data points for the BGL forms of $\{ f(w)$, $F_1(w)$, $F_2(w)$, $g(w)\}$ at $w=\{1.025$, $1.060$, $1.100\}$. 

It is interesting to note that the precision of these recent lattice inputs has reached the $5\%$ level.

\subsubsection[LCSR input of \texorpdfstring{$\boldsymbol{\bar{B} \to \Dgen}$}{B2Dstar} form-factors]{LCSR input of $\boldsymbol{\bar{B} \to \Dgen}$ form-factors}
%%%%%%%%%%%%%%%%%%%%%%%%

There are currently two independent LCSR studies of the $\bar{B} \to \Dgen$ transition form factors, LCSR18~\cite{Gubernari:2018wyi} within the framework of QCD LCSR with $B$-meson light-cone distribution amplitudes including higher-twist corrections up to twist-4 accuracy, and LCSR23~\cite{Cui:2023jiw} that refines the large-recoil treatment through a reformulation based on the soft-collinear effective theory with the complete next-to-leading QCD corrections. The LCSR18 study shows their results for the standard forms of $\{ f_+(q^2)$, $f_0(q^2)$, $f_T(q^2)$, $A_0(q^2)$, $A_1(q^2)$, $A_{12}(q^2)$, $V(q^2)$, $T_1(q^2)$, $T_2(q^2)$, $T_{23}(q^2) \}$ at $q^2/\text{GeV}^2 = \{-15$, $-10$, $-5$, $0\}$, while the LCSR23 study gives the results for the helicity form factors $\{ \mathcal F_+(q^2)$, $\mathcal F_0(q^2)$, $\mathcal A_0(q^2)$, $\mathcal A_1(q^2)$, $\mathcal A_{12}(q^2)$, $\mathcal V(q^2) \}$ at $q^2/\text{GeV}^2 = \{-3$, $-2$, $-1$, $0$, $1$, $2$, $3\}$, with their relations to the standard forms given by 
\begin{align}
 & \mathcal F_+  = f_+\,, \quad \mathcal F_0 = \frac{m_B}{2E_D} f_0 \,, \\[0.5em]
 & \mathcal V  = \frac{m_B}{m_B+m_\Dst} V \,, ~~  \mathcal A_0 = \frac{m_\Dst}{E_\Dst} A_0 \,, ~~ \mathcal A_1 = \frac{m_B+m_\Dst}{2E_\Dst} A_1\,, ~~ \mathcal A_2 = \frac{m_B-m_\Dst}{m_B} A_2 \,, 
\end{align}
where $\mathcal A_{12} = \mathcal A_1 - \mathcal A_2$ and $E_\Dgen = (m_B^2+m_\Dgen^2-q^2)/(2m_B)$. The uncertainties of the LCSR inputs are at the level of $20\%$.

\subsection{Unitarity bounds}
%%%%%%%%%%%%%%%%%%%%%%%%

The $\bar{B} \to \Dgen$ transition form-factors have to satisfy the unitarity bounds, also known as the dispersive bounds, which are often implemented in a broader range of contexts. 
For the application to the form-factors in $B$-meson decays, see, \textit{e.g.}, Ref.~\cite{Gubernari:2026sqc} and the references therein.
In practice, the unitarity bounds provide a model-independent way to estimate the size of the neglected higher-order terms for a truncated expansion.
In certain cases, it may further restrict the allowed regions of the expansion coefficients, by turning analyticity and dispersion relations into quantitative constraints on the form-factors~\cite{deRafael:1992tu,deRafael:1993ib,Boyd:1997kz,Caprini:1997mu}.

Specific to the BGL parameterization, we have, by construction, the following simple requirements~\cite{Boyd:1997kz,Bigi:2016mdz,Bigi:2017jbd}:
\begin{align}
 &\sum_{n=0}^{N_F} (b_n^{f_+})^2 + (b_n^g)^2 <1 \,,&
 &\sum_{n=0}^{N_F} (b_n^{f_0})^2 <1 \,,& \\[0.5em]
 &\sum_{n=0}^{N_F} \left[ (b_n^f)^2 + (b_n^{F_1})^2 \right] < 1 \,,&
 &\sum_{n=0}^{N_F} (b_n^{F_2})^2 <1 \,,& 
 %
 %&\sum_{n=0}^{N_F} (b_n^g)^2 <1 \,,& 
\end{align}
from the unitarity bounds. Let us mention in advance that our BGL results obtained in this paper satisfy these constraints; see appendix~\ref{App:FFresult_all} for sure. 

The BSZ parameterization is taken as a reference form-factor model. In this sense, we do not care about the unitarity bounds on the BSZ coefficients in this paper. 

On the other hand, the unitarity bounds for the HQET parameterization give non-trivial relations among the form-factor parameters~\cite{Caprini:1997mu,Boyd:1997kz}.\footnote{
The CLN parameterization, a simplified form based on HQET, implements the unitarity bound to reduce the free parameters in the leading functions $V_1(w)=h_+(w)-(m_B-m_D)/(m_B+m_D) h_-(w)$ and $h_{A_1}(w)$, which leads to the single-parameter forms~\cite{Caprini:1997mu}.  
However, it is argued in Refs.~\cite{Bigi:2017njr,Bigi:2017jbd,Gambino:2019sif} that the corresponding approximations in the other form-factors such as $h_{A_2}(w)$ and $h_{A_3}(w)$ are not proper regarding the heavy-quark expansion. 
Given the current experimental and theoretical accuracy, therefore, the original CLN parameterization is no longer recommended to use in phenomenological analyses. 
} 
Explicit forms of the bounds can be obtained from Refs.~\cite{Caprini:1997mu,Bigi:2017jbd}. 
Note also that the unitarity bounds for the $\bar{B}_q \to D^{(*)}_q$ tensor form-factors have been recently studied in Ref.~\cite{Bordone:2025jur}. 
However, an efficient way to include the unitarity bounds in the fit analysis for the HQET parameterization is still under discussion. 
In particular, it was pointed out that imposing the unitarity bounds in the fit could lead to biases on the fitted values; see Ref.~\cite{Belle2week2023} for a recent report. 

Nevertheless, we need to take into account the unitarity bounds for HQET, 
since we already find that the best fit results for the HQET coefficients without imposing these constraints would violate the unitarity bounds badly. 
In this work, we exactly follow the implementation from the \texttt{EOS} software~\cite{EOSAuthors:2021xpv} for both BGL and HQET.

%%%%%%%%%%%%%%%%%%%%%%%%%%%%%%%%%%%%%%%%%%%%%%%%%%%
\subsection{Fit procedure} 
\label{sec:procedure}
%%%%%%%%%%%%%%%%%%%%%%%%%%%%%%%%%%%%%%%%%%%%%%%%%%%

We will follow the Bayesian fit analysis adopted in Ref.~\cite{Iguro:2020cpg} in a way that the Markov-Chain-Monte-Carlo (MCMC) computation is performed by {\tt Stan} via {\tt Cmdstanpy} package~\cite{Carpenter:2017qjk}, 
which gives fitted values of the form-factor parameters, the CKM matrix element $|V_{cb}|$, and the NP Wilson coefficients $C_X$ in each parameterization. This is a state-of-the-art platform for statistical modeling and high-performance statistical computation, which has not been widely used in the particle physics community though. Therefore, our results can serve as a compatible check for other fit results in the literature. 

To proceed, let us first classify the data points used throughout this paper as
\begin{align*}
 & \textbf{Theory}:~ 
 \begin{cases}
 \textbf{Lattice}:~  \text{MILC15},~ \text{HPQCD15},~ \text{MILC21},~ \text{HPQCD23},~ \text{JLQCD23} \\[0.5em]
 \textbf{LCSR}:~ \text{LCSR18}, ~ \text{LCSR23} 
 \end{cases} 
 \,, \\[1em]
 & \textbf{Distribution}:~ 
 \begin{cases} 
 \textbf{(Raw)}~ \Delta\Gamma_{i, \text{exp}}^D,~ \Delta\Gamma_{i, \text{exp}}^x \\[0.5em]
 \textbf{(Combined)}~ \Delta\Gamma_{i, \text{ave}}^D/\Gamma^D_\text{ave},~  \Delta\Gamma_{i, \text{ave}}^x/\Gamma^\Dst_\text{ave} 
 \end{cases}
 \,, \\[1em] 
 & \textbf{BrRatio}:~ \text{PDG average for $B^+ \to \bar D^{(*)0} \ell^+ \nu$} \,, 
\end{align*}
where ``exp'' denotes the individual measurements of \{Belle15, Belle18, Belle23, BelleII23, BelleII25\} and ``ave'' means the averaged one, as explained in section~\ref{sec:data}. It should be noted that the form-factor parameters can be fitted to all of them, while $|V_{cb}|$ and $C_X$ are determined from \textbf{Distribution (Raw)} and \textbf{BrRatio}. 
To be precise, the Belle23 data points cannot be used to fit $|V_{cb}|$, as already mentioned in section~\ref{sec:data}. In this paper, we consider the following two cases for our fit procedure:  
\begin{itemize}
 \item[(A)] Fit $\text{form-factors}\, (\,+\, C_X\,) + |V_{cb}|$ simultaneously to \textbf{Theory} $+$ \textbf{Distribution (Raw)} $+$ \textbf{BrRatio},
 \item[(B)] Fit $\text{form-factors}\, (\,+\, C_X\,)$ to \textbf{Theory} $+$ \textbf{Distribution (Combined)}, and then obtain $|V_{cb}|$ from \textbf{BrRatio}. This corresponds to the PDG setup~\cite{ParticleDataGroup:2024cfk}. 
\end{itemize}
Regarding the kinematic distributions of $\bar{B} \to D^*\ell\bar\nu$ decays, we should point out that $\Delta\Gamma_i^x$ for each $x = \{w$, $\cos\theta_\ell$, $\cos\theta_V$, $\chi\}$ is not statistically independent, since the data points are collected from the four-fold differential decay rate $d\Gamma_\text{full}/(dw\,d\cos\theta_\ell\,d\cos\theta_V\,d\chi)$ by integrating over the other three variables. Keeping this point in mind, we take one of the $x$ data points in the fit for case~(A), while considering all the $x$ data points for case~(B). According to our understanding, the PDG setup does not care about this point. In summary, we have  
\begin{itemize}
 \item \text{Scenario (A)} with {\bf one} of $x = \{w$, $\cos\theta_\ell$, $\cos\theta_V$, $\chi\}$, 
 \item \text{Scenario (B)} with {\bf all} of $x = \{w$, $\cos\theta_\ell$, $\cos\theta_V$, $\chi\}$.
\end{itemize}
The former has four possible scenarios for each parameterization, which will be labeled as (A, $w$), (A, $\cos\theta_\ell$), (A, $\cos\theta_V$), and (A, $\chi$), respectively. 

The Bayesian fit analysis requires prior probability distributions for the fitting parameters. Concerning a typical order of each form-factor expansion, we take sufficiently broader prior distributions such as $a_n^F \sim \mathcal N (0, 100)$ and $b_n^F \sim \mathcal N (0, 1)$ for the BSZ and BGL parameterizations, respectively. Regarding the HQET parameterization, the QCD sum rule study~\cite{Neubert:1992wq,Neubert:1992pn,Ligeti:1993hw} shows narrow constraints on the subleading IW functions $\eta(w)$, $\hat\chi_2(w)$, and $\hat\chi_3(w)$. Instead of adopting them as explicit constraints, we take them as the prior distributions, with~\cite{Iguro:2020cpg} 
\begin{align}
 \label{eq:QCDSR_1}
 &c_0^{\hat\chi_2} \sim \mathcal N (-0.06, 0.02) \,,& &c_1^{\hat\chi_2} \sim \mathcal N (0.00, 0.02) \,,&  &c_2^{\hat\chi_2} \sim \mathcal N (-0.01, 0.02) \,,& \\[0.5em]
 & & & c_1^{\hat\chi_3} \sim \mathcal N (0.04, 0.03) \,,& &c_2^{\hat\chi_3} \sim \mathcal N (-0.03, 0.05) \,,& \\[0.5em]
 &c_0^{\eta} \sim \mathcal N (0.62, 0.12) \,,& &c_1^{\eta} \sim \mathcal N (0.04, 0.03) \,,&  &c_2^{\eta} \sim \mathcal N (0.05, 0.07) \,.& 
 \label{eq:QCDSR_2}
\end{align}
On the other hand, we take $c_n^F \sim \mathcal N (0, 1)$ for the IW functions $\xi$ and $\hat\ell_i$. For all the fitting scenarios considered, we will also study how these prior distributions affect our final results. 

%%%%%%%%%%%%%%%%%%%%%%%%%%%%%%%%%%%%%%%%%%%%%%%%%%%
\section{Numerical results} 
\label{sec:result}
%%%%%%%%%%%%%%%%%%%%%%%%%%%%%%%%%%%%%%%%%%%%%%%%%%%

%%%%%%%%%%%%%%%%%%%%%%%%%%%%%%%%%%%%%%%%%%%%%%%%%%%
\subsection[\texorpdfstring{$|V_{cb}|$}{Vcb} determination within the SM]{\texorpdfstring{$\boldsymbol{|V_{cb}|}$}{Vcb} determination within the SM} 
%%%%%%%%%%%%%%%%%%%%%%%%%%%%%%%%%%%%%%%%%%%%%%%%%%%

\begin{table}[t]
  \centering
  \resizebox{\textwidth}{!}{
  \renewcommand{\arraystretch}{1.5}
  \begin{tabular}{l |c c | c c | c c}
    \toprule
    & \multicolumn{6}{c}{$|V_{cb}|$ fit result ($\times 10^3$)} \\
    \cmidrule(l){2-7}
    \multirow{2}{*}{\textbf{Fit scenario}}
      & \multicolumn{2}{c| }{ \textbf{BSZ } }
      & \multicolumn{2}{c| }{ \textbf{BGL } }
      & \multicolumn{2}{c }{ \textbf{HQET } } \\
    \cmidrule(lr){2-3} \cmidrule(lr){4-5} \cmidrule(lr){6-7}
      & $N_F=1$ & $N_F=2$ & $N_F=1$ & $N_F=2$ & $(2/1/0)$ & $(3/2/1)$ \\
    \midrule
    $(\text{A},w)$
      & $39.3 \pm 0.3$ %BSZ NF=1
      & $39.5 \pm 0.4$ %BSZ NF=2
      & $40.0 \pm 0.4$ %BGL NF=1     
      & $39.5 \pm 0.4$ %BGL NF=2
      & $39.3 \pm 0.3$ %HQET210
      & $39.4 \pm 0.3$  \\[0.2em] %HQET321
    $(\text{A},\cos\theta_\ell)$
      & $39.2 \pm 0.4$ %BSZ NF=1
      & $39.8 \pm 0.4$ %BSZ NF=2
      & $40.1 \pm 0.4$ %BGL NF=1
      & $39.9 \pm 0.4$ %BGL NF=2
      & $39.4 \pm 0.4$ %HQET210
      & $39.4 \pm 0.4$ \\[0.2em] %HQET321
    $(\text{A},\cos\theta_V)$
      & $39.3 \pm 0.4$ %BSZ NF=1
      & $39.8 \pm 0.4$ %BSZ NF=2
      & $40.2 \pm 0.4$ %BGL NF=1 
      & $39.9 \pm 0.4$ %BGL NF=2     
      & $39.5 \pm 0.4$ %HQET210
      & $39.6 \pm 0.4$ \\[0.2em] %HQET321
    $(\text{A},\chi)$
      & $39.9 \pm 0.4$ %BSZ NF=1
      & $40.6 \pm 0.5$ %BSZ NF=2
      & $41.1 \pm 0.4$ %BGL NF=1
      & $40.6 \pm 0.4$ %BGL NF=2     
      & $40.3 \pm 0.4$ %HQET210
      & $40.4 \pm 0.4$ \\[0.2em] %HQET321
    \midrule
    $(\text{B}) + \mathcal B(B^+ \to \bar D^{0} \ell^+ \nu)$
      & $38.0 \pm 0.7$ %BSZ NF=1
      & $38.9 \pm 0.8$ %BSZ NF=2
      & $39.2 \pm 0.8$ %BGL NF=1
      & $38.8 \pm 0.8$ %BGL NF=2
      & $40.7 \pm 0.8$ %HQET210
      & $39.2 \pm 0.8$ \\[0.2em] %HQET321
    $(\text{B}) + \mathcal B(B^+ \to \bar D^{*0} \ell^+ \nu)$
      & $39.9 \pm 0.5$ %BSZ NF=1
      & $39.7 \pm 0.6$ %BSZ NF=2
      & $40.3 \pm 0.5$ %BGL NF=1
      & $39.8 \pm 0.6$ %BGL NF=2
      & $38.9 \pm 0.5$ %HQET210
      & $39.7 \pm 0.5$ \\[0.2em] %HQET321
    $(\text{B}) + \text{combined}$
      & $38.9 \pm 0.3$ %BSZ NF=1
      & ${\bf 39.3 \pm 0.3}$ %BSZ NF=2
      & $39.7 \pm 0.3$ %BGL NF=1
      & ${\bf 39.3 \pm 0.3}$ %BGL NF=2
      & $39.8 \pm 0.3$ %HQET210
      & ${\bf 39.5 \pm 0.3}$ \\ %HQET321 
    \bottomrule
  \end{tabular}
  }
  \caption{
  Fit results of $|V_{cb}|$ in the SM for different fit scenarios and form-factor parameterizations.
  The PDG average is almost reproduced in our fit as highlighted. 
  See main text for descriptions of the fit scenarios.   
  \label{tab:VcbFitSM}
  }
\end{table}

In Table~\ref{tab:VcbFitSM}, we show the SM fit results for $|V_{cb}|$ for the different fit scenarios and form-factor parameterizations. The main observations for each parameterization are made as follows. 
\begin{itemize}
 \item 
 \textbf{BSZ parameterization:}
 The fit scenarios $(\text{A},w)$, $(\text{A},\cos\theta_\ell)$, and $(\text{A},\cos\theta_V)$ give mutually consistent values of $|V_{cb}|$, whereas $(\text{A},\chi)$ leads to a larger value. 
 In scenario (B), the values of $|V_{cb}|$ determined from $\mathcal B(B^+ \to \bar D^{0} \ell^+ \nu)$ and $\mathcal B(B^+ \to \bar D^{*0} \ell^+ \nu)$ show a visible difference, particularly for $N_F=1$. For $N_F=2$, the two determinations are closer and compatible within the uncertainties.
 This situation is similar to that observed in the PDG values discussed in the introduction.
 Moreover, the combined result for $N_F=2$ is consistent with the PDG value given by Eq.~\eqref{eq:VcbPDGofficial}.
 \item
 \textbf{BGL parameterization:}
 The results for $N_F=2$ are very close to those obtained with the BSZ parameterization.
 This is expected because all the form-factors are treated independently in both parameterizations. 
 For $N_F=1$, BGL gives systematically larger values than BSZ. 
 This may reflect the different implementations of the kinematic constraints.
 We also confirm that the combined result for $N_F=2$ in scenario (B) is consistent with the PDG value.
 \item
 \textbf{HQET parameterization:}
 The HQET results show a similar dependence on the fit scenario to those obtained with BSZ and BGL.
 %In scenario (A), the $(\text{A},w)$, $(\text{A},\cos\theta_\ell)$, and $(\text{A},\cos\theta_V)$ fits give mutually consistent values of $|V_{cb}|$, whereas $(\text{A},\chi)$ leads to a larger value.
 Both the $(2/1/0)$ and $(3/2/1)$ setups give values close to the BSZ and BGL results for $N_F=2$.
 In scenario (B), the two $|V_{cb}|$ determinations are compatible within the uncertainties for $(3/2/1)$, and the combined value is consistent with the PDG value. 
 This indicates that the HQET description works well for the $B \to D^{(*)}$ transitions, compatible with the QCD-based BSZ and BGL descriptions.
\end{itemize} 
To summarize, we find in scenario (B) that BSZ ($N_F=2$), BGL ($N_F=2$), and HQET $(3/2/1)$ give the $|V_{cb}|$ values in very good agreements with the PDG value. 
In scenario (A), the $(\text{A},w)$ results are close to the corresponding combined results in scenario (B), whereas $(\text{A},\chi)$ gives systematically larger values for all parameterizations. 
We provide details of the BSZ fit results in section~\ref{sec:chisqbest}, but do not show any further results for this parameterization in the following study.

The good agreement between scenarios (A) and (B) suggests that the present analysis is not significantly affected by the so-called D'Agostini bias~\cite{DAgostini:1993arp}.
Our analysis propagates the fitted-parameter uncertainties within a Bayesian framework using the {\tt CmdStanPy} package~\cite{Carpenter:2017qjk}.
However, a dedicated validation of the fit procedure in scenario (A) would be needed to assess the possible impact of D'Agostini bias, which is beyond the scope of this paper. 
For instance, a normalization bias (uncertainty) in the determination of CKM matrix elements is discussed in Ref.~\cite{Ball:2009mk}.

%%%%%%%%%%%%FIG%%%%%%%%%%%%
\begin{figure}[htbp]
\begin{center}
\includegraphics[viewport=0 0 1125 1400, width=\textwidth]{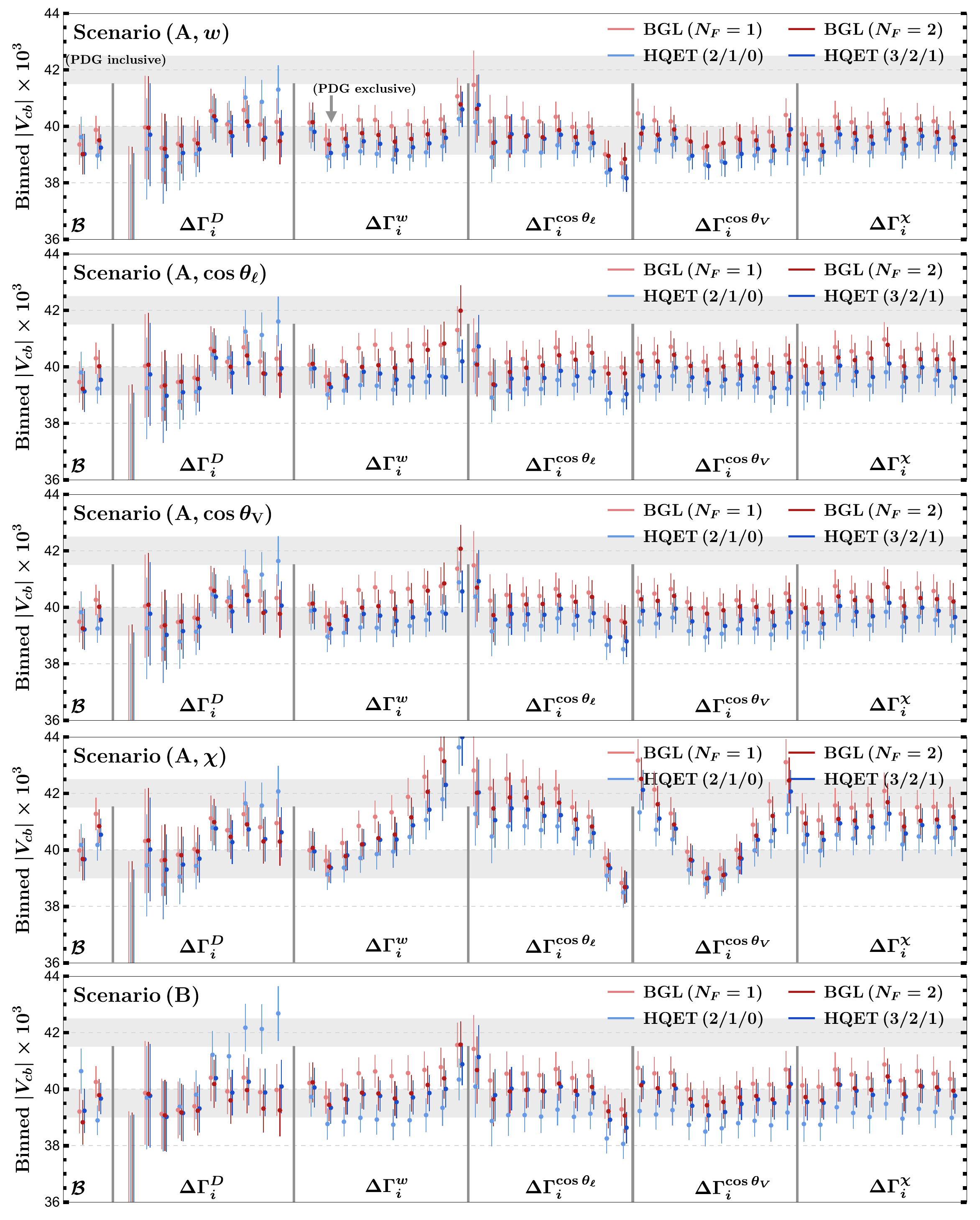}
\caption{
The binned $|V_{cb}|$ values obtained from the combined distribution data $\Delta\Gamma_i^{D}$ and $\Delta\Gamma_i^{x}$ for $x=\{w, \cos\theta_\ell, \cos\theta_V, \chi\}$ in each fit scenario. 
Those obtained from the branching ratios $\mathcal B(B^+ \to \bar D^{0} \ell^+ \nu)$ and $\mathcal B(B^+ \to \bar D^{*0} \ell^+ \nu)$ are also shown in the first and second bins, labeled as $\mathcal B$. 
The PDG averages~\cite{ParticleDataGroup:2024cfk} for the inclusive and exclusive determinations of $|V_{cb}|$ are denoted by the gray bands. See text for further details.
} 
\label{Fig:BinVcb}
\end{center}
\end{figure}
%%%%%%%%%%%%FIG%%%%%%%%%%%%

In Fig.~\ref{Fig:BinVcb}, we show the values of $|V_{cb}|$ obtained from \textbf{Distribution (Combined)} and \textbf{BrRatio} for each fit scenario. 
They are obtained by using the fitted form-factor parameters to evaluate $\Delta\Gamma_i^{D}$ for $\bar{B} \to D \ell\bar\nu$ and $\Delta\Gamma_i^{x}$ for $\bar{B} \to D^*\ell\bar\nu$, with $x=\{w, \cos\theta_\ell, \cos\theta_V, \chi\}$, 
without multiplying them by $|V_{cb}|^2$, and then confronting these quantities with $\Delta\Gamma_{i, \text{ave}}^{D}$ and $\Delta\Gamma_{i, \text{ave}}^{x}$.
In the following, we refer to the resulting values as binned $|V_{cb}|$.
The covariances of the fitted form-factor values and experimental measurements are properly taken into account during our fits.
In the figure, the first and second bins correspond to $|V_{cb}|$ determined from $\mathcal B(B^+ \to \bar D^{0} \ell^+ \nu)$ and $\mathcal B(B^+ \to \bar D^{*0} \ell^+ \nu)$, respectively. 
Furthermore, the gray bands indicate the PDG averages~\cite{ParticleDataGroup:2024cfk} for the inclusive and exclusive determinations of $|V_{cb}|$.

%%%%%%%%
%%%%%%%%
We first observe that all fit scenarios give broadly consistent binned values of $|V_{cb}|$ from $\Delta\Gamma_i^D$ in \textbf{Distribution (Combined)} for BGL ($N_F=1,2$) and HQET $(3/2/1)$.\footnote{
Due to phase-space suppression, the central value of $\Delta\Gamma_{i=1}^D$ is small and has a relatively large uncertainty. As a result, the corresponding binned value of $|V_{cb}|$ is poorly constrained.
}
These values also agree with those obtained from $\mathcal B(B^+ \to \bar D^{0} \ell^+ \nu)$.
This supports a consistent description of $\bar{B} \to D \ell\bar\nu$ for these parameterizations.
In contrast, HQET $(2/1/0)$ gives somewhat less consistent binned values in the $D$ mode, particularly in scenario (B).
Regarding the $\bar{B} \to D^* \ell\bar\nu$ mode, we observe the following informative features:
\begin{itemize}
 \item 
 Scenario (B) gives broadly consistent binned values of $|V_{cb}|$ among the 40 $D^*$ bins, with some deviations, particularly in the last $w$ bin and the first/last $\cos\theta_\ell$ bins.
 These values are also consistent with those obtained from $\mathcal B(B^+ \to \bar D^{*0} \ell^+ \nu)$.
 However, their values depend on the form-factor parameterizations.
 The BGL ($N_F=1,2$) and HQET $(3/2/1)$ results are generally consistent with the PDG exclusive average. 
 Moreover, the $D$ and $D^*$ modes give consistent binned values. 
 On the other hand,  HQET $(2/1/0)$ yields systematically smaller $D^*$ values and shows a clear separation especially with the higher-$w$ $D$ bins.
 \item
 In scenario (A), each $(\text{A},x)$ fit directly includes the corresponding $x$-distribution dataset.
 As expected from such a fit setup, Fig.~\ref{Fig:BinVcb} shows that the binned values extracted from $\Delta\Gamma_i^x$ are relatively stable for the fitted choice of $x=\{w,\cos\theta_\ell,\cos\theta_V,\chi\}$.
 \item
 The $(\text{A},w)$ fit shows fluctuations similar to those in the (B) fit for both the BGL and HQET parameterizations.
 In particular, the BGL ($N_F=2$) and HQET $(3/2/1)$ results remain compatible with the PDG exclusive average. 
 BGL ($N_F=1$) and HQET $(2/1/0)$ give relatively larger and smaller binned values for the $D^*$ mode, respectively.
 \item
 The $(\text{A},\cos\theta_\ell)$ and $(\text{A},\cos\theta_V)$ fits give relatively stable values from the angular distribution data $\Delta\Gamma_i^{x=\cos\theta_\ell,\cos\theta_V,\chi}$, 
 whereas the binned values from $\Delta\Gamma_i^w$ increase in the last bins, most visibly for BGL ($N_F=2$). 
 The corresponding variation is milder for HQET $(3/2/1)$.
 The stable BGL results remain consistent with the PDG exclusive average, those from the $D$ mode, and the (B) fit result.
 \item
 The $(\text{A},\chi)$ fit does not give stable values for the other distribution observables in either the BGL or HQET parameterization.
 In particular, sizable variations across the $w$, $\cos\theta_\ell$, and $\cos\theta_V$ bins are observed, even though the values from the fitted $\chi$ distribution are relatively stable.
\end{itemize}
As an additional overall feature, we observe that the HQET $(2/1/0)$ parameterization generally yields smaller binned values of $|V_{cb}|$ in the $D^*$ mode than the other parameterizations.  
In scenario (B), comparing these binned values with the \textbf{BrRatio} determinations provides a consistency check, since the \textbf{Distribution} inputs to the fit are normalized.
We also find that the difference between BGL ($N_F=1$) and BGL ($N_F=2$) is relatively small in all fit scenarios. 
This indicates that the BGL results are qualitatively stable against increasing the number of form-factor parameters.
By contrast, the difference between HQET $(2/1/0)$ and HQET $(3/2/1)$ becomes more visible in the binned values from the $D$ mode, most clearly in scenario (B).
This suggests that the HQET results are more sensitive to the treatment of higher-order contributions and additional form-factor parameters.
%%%%%%%%
%%%%%%%%

From the above findings, we conclude that the $(\text{A},w)$ and (B) scenarios provide the most reliable fit setups for both the BGL and HQET parameterizations.
On the other hand, the angular-distribution fits $(\text{A},\cos\theta_\ell)$, $(\text{A},\cos\theta_V)$, and $(\text{A},\chi)$ lead to discrepancies with the other distribution data.
These results suggest that the $w$ distribution provides a particularly robust constraint on the form-factor shapes and plays an essential role in the $|V_{cb}|$ determination.
In addition, the angular distribution data contain complementary information, which becomes important when they are combined with the other distributions, as in scenario (B).
This supports the use of scenario (B), where all distribution data are included simultaneously to obtain a more comprehensive fit.

%%%%%%%%%%%%%%%%%%%%%%%%%%%%%%%%%%%%%%%%%%%%%%%%%%%
\subsection{Closer look at the fit results within the SM} 
%%%%%%%%%%%%%%%%%%%%%%%%%%%%%%%%%%%%%%%%%%%%%%%%%%%

\subsubsection[\texorpdfstring{$\chi^2_\mathrm{best}$}{chisqbest} values: decomposition and discussion]{\texorpdfstring{$\boldsymbol{\chi^2_\mathrm{best}}$}{chisqbest} values: decomposition and discussion} 
\label{sec:chisqbest}
%%%%%%%%%%%%%%%%%%%%%%%%%%%%%%%%

Since the fit scenarios (A) and (B) involve different experimental datasets, a direct comparison of the standard deviations of the corresponding fit results is not appropriate.
Instead, we construct $\chi^2_\text{best}$ with respect to
\begin{align}
 \big\{ \textbf{Distribution (Combined)} \,,~ \textbf{BrRatio} \,,~ \textbf{Lattice} \,,~ \textbf{LCSR} \big\} \,, 
 \label{eq:chi_data}
\end{align}
using the best-fit parameters obtained for each scenario.
Although \textbf{BrRatio} is not included in scenario (B), we construct its $\chi^2_\text{best}$ using the combined value of $|V_{cb}|$, because $|V_{cb}|$ should be unique within the SM.
This enables a consistent comparison of the agreement with the data across different form-factor parameterizations. For the analysis within the SM, the datasets in Eq.~\eqref{eq:chi_data} contain $\{50, 2, 54, 56\}$ data points, respectively. 

\begin{table}[t]
  \centering
  \resizebox{\textwidth}{!}{
  \renewcommand{\arraystretch}{1.5}
  \begin{tabular}{l |c c | c c | c c}
    \toprule
    & \multicolumn{6}{c}{$\chi^2_\text{best}$ for \textbf{Distribution (Combined)} / \textbf{BrRatio} / \textbf{Lattice} } \\
    \cmidrule(l){2-7}
    \multirow{2}{*}{\textbf{Fit scenario}}
      & \multicolumn{2}{c| }{ \textbf{BSZ } }
      & \multicolumn{2}{c| }{ \textbf{BGL } }
      & \multicolumn{2}{c }{ \textbf{HQET } } \\
    \cmidrule(lr){2-3} \cmidrule(lr){4-5} \cmidrule(lr){6-7}
      & $N_F=1$ & $N_F=2$ & $N_F=1$ & $N_F=2$ & $(2/1/0)$ & $(3/2/1)$ \\
    \midrule
    (A, $w$)
      & $ 25.0 / 2.4 / 63.9 $ %BSZ NF=1  $ 16.7 / 4.3 / 71.5 $ %BSZ NF=1
      & $ 20.4 / 0.7 / 63.7 $ %BSZ NF=2  $ 16.6 / 1.5 / 61.9 $ %BSZ NF=2
      & $ 26.8 / 1.0 / 44.4 $ %BGL NF=1  $ 22.0 / 0.7 / 47.3 $ %BGL NF=1
      & $ 20.8 / 0.8 / 62.8 $ %BGL NF=2  $ 18.2 / 0.4 / 68.1 $ %BGL NF=2
      & $ 33.2 / 1.1 / 89 $ %HQET210  $ 26.7 / 1.1 / 92.7 $ %HQET210
      & $ 27.9 / 0.4 / 50.8 $  \\[0.2em] %HQET321  $ 24.2 / 0.6 / 67.7 $  \\[0.2em] %HQET321
    (A, $\cos\theta_\ell$)
      & $ 15.7 / 2.6 / 83.7 $ %BSZ NF=1  $ 14.6 / 5.2 / 73.1 $ %BSZ NF=1
      & $ 29.0 / 1.1 / 79.6 $ %BSZ NF=2  $ 16.4 / 2.9 / 68.5 $ %BSZ NF=2
      & $ 21.5 / 1.4 / 66.7 $ %BGL NF=1  $ 14.3 / 1.5 / 53.1 $ %BGL NF=1
      & $ 32.4 / 1.3 / 84.8 $ %BGL NF=2  $ 23.1 / 2.2 / 69.4 $ %BGL NF=2
      & $ 35.9 / 0.6 / 103 $ %HQET210  $ 17.8 / 1.2 / 123 $ %HQET210
      & $ 16.6 / 0.3 / 69.1 $  \\[0.2em] %HQET321  $ 23.2 / 0.3 / 73.0 $  \\[0.2em] %HQET321
    (A, $\cos\theta_V$)
      & $ 17.5 / 2.6 / 75.8 $ %BSZ NF=1  $ 14.5 / 3.7 / 73.0 $ %BSZ NF=1
      & $ 28.7 / 1.0 / 75.1 $ %BSZ NF=2  $ 15.5 / 0.6 / 69.9 $ %BSZ NF=2
      & $ 22.1 / 1.4 / 61.4 $ %BGL NF=1  $ 19.7 / 0.2 / 45.5 $ %BGL NF=1
      & $ 32.6 / 1.3 / 79.8 $ %BGL NF=2  $ 22.1 / 0.7 / 70.9 $ %BGL NF=2
      & $ 39.6 / 0.8 / 89 $ %HQET210  $ 17.7 / 0.1 / 112 $ %HQET210
      & $ 19.9 / 0.4 / 57.9 $  \\[0.2em] %HQET321  $ 24.0 / 1.1 / 69.4 $  \\[0.2em] %HQET321
    (A, $\chi$)
      & $ 112 / 5.0 / 56.0 $ %BSZ NF=1  $ 18.6 / 3.3 / 74.2 $ %BSZ NF=1
      & $ 158 / 2.5 / 48.3 $ %BSZ NF=2  $ 23.2 / 1.1 / 75.3 $ %BSZ NF=2
      & $ 173 / 3.4 / 37.2 $ %BGL NF=1  $ 41.2 / 0.5 / 32.4 $ %BGL NF=1
      & $ 168 / 2.8 / 49.5 $ %BGL NF=2  $ 31.6 / 0.7 / 71.6 $ %BGL NF=2
      & $ 111 / 0.3 / 64.6  $ %HQET210  $ 45.3 / 0.1 / 98.8 $ %HQET210
      & $ 109 / 1.7 / 36.5 $  \\[0.2em] %HQET321  $ 40.6 / 0.8 / 63.0 $  \\[0.2em] %HQET321
    \midrule
    (B)
      & $ 22.0 / 8.5 / 67.8 $ %BSZ NF=1  $ 19.8 / 5.8 / 65.7 $ %BSZ NF=1
      & $ 26.7 / 1.6 / 65.9 $ %BSZ NF=2  $ 22.3 / 0.7 / 64.6 $ %BSZ NF=2
      & $ 25.4 / 2.6 / 50.2 $ %BGL NF=1  $ 21.5 / 1.3 / 42.9 $ %BGL NF=1        
      & $ 28.0 / 2.2 / 67.6 $ %BGL NF=2  $ 21.4 / 0.6 / 72.5 $ %BGL NF=2
      & $ 42.6 / 7.6 / 76.6 $ %HQET210  $ 34.4 / 3.9 / 86.2 $ %HQET210
      & $ 26.0 / 0.4 / 50.1 $  \\ %HQET321  $ 22.5 / 2.6 / 71.0 $  \\ %HQET321
   \bottomrule
  \end{tabular}
  }
  \caption{
  The $\chi^2_\text{best}$ values with respect to \textbf{Distribution (Combined)}, \textbf{BrRatio}, and \textbf{Lattice}, based on our fit results for all form-factor parameterizations and fit scenarios. 
  \label{tab:ChiSqSM_exp}
  }
\end{table}

In Table~\ref{tab:ChiSqSM_exp}, we summarize the $\chi^2_\text{best}$ values of \textbf{Distribution (Combined)}, \textbf{BrRatio}, and \textbf{Lattice} for all fit scenarios and form-factor parameterizations within the SM.
The main observations are made below.
\begin{itemize}
\item \textbf{Distribution (Combined):}
The values of $\chi^2_\text{best}$ per bin are sufficiently small, with $\chi^2_\text{best}/\mathrm{bin} < 1$ for all form-factor parameterizations and fit scenarios except for the $(\text{A},\chi)$ scenario.
This indicates that the experimental measurements are reproduced well in the other scenarios.
The poor fit in the $(\text{A},\chi)$ scenario is seen for all parameterizations, with $\chi^2_\text{best}/\mathrm{bin} \approx 2.2$ -- $3.5$.
\item \textbf{BrRatio:}
In scenario (A), the $(\text{A},w)$, $(\text{A},\cos\theta_\ell)$, and $(\text{A},\cos\theta_V)$ fits reproduce the experimental data with $\chi^2_\text{best}/\mathrm{bin} < 1$ for all form-factor parameterizations, except for BSZ ($N_F=1$).
The $(\text{A},\chi)$ fit gives a somewhat poorer description of \textbf{BrRatio} for BSZ and BGL. For HQET, the fit describes \textbf{BrRatio} well but poorly reproduces \textbf{Distribution}, which may suggest a trade-off between the descriptions of the two datasets.
In scenario (B), the $\chi^2_\text{best}$ values are generally larger, except for HQET $(3/2/1)$. 
Since \textbf{BrRatio} is not included in the fit, this behavior is natural and illustrates the importance of a simultaneous determination of $|V_{cb}|$.
Nevertheless, the fit results for the BSZ ($N_F=2$), BGL ($N_F=2$), and HQET $(3/2/1)$ parameterizations are sufficiently consistent with the experimental data at the level of $\chi^2_\text{best}/\mathrm{bin} \lesssim 1$.
\item \textbf{Lattice:}
The values of $\chi^2_\text{best}/\mathrm{bin}$ range from about $0.7$ to $1.9$. 
However, the lattice data are generally less well described than the distribution data, except in the $(\text{A},\chi)$ fit.
This issue will be examined in more detail below.
\end{itemize}

%%%%%
%%%%%
%%%%%

\begin{table}[t]
  \centering
  \resizebox{\textwidth}{!}{
  \renewcommand{\arraystretch}{1.5}
  \begin{tabular}{l |c c | c c | c c}
    \toprule
    & \multicolumn{6}{c}{$\chi^2_\text{best}$ for MILC15 (6) / HPQCD15 (4) / MILC21 (12) / JLQCD23 (12) / HPQCD23 (20) } \\
    \cmidrule(l){2-7}
    \cmidrule(l){2-7}
    \multirow{2}{*}{\textbf{Fit scenario}}
      & \multicolumn{2}{c| }{ \textbf{BSZ } }
      & \multicolumn{2}{c| }{ \textbf{BGL } }
      & \multicolumn{2}{c }{ \textbf{HQET } } \\
    \cmidrule(lr){2-3} \cmidrule(lr){4-5} \cmidrule(lr){6-7}
      & $N_F=1$ & $N_F=2$ & $N_F=1$ & $N_F=2$ & $(2/1/0)$ & $(3/2/1)$ \\
    \midrule
    (A, $w$)
      & $ 23 / 1 / 27 / \phantom{0}9 / 4 $ %BSZ NF=1
      & $ 1 / 3 / 40 / 14 / 5 $ %BSZ NF=2
      & $ 5/4/22/ \phantom{0}9 / 4 $ %BGL NF=1     
      & $ 3/2/38/14 / 5 $ %BGL NF=2
      & $ 24/4/37/16 / 8 $ %HQET210
      & $ 1/3/25/15 / 7 $  \\[0.2em] %HQET321
    (A, $\cos\theta_\ell$)
      & $ 22 / 1 / 41 / 13 / 6 $ %BSZ NF=1
      & $ 1 / 3 / 51 / 19 / 6 $ %BSZ NF=2
      & $ 4/4/37/15 / 6 $ %BGL NF=1     
      & $ 3/2/52/22 / 6 $ %BGL NF=2
      & $ 24/4/44/23 / 9 $ %HQET210
      & $ 2/3/39/18 / 8 $  \\[0.2em] %HQET321
    (A, $\cos\theta_V$)
      & $ 23 / 1 / 34 / 13 / 5 $ %BSZ NF=1
      & $ 1 / 3 / 47 / 19 / 5 $ %BSZ NF=2
      & $ 4/4/32/15 / 6 $ %BGL NF=1     
      & $ 4/2/47/22 / 5 $ %BGL NF=2
      & $ 24/4/38/16 / 8 $ %HQET210
      & $ 1/3/33/15 / 7 $  \\[0.2em] %HQET321
    (A, $\chi$)
      & $ 25 / 1 / 17 / 10 / 3 $ %BSZ NF=1
      & $ 2 / 4 / 21 / 18 / 3 $ %BSZ NF=2
      & $ 4/4/12/13 / 3 $ %BGL NF=1     
      & $ 5/3/21/18 / 3 $ %BGL NF=2
      & $ 20/3/24/13 / 6 $ %HQET210
      & $ 1/3/15/12 / 5 $  \\[0.2em] %HQET321
    \midrule
    (B)
      & $ 21 / 2 / 30 / 10 / 5 $ %BSZ NF=1
      & $ 1 / 3 / 41 / 15 / 5 $ %BSZ NF=2
      & $ 5/4/26/11 / 5 $ %BGL NF=1     
      & $ 3/2/40/16 / 6 $ %BGL NF=2
      & $ 15/3/36/15 / 8 $ %HQET210
      & $ 1/3/28/12 / 7 $  \\ %HQET321
   \bottomrule
  \end{tabular}
  }
  \caption{
  The $\chi^2_\text{best}$ values with respect to the lattice data, based on the fit results for each scenario. 
  The number of the data points for each lattice study is given in the bracket. 
  \label{tab:ChiSqSM_lat}
  }
\end{table}

In Table~\ref{tab:ChiSqSM_lat}, we show the decomposition of $\chi^2_\text{best}$ for the individual lattice results, 
namely \{MILC15, HPQCD15\} for the $\bar{B} \to D$ and \{MILC21, JLQCD23, HPQCD23\} for the $\bar{B} \to D^*$ transition form-factors.
The following observations can be made. 
\begin{itemize}
 \item
 The BSZ ($N_F=2$), BGL ($N_F=1,2$), and HQET $(3/2/1)$ parameterizations reproduce the MILC15 dataset with $\chi^2_\text{best}/\text{bin} \lesssim 1$.
 By contrast, the BSZ ($N_F=1$) and HQET $(2/1/0)$ parameterizations provide a poor description of the same dataset.
 The HPQCD15 dataset is well described across all fit scenarios and form-factor parameterizations.
 \item 
 The MILC21 dataset is generally poorly described, with $\chi^2_\text{best}/\text{bin} \approx 1$ -- $4.3$, which may indicate a tension with the other datasets included in our fit.
 A similar feature has also been reported in other studies, such as Refs.~\cite{Martinelli:2023fwm,Martinelli:2024bov}.
 \item
 The JLQCD23 data points are reproduced at the level of $\chi^2_\text{best}/\text{bin} \approx 0.8$ -- $1.9$, while the HPQCD23 data points are well described per bin.
 This feature is consistent with the fact that the latter have relatively larger uncertainties.
\end{itemize}
Regarding the goodness of fit, the HQET $(2/1/0)$ parameterization generally gives a worse description of the lattice data.
This may reflect the limited parameter freedom that results in a tension between the distribution and lattice inputs within this framework.
In contrast, the HQET $(3/2/1)$ parameterization gives smaller $\chi^2_\text{best}$ values for the lattice data than HQET $(2/1/0)$, with values comparable to or smaller than those obtained with BGL and BSZ.
This suggests that additional parameter freedom is required to accommodate the lattice and distribution data simultaneously.
We also find that scenarios (A) and (B) give qualitatively similar patterns for the individual lattice datasets, although their $\chi^2_\text{best}$ values depend on the chosen distribution.

%%%%%
%%%%%
%%%%%

\begin{table}[t]
  \centering
  \resizebox{\textwidth}{!}{
  \renewcommand{\arraystretch}{1.5}
  \begin{tabular}{l |c c | c c | c c}
    \toprule
    & \multicolumn{6}{c}{$\chi^2_\text{best}$ for LCSR18 (22) / LCSR23 (34)  } \\
    \cmidrule(l){2-7}
    \cmidrule(l){2-7}
    \multirow{3}{*}{\textbf{Fit scenario}}
      & \multicolumn{2}{c| }{ \textbf{BSZ } }
      & \multicolumn{2}{c| }{ \textbf{BGL } }
      & \multicolumn{2}{c }{ \textbf{HQET } } \\
    \cmidrule(lr){2-3} \cmidrule(lr){4-5} \cmidrule(lr){6-7}
      & $N_F=1$ & $N_F=2$ & $N_F=1$ & $N_F=2$ & $(2/1/0)$ & $(3/2/1)$ \\
    \midrule
    (A, $w$)
      & $ 22 (\phantom{0}8) / 181 (32) $ %BSZ NF=1
      & $ 12 (\phantom{0}7) / 112 (32) $ %BSZ NF=2
      & $ 13 (\phantom{0}6) / 190 (31) $ %BGL NF=1     
      & $ \phantom{0}9 (\phantom{0}6) / 121 (32) $ %BGL NF=2
      & $ 20 (\phantom{0}6) / 147 (28) $ %HQET210
      & $ 4 (\phantom{0}8) / 110 (33) $  \\[0.2em] %HQET321
    (A, $\cos\theta_\ell$)
      & $ 25 (\phantom{0}9) / 179 (30) $ %BSZ NF=1
      & $ 16 (11) / \phantom{0}88 (36) $ %BSZ NF=2
      & $ 19 (11) / 176 (33) $ %BGL NF=1     
      & $ 13 (10) / \phantom{0}89 (36) $ %BGL NF=2
      & $ 24 (\phantom{0}9) / 136 (30) $ %HQET210
      & $ 5 (\phantom{0}6) / 111 (33) $  \\[0.2em] %HQET321
    (A, $\cos\theta_V$)
      & $ 25 (10) / 175 (32) $ %BSZ NF=1
      & $ 16 (11) / \phantom{0}87 (36) $ %BSZ NF=2
      & $ 19 (11) / 173 (33) $ %BGL NF=1     
      & $ 13 (10) / \phantom{0}89 (36) $ %BGL NF=2
      & $ 24 (\phantom{0}9) / 137 (31) $ %HQET210
      & $ 4 (\phantom{0}5) / 107 (33) $  \\[0.2em] %HQET321
    (A, $\chi$)
      & $ 32 (20) / 160 (44) $ %BSZ NF=1
      & $ 19 (27) / \phantom{0}82 (52) $ %BSZ NF=2
      & $ 25 (23) / 150 (51) $ %BGL NF=1     
      & $ 19 (28) / \phantom{0}82 (53) $ %BGL NF=2
      & $ 27 (14) / 131 (41) $ %HQET210
      & $ 8 (16) / \phantom{0}92 (48) $  \\[0.2em] %HQET321
    \midrule
    (B)
      & $ 24 (\phantom{0}9) / 174 (35) $ %BSZ NF=1
      & $ 13 (10) / \phantom{0}98 (35) $ %BSZ NF=2
      & $ 15 (\phantom{0}7) / 180 (33) $ %BGL NF=1     
      & $ 11 (\phantom{0}9) / 103 (35) $ %BGL NF=2
      & $ 19 (\phantom{0}6) / 146 (27) $ %HQET210
      & $ 4 (\phantom{0}6) / 108 (35) $  \\ %HQET321
   \bottomrule
  \end{tabular}
  }
  \caption{
  The $\chi^2_\text{best}$ values with respect to the LCSR data. 
  The values in the brackets are obtained by neglecting correlation among the data points. 
  \label{tab:ChiSqSM_lcsr}
  }
\end{table}

%%%%%
%%%%%
%%%%%
A similar decomposition of $\chi^2_\text{best}$ for the individual LCSR results is shown in Table~\ref{tab:ChiSqSM_lcsr}. 
We find poor fits to LCSR23, while LCSR18 is generally described with $\chi^2_\text{best}/\text{bin} \lesssim 1$.
In particular, the LCSR23 data points are not well reproduced in any fit scenario or form-factor parameterization.
To further investigate this issue, we also evaluate $\chi^2_\text{best}$ by using the same LCSR data points but neglecting their correlations. 
In this case, we find that the LCSR23 values decrease from $\chi^2_\text{best}/\text{bin} \simeq 2.4$ -- $5.6$ to about $0.8$ -- $1.6$, as shown in the brackets in Table~\ref{tab:ChiSqSM_lcsr}. 
This observation suggests that the poor fit is due to the strong correlations among the 34 data points in LCSR23, rather than solely by the large deviations of the individual data points themselves.
A more detailed investigation of the LCSR covariance matrices and their compatibility with the other theoretical and experimental inputs is beyond the scope of the present work.

\subsubsection{Form-factor parameters} 
%%%%%%%%%%%%%%%%%%%%%%%%%%%%%%%%

%%%%%%%%%%%%FIG%%%%%%%%%%%%
\begin{figure}[t]
\begin{center}
\includegraphics[viewport=0 0 563 507, width=0.49\textwidth]{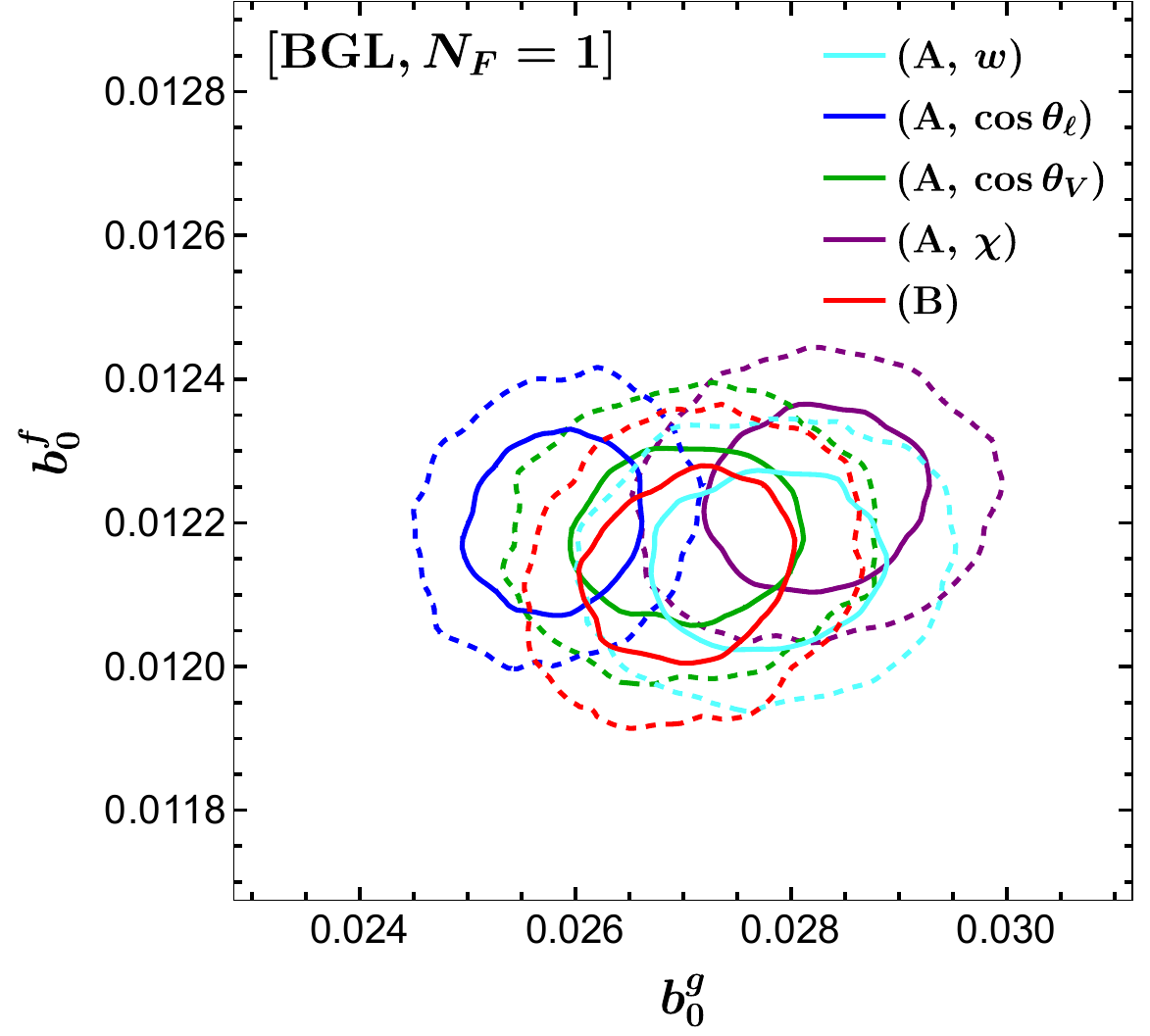}\hfill
\includegraphics[viewport=0 0 563 507, width=0.49\textwidth]{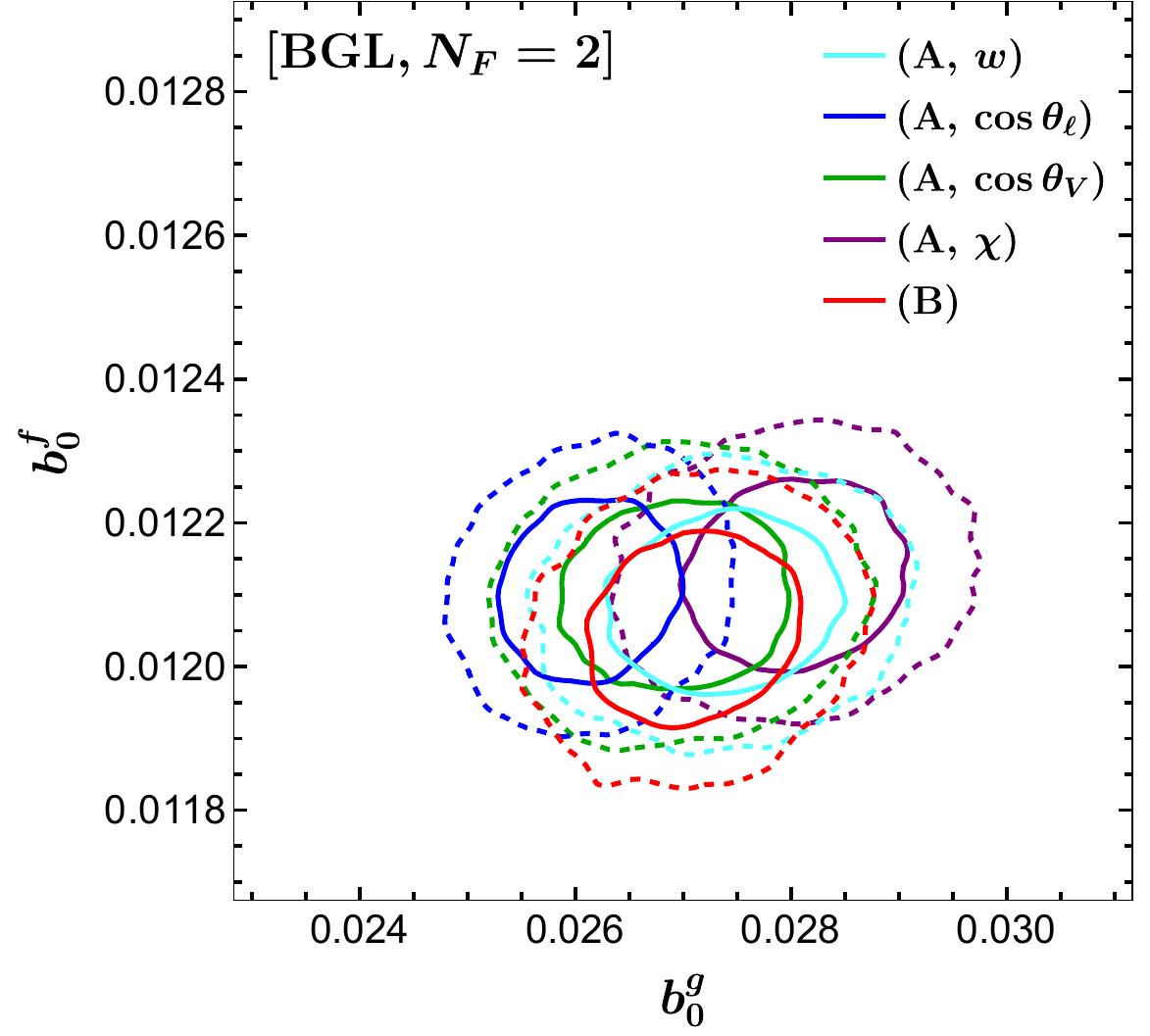}\\[0.5em]
\,\,\,
\includegraphics[viewport=0 0 563 562, width=0.44\textwidth]{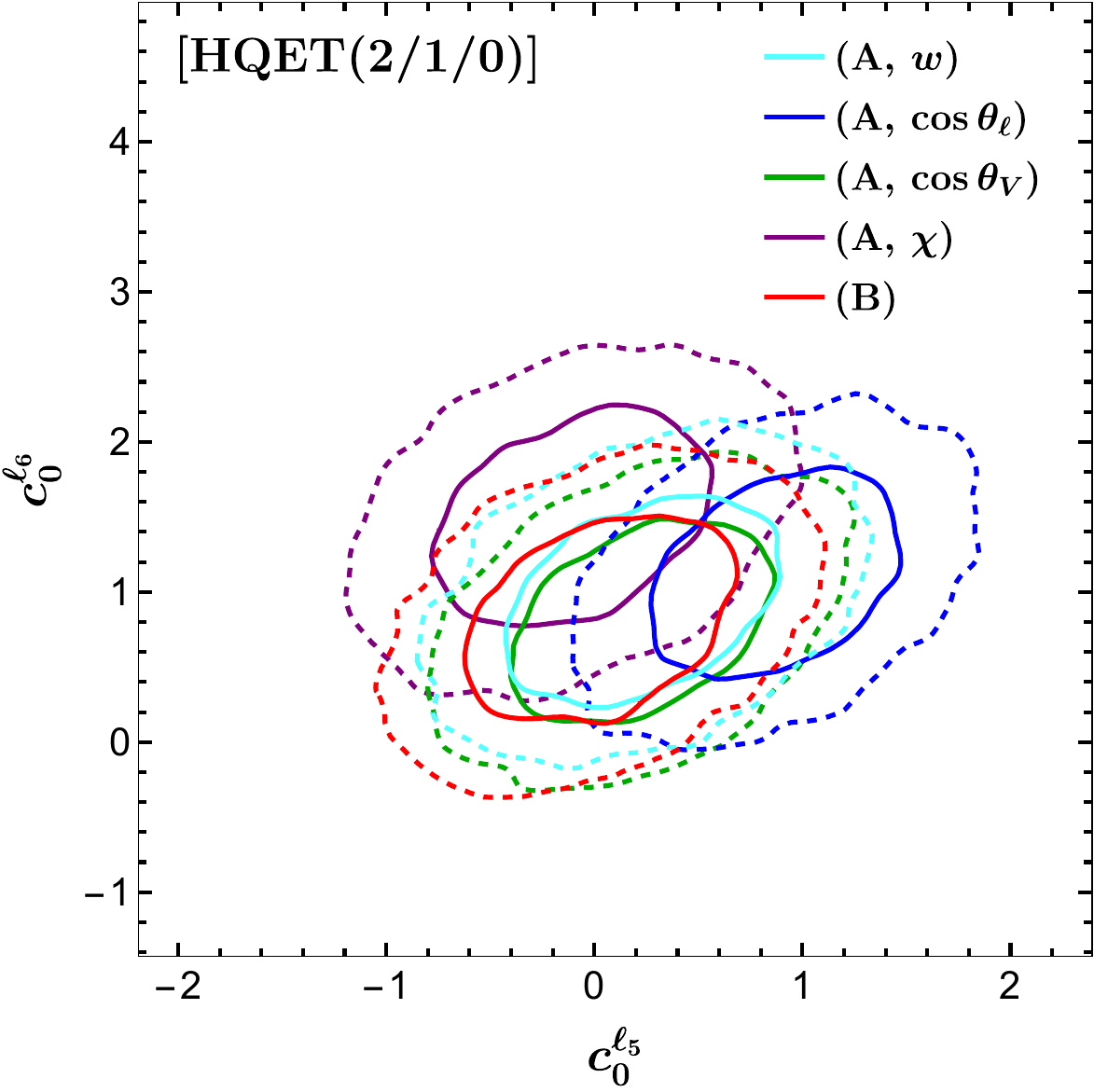}\qquad\quad
\includegraphics[viewport=0 0 563 574, width=0.44\textwidth]{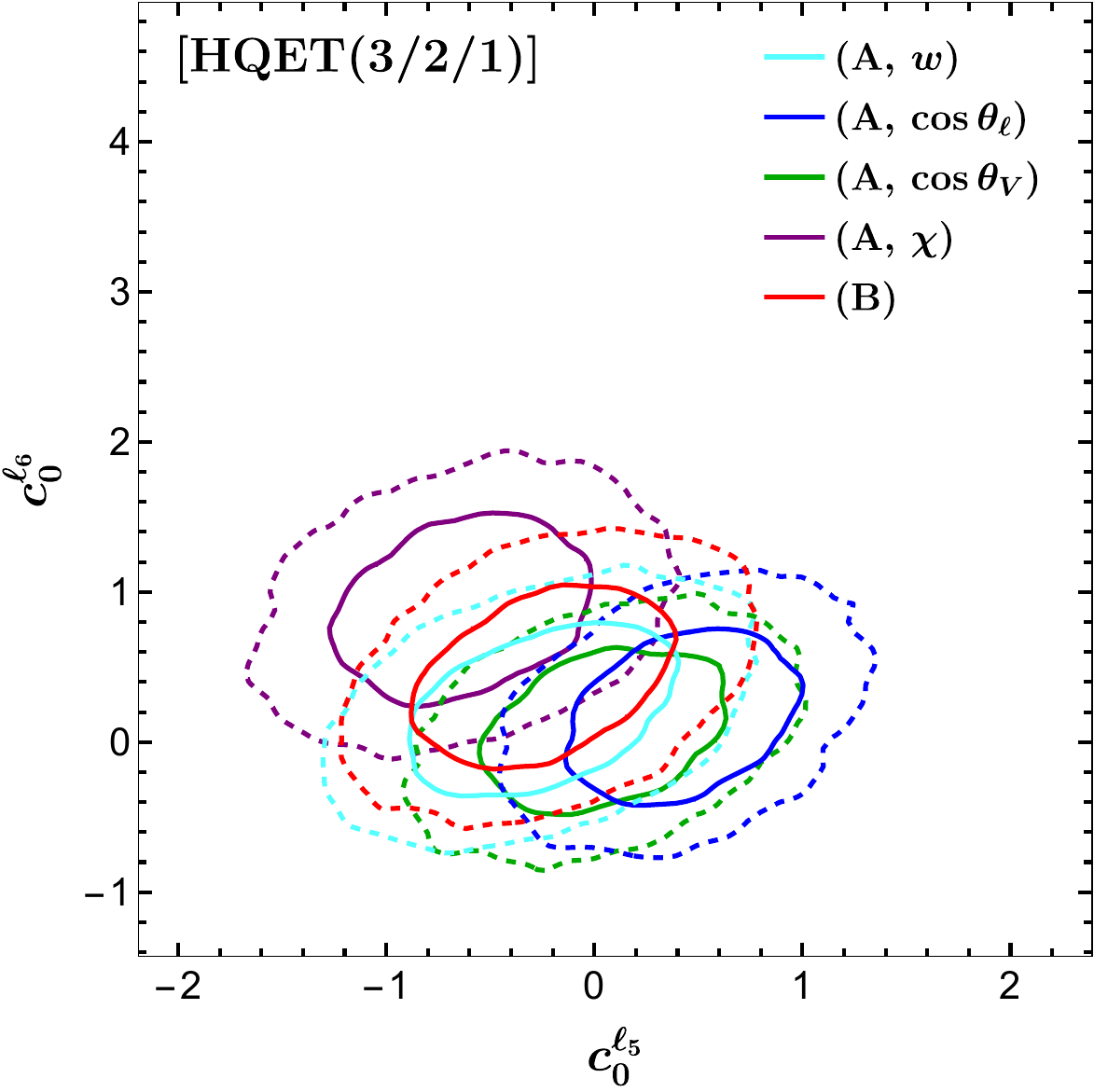}
\caption{
The contour plots for the fit results of the two form-factor parameters in the BGL and HQET parameterizations. 
The form-factor parameters chosen in the figure are those showing the largest differences in the fitted values across different fit scenarios.
\label{Fig:FFcontour}
}
\end{center}
\end{figure}
%%%%%%%%%%%%FIG%%%%%%%%%%%%

We have also obtained our fit results for the form-factor parameters for all fit scenarios and form-factor parameterizations.
All numerical results are shown in appendix~\ref{App:FFresult_all}.
Here, we discuss several interesting features of the form-factor fits.

Among the fitted form-factor parameters, we display in Fig.~\ref{Fig:FFcontour} the contour plots for two representative parameters across different fit scenarios, where the $68\%$ and $95\%$ confidence levels (CLs) are shown with solid and dashed contours, respectively.
From the figure, we observe that the BGL parameterizations show only minor differences between $N_F=1$ and $N_F=2$, while moderate variations appear among the different fit scenarios. Furthermore, the result obtained in scenario (B) appears to be close to an average of those obtained in the four individual $(\text{A},x)$ scenarios.
This behavior is expected because scenario (B) includes all four distribution variables simultaneously.

The HQET parameterizations show visible variations among the different fit scenarios for the subsubleading form-factor parameters $\bl{c_0^{\ell_5}}$ and $\bl{c_0^{\ell_6}}$.
Nevertheless, the 95\% CL regions overlap among the fit scenarios within each parameterization.
In particular, the allowed region in scenario (B) overlaps with those obtained from the individual $(\text{A},x)$ fits.
Similar scenario dependence is also observed in some of the other subsubleading form-factor parameters $\bl{c_n^{\ell_i}}$, although its magnitude varies among the parameters.
By contrast, the leading form-factor parameters $\bl{c_n^{\xi}}$ and the subleading form-factor parameters $\bl{c_n^{\chi_{2,3}}}$ and $\bl{c_n^{\eta}}$ remain relatively stable across the fit scenarios.
These features can also be seen in Table~\ref{Tab:SMfit_HQET}.

Let us recall the key difference between scenarios (A) and (B): in scenario (A), both the form-factor parameters and the CKM matrix element $|V_{cb}|$ are fitted simultaneously; in scenario (B), on the other hand, all the four individual distribution datasets are normalized, and $|V_{cb}|$ is removed from the distribution fit.
The observed consistency between the two scenarios suggests that this normalization procedure has little impact on the extracted form-factor parameters.
This again indicates that the D'Agostini bias effect is not significant in the present analysis, as also seen in the $|V_{cb}|$ determination.

\subsubsection{Prior distributions} 
%%%%%%%%%%%%%%%%%%%%%%%%%%%%%%%%

We now briefly discuss the effects of the prior distributions adopted in the present MCMC fit analysis, as described in section~\ref{sec:procedure}.

For the HQET parameterization, we have additionally performed an analysis with $c_n^F \sim \mathcal N(0,1)$ for all the IW functions, including the subleading ones, instead of the QCD-sum-rule-based setup given in Eqs.~\eqref{eq:QCDSR_1}--\eqref{eq:QCDSR_2}.
We found that the change in the extracted values of $|V_{cb}|$ does not exceed about $2\%$, although the changes in some form-factor parameters are significantly larger.
This indicates that the individual HQET parameters, especially the higher-order ones, can be sensitive to the choice of prior distributions.
However, such variations are largely absorbed by correlations among the form-factor parameters and do not directly propagate to the extracted values of $|V_{cb}|$.
Thus, we conclude that the prior dependence is visible at the level of individual form-factor parameters, but its impact on the $|V_{cb}|$ determination is limited.

For the BSZ and BGL parameterizations, we have also confirmed that the prior distributions do not affect the fit results, unless the favored form-factor regions are close to those constrained by the priors.

%%%%%%%%%%%%%%%%%%%%%%%%%%%%%%%%%%%%%%%%%%%%%%%%%%%
\subsection[New physics effect on \texorpdfstring{$|V_{cb}|$}{Vcb} determination]{New physics effect on \texorpdfstring{$\boldsymbol{|V_{cb}|}$}{Vcb} determination} 
\label{sec:NPstudy}
%%%%%%%%%%%%%%%%%%%%%%%%%%%%%%%%%%%%%%%%%%%%%%%%%%%

We now investigate NP effects on the determination of $|V_{cb}|$ using a simultaneous fit to all normalized distributions and branching ratios.

Scenario (B) includes all four normalized $D^*$-mode distributions in $x = \{w$, $\cos\theta_\ell$, $\cos\theta_V$, $\chi\}$,
but uses \textbf{BrRatio} only in the subsequent extraction of $|V_{cb}|$.
This can lead to poor agreement with \textbf{BrRatio}
when the branching ratios are evaluated using the combined value of $|V_{cb}|$.
In the SM, the deviations are generally modest,
as shown in Table~\ref{tab:ChiSqSM_exp}.
In some NP cases, however, we find substantially larger deviations.
%Our comparisons show, however, that substantially larger deviations can arise in some NP cases.
%
Conversely, scenario $(\text{A},w)$ includes the absolute normalization but does not include the angular distribution data.
Our comparisons show that, in some NP cases, this can result in substantial deviations from the angular distributions,
even when the branching ratios are well reproduced.

For the NP analysis, we therefore extend scenario (B) by including \textbf{BrRatio} directly in the fit likelihood.
In this setup, denoted by scenario (B) + \textbf{BrRatio}, the form-factor parameters, $C_X$, and $|V_{cb}|$ are fitted simultaneously to \textbf{Theory}, \textbf{Distribution (Combined)}, and \textbf{BrRatio}.
We retain the normalized $D$-mode distribution and all four $D^*$-mode distributions, thereby combining their shape information with the normalization constraints from both branching ratios.
We adopt this common setup for the BGL and HQET parameterizations in the following NP analysis.\footnote{
\bl{This is the new setup updated from that adopted in arXiv version~1}.
}

%%%%%%%
%%%%%%%
%%%%%%%

Let us recall that the scalar ($S_{L,R}$) and tensor ($T$) NP currents do not interfere with the SM current in $\bar{B} \to D^{(*)}\ell\bar\nu$ decays in the massless limit of $\ell$, due to spin conservation and negligible light-lepton masses~\cite{Duan:2024ayo}. We also recall that the NP Wilson coefficients $C_X$ are defined as in Eq.~\eqref{eq:Hamiltonian}, where the overall normalization is given by $2\sqrt 2 G_F V_{cb}$. Thus, the total contribution can be schematically written as
\begin{align}
 \label{eq:NPgamma}
 d\Gamma_\text{total}^{D^{(*)}} = |V_{cb}|^2 \left( d\hat\Gamma_\text{SM}^{D^{(*)}} + |C_X|^2 d\hat\Gamma_{X}^{D^{(*)}} \right) \,,  
 \qquad 
 (X=S_{L},\,S_{R},\,T) \,, 
\end{align}
where $d\hat\Gamma_{\text{SM},X}^{D^{(*)}}$ are described in terms of the form-factor parameters. 
This indicates that the relative signs of $C_{S_{L,R}}$ and $C_T$ are not determined. 
On the other hand, the right-handed quark current $V_R$ receives a non-trivial contribution from the SM--$V_R$ interference term, and the total contribution is given by
\begin{align} 
 d\Gamma_\text{total}^{D^{(*)}} = |V_{cb}|^2 \left( d\hat\Gamma_\text{SM}^{D^{(*)}} \pm \text{Re}\, \left[C_{V_R}\right] \, d\hat\Gamma_{\text{SM--}V_R}^{D^{(*)}}  + |C_{V_R}|^2 d\hat\Gamma_{V_R}^{D^{(*)}} \right) \,,
\end{align}
where $\pm$ applies for $\bar{B} \to D\ell\bar\nu$ and $\bar{B} \to D^{*}\ell\bar\nu$ decays, respectively.

%%%%%%%%%%%%FIG%%%%%%%%%%%%
\begin{figure}[t]
\begin{center}
\includegraphics[viewport=0 0 473 466, width=0.45\textwidth]{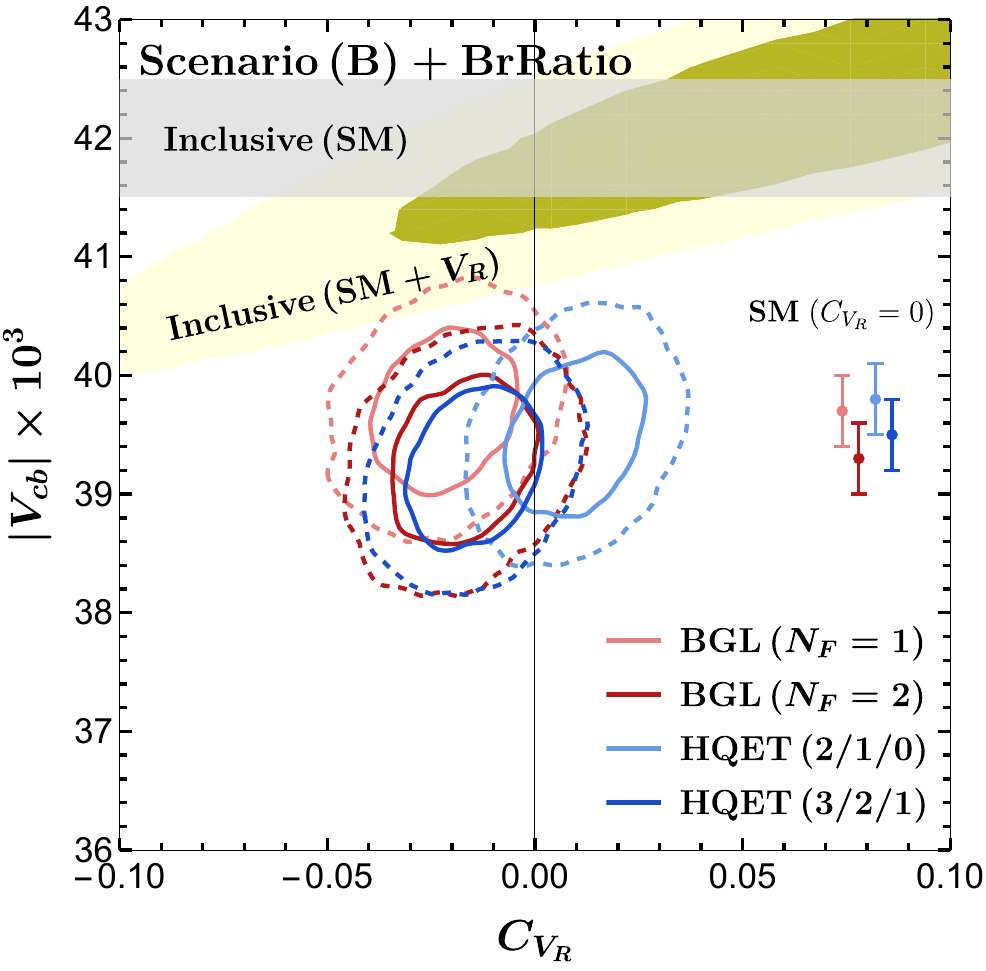}\\
\includegraphics[viewport=0 0 473 473, width=0.45\textwidth]{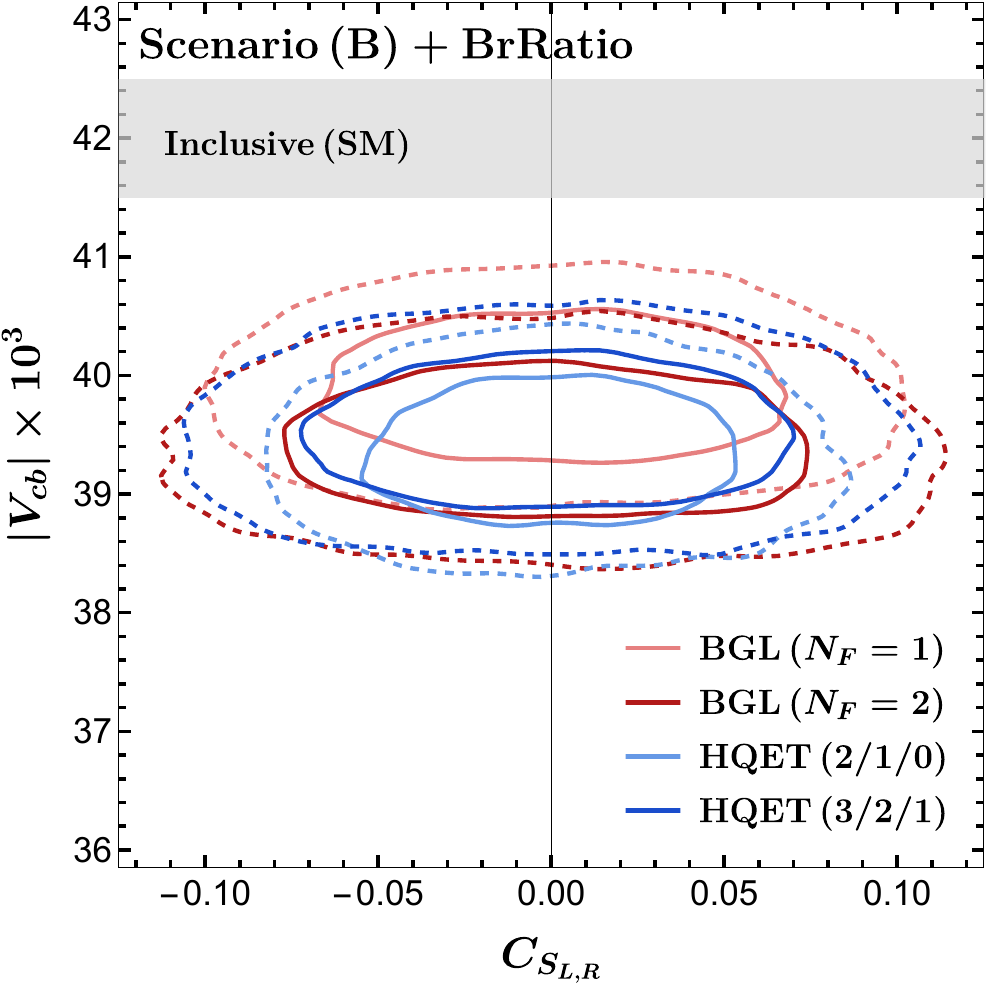}\hfill
\includegraphics[viewport=0 0 473 482, width=0.45\textwidth]{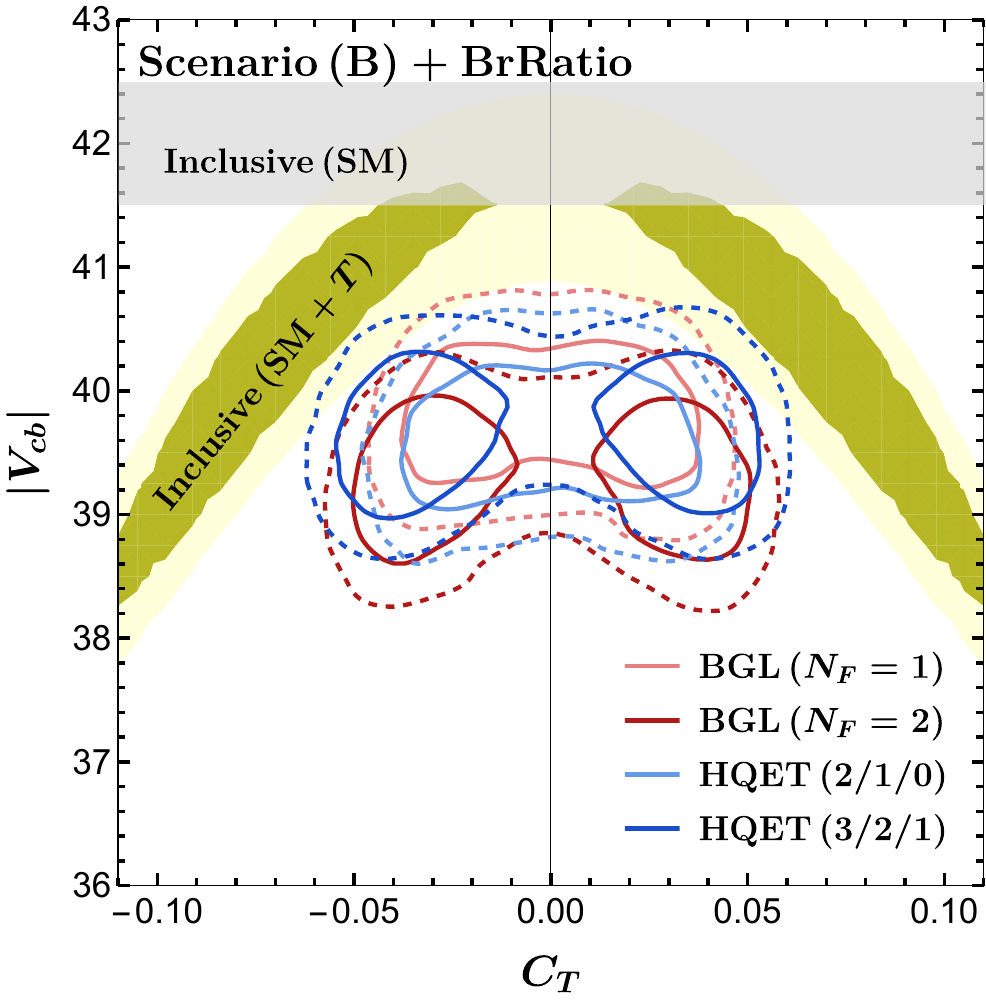}
\caption{
The BGL and HQET fit results in the $(C_X, |V_{cb}|)$ plane for the $V_R$ (top), $S_{L,R}$ (left), and $T$ (right) cases in scenario (B) + \textbf{BrRatio}.
The solid and dashed contours enclose the $68\%$ and $95\%$ CL regions, respectively.
The points with $1\sigma$ error bars at the right edge of the left panel show the combined $|V_{cb}|$ value in the SM from scenario (B) as in Table~\ref{tab:VcbFitSM}.  
%their horizontal positions are shifted for visibility and do not represent non-zero $C_{V_R}$.
The gray bands show the SM inclusive reference for $|V_{cb}|$.
The yellow regions show the inclusive $\text{SM}+V_R$ and $\text{SM}+T$ results of Ref.~\cite{Carvunis:2025vab}, with darker and lighter shading denoting the $68\%$ and $95\%$ CL regions, respectively.
\label{Fig:VcbCVR}
}
\end{center}
\end{figure}
%%%%%%%%%%%%FIG%%%%%%%%%%%%

In Fig.~\ref{Fig:VcbCVR}, we show the $68\%$ and $95\%$ CL regions in the $(C_X, |V_{cb}|)$ plane for the BGL and HQET parameterizations.
The left, middle, and right panels correspond to the $V_R$, $S_{L,R}$, and $T$ type NP contributions, respectively, with one real Wilson coefficient varied at a time.
All results are obtained in scenario (B) + \textbf{BrRatio}, and the allowed regions for the four form-factor parameterizations substantially overlap.
For $V_R$, the favored regions obtained with BGL ($N_F=2$) and HQET $(3/2/1)$ almost coincide in the $(C_{V_R}, |V_{cb}|)$ plane.
It is also interesting that both parameterizations have their best-fit points at $C_{V_R} \neq 0$,
although the $95\%$ CL regions include $C_{V_R}=0$ for all four parameterizations. 
The $S_{L,R}$ results remain consistent with the SM, namely, $C_{S_{L,R}}=0$.
When only one scalar coefficient is non-zero, $S_L$ and $S_R$ give identical decay distributions in the massless-lepton limit, so the middle panel represents either case.
The $T$ results exhibit two branches in the contours at opposite signs of $C_T$, reflecting the dependence on $|C_T|^2$ in Eq.~\eqref{eq:NPgamma}.
Although non-zero tensor contributions are favored around the best-fit regions, the $95\%$ CL regions include $C_T=0$ as well as the $V_R$ results.

\begin{table}[t]
  \centering
  \setlength{\tabcolsep}{12pt}
  \renewcommand{\arraystretch}{1.2}
  \begin{tabular}{c | c c | c c}
    \toprule
      & \multicolumn{4}{c}{$\Delta\chi^2_{\mathrm{data}}$} \\
    \cmidrule(l){2-5}
    \multirow{2}{*}{\textbf{NP}}
      & \multicolumn{2}{c|}{\textbf{BGL}}
      & \multicolumn{2}{c}{\textbf{HQET}} \\
    \cmidrule(lr){2-3} \cmidrule(lr){4-5}
      & $N_F=1$ & $N_F=2$ & $(2/1/0)$ & $(3/2/1)$ \\
    \midrule
    $V_R$ & $3.58$ & $2.19$ & $0.45$ & $2.10$ \\
    $S_L$ & $0.00$ & $0.00$ & $0.00$ & $0.00$ \\
    $T$   & $0.61$ & $3.13$ & $0.32$ & $2.19$ \\
    \bottomrule
  \end{tabular}
  \caption{
    Changes in the data contribution to $\chi^2$ in
    scenario (B) + \textbf{BrRatio}, defined as
    $\Delta\chi^2_{\mathrm{data}}
    = \chi^2_{\mathrm{data},\,C_X=0}
    - \chi^2_{\mathrm{data},\,\mathrm{NP}}$.
    Both terms are evaluated at the respective
    maximum a posteriori points using the same input datasets,
    with the form-factor parameters and $|V_{cb}|$ refitted
    in the $C_X=0$ reference fit.
    For the tensor case, the tensor form-factor inputs
    are retained in the reference fit.
  }
  \label{tab:NPDeltaChiSq}
\end{table}

To assess the improvement due to NP, we compare each NP fit with a reference fit in the same scenario (B) + \textbf{BrRatio}, 
fixing $C_X=0$ and refitting the form-factor parameters and $|V_{cb}|$. 
The same input datasets are used in each comparison, including the tensor form-factor inputs for the tensor case.
Then, we quantify the change in agreement with the fitted data by
\begin{align}
  \Delta\chi^2_{\mathrm{data}}
  \equiv \chi^2_{\mathrm{data},\,C_X=0}
       - \chi^2_{\mathrm{data},\,\mathrm{NP}} \,.
  \label{eq:NPDeltaChiSq}
\end{align}
Both terms are evaluated at the respective maximum a posteriori (MAP) points.
Here, $\chi^2_{\mathrm{data}}$ includes the contributions from the selected \textbf{Theory}, \textbf{Distribution (Combined)}, and \textbf{BrRatio} datasets. 
%but excludes the prior and unitarity-bound penalties.
Positive values indicate better agreement with the data at the MAP point in the presence of NP.
We use $\Delta\chi^2_{\mathrm{data}}$ to quantify the improvement in agreement with the data.
Assessing the statistical significance of this improvement would require a dedicated hypothesis test, which is beyond the scope of this paper.

The results are summarized in Table~\ref{tab:NPDeltaChiSq}.
Focusing on BGL ($N_F=2$) and HQET $(3/2/1)$, we find similar improvements for $V_R$, with $\Delta\chi^2_{\mathrm{data}} \approx 2$.
The two parameterizations thus give similar improvements as well as similar favored regions of $(C_{V_R}, |V_{cb}|)$ as in Fig.~\ref{Fig:VcbCVR}.
For the $T$ case, the corresponding reductions are $\approx 3$ and $2$, while the $S_{L,R}$ case gives no improvement in either parameterization.
Considering the fitted values of $\chi^2_{\mathrm{data},\,C_X=0}$ and $\chi^2_{\mathrm{data},\,\mathrm{NP}}$, we find that the reductions amount to less than $1.5\%$ of the reference values, indicating only a small relative improvement in the overall agreement with the fitted datasets.

The left and right panels of Fig.~\ref{Fig:VcbCVR} also allow a comparison with the inclusive analyses of Ref.~\cite{Carvunis:2025vab}.
The preferred exclusive regions lie at smaller $|V_{cb}|$ than the corresponding inclusive regions, although the outer allowed regions can overlap.
Together with the modest improvements in Table~\ref{tab:NPDeltaChiSq}, these results indicate that allowing $V_R$ or tensor contributions does not provide a clear resolution of the inclusive--exclusive $|V_{cb}|$ discrepancy in the present analysis.

%%%%%%%%%%%%%%%%%%%%%%%%%%%%%%%%%%%%%%%%%%%%%%%%%%%
\subsection[Predictions on \texorpdfstring{$R_D$}{RD} and \texorpdfstring{$R_{D^*}$}{RDstar}]{Predictions on \texorpdfstring{$\boldsymbol{R_D}$}{RD} and \texorpdfstring{$\boldsymbol{R_{D^*}}$}{RDstar}} 
%%%%%%%%%%%%%%%%%%%%%%%%%%%%%%%%%%%%%%%%%%%%%%%%%%%

With our fitted results for the $\bar{B} \to D^{(*)}$ form-factors as input, we now investigate the LFU ratios $R_D$ and $R_{D^*}$ introduced in section~\ref{sec:Intro}. 
%\begin{align}
% R_D = \frac{\Gamma(\bar{B} \to D\tau\bar\nu)}{\Gamma(\bar{B} \to D\ell\bar\nu)} \,, 
% \qquad 
% R_{D^*} = \frac{\Gamma(\bar{B} \to D^*\tau\bar\nu)}{\Gamma(\bar{B} \to D^*\ell\bar\nu)} \,, 
%\end{align}
%for $\ell = e, \mu$. 
As is widely known, these two observables have been receiving a lot of attention since the BaBar measurements~\cite{BaBar:2012obs,BaBar:2013mob} showed significant deviations from the SM predictions. 
The most recent world averages are reported as~\cite{HFLAV2025CKM}
\begin{align} \label{eq:exp-HFLAV}
 R_D^\text{ave} = 0.358 \pm 0.024 \,,  
 \qquad 
 R_{D^*}^\text{ave} = 0.281 \pm 0.011\,, 
\end{align}
while the HFLAV arithmetic averages of the SM predictions are~\cite{HFLAV2025CKM}
\begin{align} \label{eq:SM-HFLAV}
 R_D^\text{SM-HFLAV} = 0.296 \pm 0.004 \,,  
 \qquad 
 R_{D^*}^\text{SM-HFLAV} = 0.254 \pm 0.005\,. 
\end{align}

\begin{table}[ht!]
  \centering
  \resizebox{\textwidth}{!}{
  \renewcommand{\arraystretch}{1.3}
  \begin{tabular}{lcccc}
    \toprule
    \textbf{Fit scenario}
    & \textbf{BGL ($N_F=1$)}
    & \textbf{BGL ($N_F=2$)}
    & \textbf{HQET ($2/1/0$)}
    & \textbf{HQET ($3/2/1$)}
    \\
    \midrule
    \multicolumn{5}{c}{$R_D$} \\
    \midrule
    (A, $w$)
      & $0.296 \pm 0.002$ %D: BGL NF=1
      & $0.288 \pm 0.003$ %D: BGL NF=2
      & $0.310 \pm 0.002$ %D: HQET210
      & $0.297 \pm 0.003$ %D: HQET321
      \\
    (A, $\cos\theta_\ell$)
      & $0.296 \pm 0.002$ %D: BGL NF=1
      & $0.289 \pm 0.003$ %D: BGL NF=2
      & $0.312 \pm 0.002$ %D: HQET210
      & $0.298 \pm 0.003$ %D: HQET321
      \\
    (A, $\cos\theta_V$)
      & $0.296 \pm 0.002$ %D: BGL NF=1
      & $0.289 \pm 0.003$ %D: BGL NF=2
      & $0.312 \pm 0.002$ %D: HQET210
      & $0.298 \pm 0.002$ %D: HQET321
      \\
    (A, $\chi$)
      & $0.298 \pm 0.002$ %D: BGL NF=1
      & $0.291 \pm 0.003$ %D: BGL NF=2
      & $0.313 \pm 0.002$ %D: HQET210
      & $0.300 \pm 0.003$ %D: HQET321
      \\
    (B)
      & $0.295 \pm 0.002$ %D: BGL NF=1
      & $0.288 \pm 0.003$ %D: BGL NF=2
      & $0.315 \pm 0.003$ %D: HQET210
      & $0.298 \pm 0.003$ %D: HQET321
      \\
    \midrule
    \multicolumn{5}{c}{$R_{D^*}$} \\
    \midrule
    (A, $w$)
      & $0.252 \pm 0.001$ %Dst: BGL NF=1
      & $0.252 \pm 0.001$ %Dst: BGL NF=2
      & $0.253 \pm 0.001$ %Dst: HQET210
      & $0.253 \pm 0.001$ %Dst: HQET321
      \\
    (A, $\cos\theta_\ell$)
      & $0.254 \pm 0.002$ %Dst: BGL NF=1
      & $0.256 \pm 0.002$ %Dst: BGL NF=2
      & $0.254 \pm 0.002$ %Dst: HQET210
      & $0.254 \pm 0.002$ %Dst: HQET321
      \\
    (A, $\cos\theta_V$)
      & $0.254 \pm 0.002$ %Dst: BGL NF=1
      & $0.256 \pm 0.002$ %Dst: BGL NF=2
      & $0.255 \pm 0.001$ %Dst: HQET210
      & $0.255 \pm 0.002$ %Dst: HQET321
      \\
    (A, $\chi$)
      & $0.262 \pm 0.002$ %Dst: BGL NF=1
      & $0.262 \pm 0.002$ %Dst: BGL NF=2
      & $0.260 \pm 0.002$ %Dst: HQET210
      & $0.260 \pm 0.002$ %Dst: HQET321
      \\
    (B)
      & $0.253 \pm 0.001$ %Dst: BGL NF=1
      & $0.253 \pm 0.002$ %Dst: BGL NF=2
      & $0.254 \pm 0.001$ %Dst: HQET210
      & $0.254 \pm 0.001$ %Dst: HQET321
      \\
    \bottomrule
  \end{tabular}
  }
  \caption{
  Fit results for the ratios $R_D$ and $R_{D^*}$ for all fit scenarios in the BGL and HQET parameterizations.
  \label{tab:RDsummary}
  }
\end{table}

In Table~\ref{tab:RDsummary}, we summarize our SM predictions for the ratios $R_D$ and $R_{D^*}$.
For $R_D$, the dependence on the form-factor parameterization is more evident than that on the fit scenario.
It may be interesting that the BGL ($N_F=2$) and HQET $(3/2/1)$ predictions lie in the different ranges $0.288$--$0.291$ and $0.297$--$0.300$, respectively, 
in spite of the good agreement between their fitted values of $|V_{cb}|$. 
The HQET $(3/2/1)$ predictions are then consistent with the HFLAV SM average.
For $R_{D^*}$, the four parameterizations give similar results within each fit scenario and their values are consistent with the HFLAV SM average, 
while the $(\text{A},\chi)$ scenario yields systematically higher values than the other scenarios.
The difference in $R_D$ may reflect the treatment of $f_0$ ($B \to D$) and $F_2$ ($B \to \Dst$), 
whose contributions are suppressed for light leptons but important for $\tau$ leptons. 
HQET relates these form-factors to the others, whereas BGL parameterizes them separately.
In any case, the tension with the experimental average persists in the present SM analysis.

%%%%%%%%%%%%FIG%%%%%%%%%%%%
\begin{figure}[ht!]
\begin{center}
\includegraphics[viewport=0 0 473 451, width=0.32\textwidth]{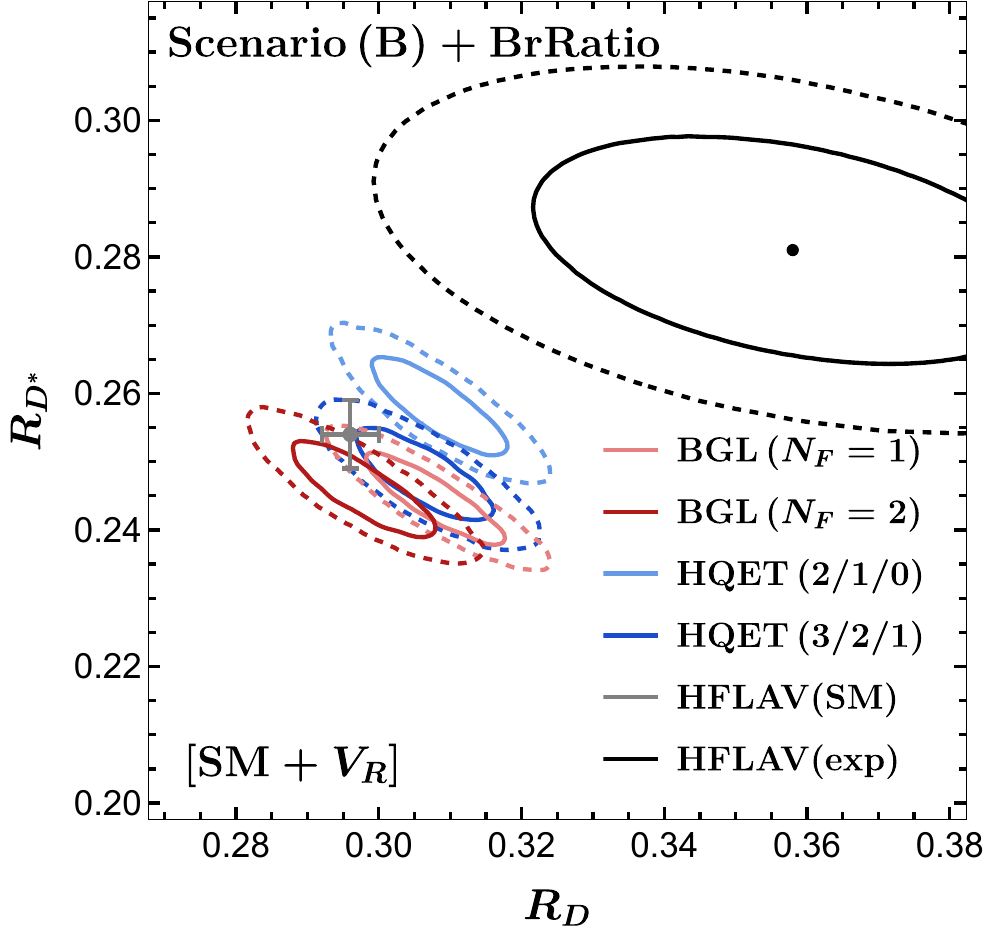}\,
\includegraphics[viewport=0 0 473 451, width=0.32\textwidth]{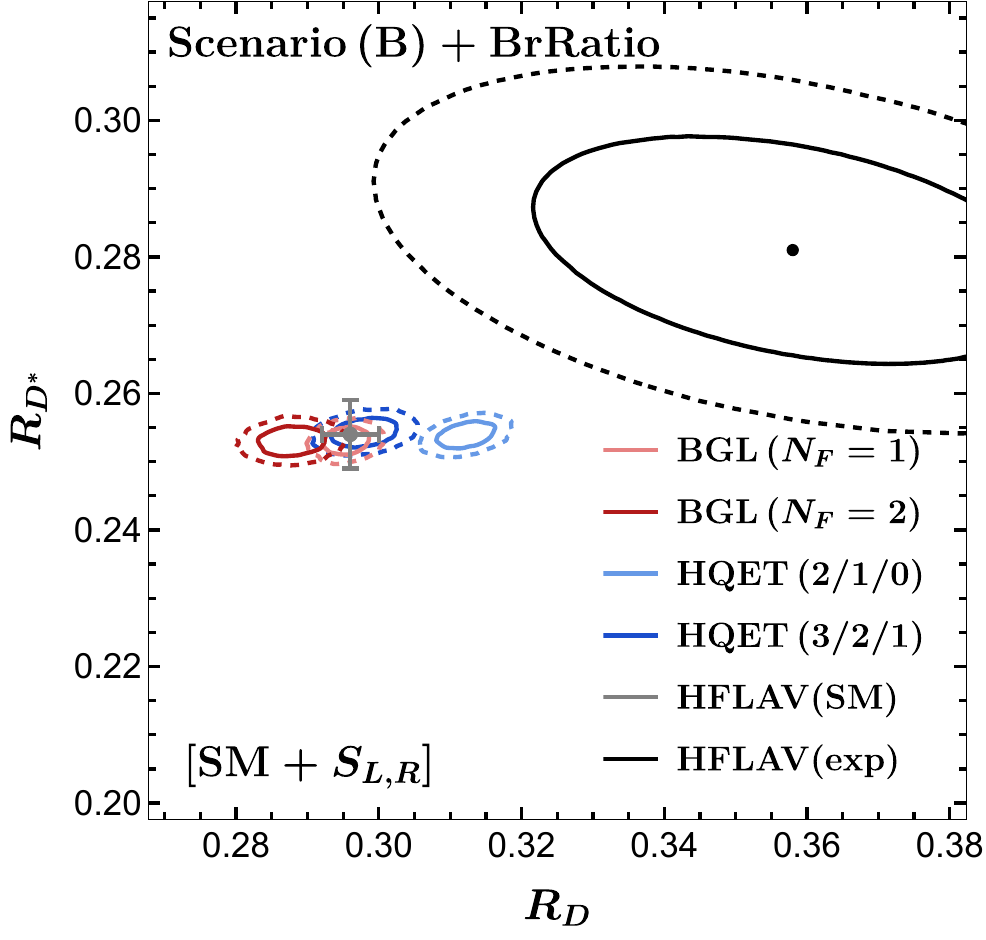}\,
\includegraphics[viewport=0 0 473 451, width=0.32\textwidth]{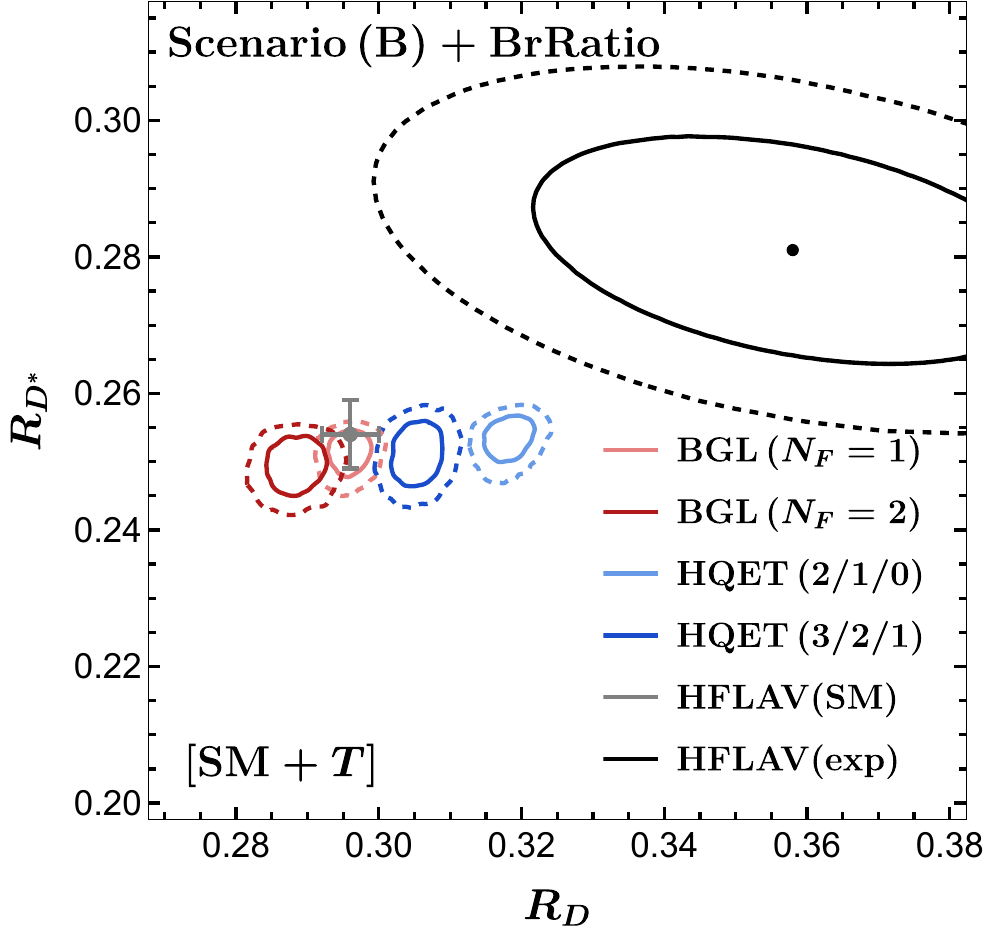}
\caption{
 Fit contours on the $(R_D, R_{D^*})$ plane in each fit scenario for SM $+$ NP studies,  
 where only $\Gamma(\bar{B} \to D^{(*)}\ell\bar\nu)$ of $R_{D^{(*)}}$ is assumed to be affected by the NP contribution, whereas $\Gamma(\bar{B} \to D^{(*)}\tau\bar\nu)$ is given by the SM. 
 The $68\%$ and $95\%$ CLs are represented by the solid and dashed contours, respectively. 
\label{Fig:RDplot}
}
\end{center}
\end{figure}
%%%%%%%%%%%%FIG%%%%%%%%%%%%

Figure~\ref{Fig:RDplot} presents the $68\%$ and $95\%$ CL regions in the $(R_D,R_{D^*})$ plane for the $V_R$, $S_{L,R}$, and $T$ cases, all obtained in scenario (B) + \textbf{BrRatio}.
The NP contributions are included only in the light-lepton decay rates in the denominators, while the $\tau$-mode rates in the numerators are evaluated in the SM using the fitted form factors.
The $V_R$ panel exhibits a strong negative correlation between $R_D$ and $R_{D^*}$ for all four parameterizations: larger values of $R_D$ are accompanied by smaller values of $R_{D^*}$.
This direction of variation does not bring both ratios closer to their experimental averages simultaneously.
Although BGL ($N_F=2$) and HQET $(3/2/1)$ give nearly coincident favored regions in the $(C_{V_R},|V_{cb}|)$ plane, 
their predictions for $R_D$ differ, while those for $R_{D^*}$ remain similar. 
This shows that the dependence of form-factor parameterization remains relevant for $R_D$ as also seen in the SM.
The $S_{L,R}$ predicted regions are compact and remain close to the corresponding SM predictions in Table~\ref{tab:RDsummary}.
In the $T$ panel, the predictions for $R_D$ span a wide range depending on the form-factor parameterization.

%%%%%%%%%%%%%%%%%%%%%%%%%%%%%%%%%%%%%%%%%%%%%%%%%%%
\section{Conclusion} 
\label{sec:summary}
%%%%%%%%%%%%%%%%%%%%%%%%%%%%%%%%%%%%%%%%%%%%%%%%%%%

We have studied the $|V_{cb}|$ determinations from the exclusive semi-leptonic $\bar{B}\to D^{(*)}\ell\bar\nu$ decays by performing comprehensive Bayesian fit analyses of the form-factors and decay distributions.
In particular, we have investigated how the extracted values of $|V_{cb}|$ depend on the choice of form-factor parameterization and on the treatment of the experimental distribution data.
For this purpose, we considered the BSZ, BGL and HQET parameterizations and compared two classes of SM fit scenarios:
in scenario (A), the decay distributions and branching ratios are fitted simultaneously, with one $D^*$-mode distribution variable included at a time;
and in scenario (B), corresponding to the PDG-type setup, the combined normalized distribution data including all four $D^*$-mode variables are used to constrain the form-factors and $|V_{cb}|$ is determined only from the branching ratios.

Our main results for the SM fit are summarized in Table~\ref{tab:VcbFitSM}.
We find that the BSZ and BGL parameterizations give very similar results for $N_F=2$, as expected from the fact that the form-factors are treated independently in both parameterizations.
In scenario (B), their combined results, as well as that obtained with HQET $(3/2/1)$, are consistent with the official PDG determination of $|V_{cb}|$.
In scenario (A), the $(\text{A},w)$, $(\text{A},\cos\theta_\ell)$, and $(\text{A},\cos\theta_V)$ fits give mutually consistent $|V_{cb}|$ values across all the parameterizations, whereas $(\text{A},\chi)$ gives a larger $|V_{cb}|$.

We have then examined the consistency of the fitted form-factors by studying the binned determinations of $|V_{cb}|$.
As shown in Fig.~\ref{Fig:BinVcb}, the $\bar{B}\to D\ell\bar\nu$ distribution gives broadly consistent binned values for all fit scenarios with BGL ($N_F=1,2$) and HQET $(3/2/1)$. 
For the $\bar{B}\to D^*\ell\bar\nu$ mode, BGL ($N_F=2$) and HQET $(3/2/1)$ give broadly consistent binned values in the $(\text{A},w)$ and (B) scenarios 
and remain compatible with the PDG exclusive average, although some deviations are visible in individual bins.
The $(\text{A},\cos\theta_\ell)$ and $(\text{A},\cos\theta_V)$ scenarios provide stable values for the corresponding angular distributions, 
but show less stable binned values in the $w$ distribution, which are visible even for BGL ($N_F=2$) and HQET $(3/2/1)$.
The $(\text{A},\chi)$ scenario does not give stable binned values for the other distribution observables in either the BGL or the HQET parameterization.
These observations suggest that the $w$ distribution provides a particularly robust constraint on the form-factor shapes, while the angular distributions contain complementary information that becomes especially useful when all these distributions are combined, as discussed in scenario (B).

Taking these observations together, and also considering the standard fit strategies adopted by PDG and HFLAV, we quote the following values from scenario (B):
\begin{align*}
 |V_{cb}| \times 10^3 =
 \begin{cases}
 39.3 \pm 0.3 & \text{BGL}~(N_F=2)  \\[0.3em]
 39.5 \pm 0.3 & \text{HQET}~(3/2/1)
 \end{cases}\,,
\end{align*}
as our representative results of the CKM matrix element $|V_{cb}|$.
We also emphasize that the other fit scenarios summarized in Table~\ref{tab:VcbFitSM} are comparably important to assess the systematic dependence associated with the treatment of the distribution data and the choice of form-factor parameterization.
With nominal uncertainties of $0.3\times10^{-3}$, corresponding to less than $1\%$, our combined analyses using both BGL ($N_F=2$) and HQET $(3/2/1)$ place the exclusive extraction of $|V_{cb}|$ in the percent-precision regime. 
Further progress will require dedicated statistical tests to assess the consistency between the $D$- and $D^*$-mode determinations, together with improved control of parameterization dependence, HQET truncation effects, and correlations among the inputs.

We have further studied possible NP effects in the light-lepton modes in terms of model-independent four-fermion operators given by Eq.~\eqref{eq:Hamiltonian}, 
using scenario (B) + \textbf{BrRatio}, in which the form-factor parameters, $|V_{cb}|$, and one real Wilson coefficient are fitted simultaneously to
\textbf{Theory}, \textbf{Distribution}, and \textbf{BrRatio} inputs.\footnote{
\bl{Let us repeat that this is the new setup updated from that adopted in arXiv version~1}.
}
Our main results are summarized in Fig.~\ref{Fig:VcbCVR} and Table~\ref{tab:NPDeltaChiSq}.
The allowed regions substantially overlap among the BGL and HQET parameterizations; in particular, BGL ($N_F=2$) and HQET $(3/2/1)$ give nearly coincident favored regions for $V_R$.
Although non-zero $V_R$ and $T$ contributions are favored near the best-fit regions, the $95\%$ CL regions include zero ($C_X=0$) for all four parameterizations.
The $S_{L,R}$ contributions also remain consistent with zero and give no improvement in agreement with the data.
%\rd{Relative to the $C_X=0$ reference fits with the same inputs and all other parameters refitted,} 
The BGL ($N_F=2$) and HQET $(3/2/1)$ fits give modest reductions of $\Delta\chi^2_{\mathrm{data}}\approx 2, 3$ for $V_R$ and for $T$.
These reductions quantify the improvement at the fitted points but a dedicated hypothesis test would be needed to assess their statistical significance.
In the end, the $|V_{cb}|$ puzzle remains unresolved in the present analysis.

Finally, using the fitted form-factors, we have evaluated the SM predictions for the two interesting ratios $R_D$ and $R_{D^*}$, which are summarized in Table~\ref{tab:RDsummary}.
For $R_D$, our predictions point to a more visible dependence on the form-factor parameterization than on the fit scenario.
Despite their similar $|V_{cb}|$ determinations, BGL ($N_F=2$) and HQET $(3/2/1)$ give $R_D$ values in the distinct ranges as seen in the table.
%$0.288$--$0.291$ and $0.297$--$0.300$, respectively.
The latter agree with the HFLAV SM average, while the BGL ($N_F=2$) values are somewhat lower.
For $R_{D^*}$, the four parameterizations give similar predictions within each fit scenario (except for the $(\text{A},\chi)$ fit), which remains consistent with the HFLAV SM average.
The deviation with experimental averages still persists within the present SM analysis. 
Following the discussions of the $|V_{cb}|$ determinations, we quote the values obtained in scenario (B) as our representative SM predictions: 
\begin{align*}
 R_D =
 \begin{cases}
 0.288 \pm 0.003 & \text{BGL}~(N_F=2)  \\[0.3em]
 0.298 \pm 0.003 & \text{HQET}~(3/2/1)
 \end{cases}\,,
\quad
 R_{D^*} =
 \begin{cases}
 0.253 \pm 0.002 & \text{BGL}~(N_F=2) \\[0.3em]
 0.254 \pm 0.001 & \text{HQET}~(3/2/1)
 \end{cases} \,.
\end{align*}

Future improvements in both experimental distribution measurements and theoretical form-factor inputs will be important for further clarifying the exclusive determination of $|V_{cb}|$ and its possible connection to NP effects.
In particular, the agreement between the $D$- and $D^*$-mode determinations within HQET $(3/2/1)$ supports a simultaneous description of the two transitions, while the differences observed for HQET $(2/1/0)$ highlight the importance of the truncation of the heavy-quark expansion.
Future high-precision data~\cite{LHCb:2018roe,Belle-II:2018jsg} will allow more stringent tests of this description and help clarify the effects of higher-order radiative and power corrections and of correlations among the theoretical and experimental inputs.

%%%%%%%%%%%%%%%%%%%%%%%%%%%%%
\acknowledgments

The authors thank Paolo Gambino and Martin Jung for providing numerical plots for their study of the inclusive process.
The authors also thank Akimasa Ishikawa for a discussion about the $\bar B \to D \ell\bar\nu$ distribution data.
The work of Ryoutaro Watanabe is supported by the National Natural Science Foundation of China under Grant Nos.~12635005, and 12675116.
The work of Xin-Qiang Li is supported by the National Natural Science Foundation of China under Grant Nos.~12635005, 12475094, and 12135006.
The work of Syuhei Iguro is supported by Toyoaki scholarship foundation and also by JSPS KAKENHI Grant Numbers 22K21347, 24K22879, 24K23939, and 25K17385.

%%%%%%%%%%%%%%%%%%%%%%%%%%%%%
\appendix
\section{\boldmath Analytical formulae of \texorpdfstring{$\alpha_s$}{alphas} corrections in HQET}
\label{App:FFinput}
%%%%%%%%%%%%%%%%%%%%%%%%%%%%%

In this appendix, we collect the explicit expressions of the one-loop $\mathcal{O}(\alpha_s)$ correction terms $\delta \hat{h}_{X}^{(\alpha_s)}$ and $\delta \hat{h}_{X}^{(\alpha_sm_c)}$ for the HQET form-factors, as defined by Eq.~\eqref{eq:HQET-hat-hX}. Following the direct calculations performed in Refs.~\cite{Bernlochner:2017jka,Bernlochner:2022ywh,Neubert:1992qq}, we have
\begin{align}
\delta \hat h_{+}^{(\alpha_s)} &= C_{V_1}+\frac{w+1}{2}(C_{V_2}+C_{V_3}) \,, \\[0.3em]
\delta \hat h_{-}^{(\alpha_s)} &= \frac{w+1}{2}(C_{V_2}-C_{V_3}) \,, \\[0.3em]
\delta \hat h_{S}^{(\alpha_s)} &= C_S \,, \\[0.3em]
\delta \hat h_{T}^{(\alpha_s)} &= C_{T_1}-C_{T_2}+C_{T_3} \,, \\[0.3em]
\delta \hat h_{V}^{(\alpha_s)} &= C_{V_1} \,,\\[0.3em]
\delta \hat h_{A_1}^{(\alpha_s)} &= C_{A_1} \,,\\[0.3em]
\delta \hat h_{A_2}^{(\alpha_s)} &= C_{A_2} \,, \\[0.3em]
\delta \hat h_{A_3}^{(\alpha_s)} &= C_{A_1}+C_{A_3} \,, \\[0.3em]
\delta \hat h_{P}^{(\alpha_s)} &= C_{P} \,, \\[0.3em]
\delta \hat h_{T_1}^{(\alpha_s)} &= C_{T_1}+\frac{w-1}{2}(C_{T_2}-C_{T_3}) \,, \\[0.3em]
\delta \hat h_{T_2}^{(\alpha_s)} &= \frac{w+1}{2}(C_{T_2}+C_{T_3}) \,, \\[0.3em]
\delta \hat h_{T_3}^{(\alpha_s)} &= C_{T_2} \,,
\end{align}
for the $\alpha_s$ corrections, and  
\begin{align}
    \delta \hat h_{+}^{(\alpha_sm_c)}
    =
    &\, C_{V_1}L_1+\frac{w+1}{2}(C_{V_2}+C_{V_3})\left(L_1-\frac{w-1}{w+1}L_4\right) \nonumber\\[0.2em]
    & +2(w-1)\left[C_{V_1}^\prime+\frac{w+1}{2}(C_{V_2}^\prime+C_{V_3}^\prime)\right] +(w-1)C_{V_3}L_5+C_g^c L_1 \,, \\[0.5em]
%%%%%%%%%%%%%%%%%%%%%%%%%%
   \delta \hat h_{-}^{(\alpha_sm_c)}
   =
   &\, \frac{w+1}{2}\left[(C_{V_2}-C_{V_3})\left(L_1-\frac{w-1}{w+1}L_4\right)+2(w-1)(C_{V_2}^\prime-C_{V_3}^\prime)\right] \nonumber\\[0.2em]
   & +C_{V_1}L_4+(w+1)C_{V_3}L_5 \,, \\[0.5em]
%%%%%%%%%%%%%%%%%%%%%%%%%%
    \delta \hat h_{S}^{(\alpha_sm_c)}
    = 
    &\, C_S\left(L_1-\frac{w-1}{w+1}L_4\right)+2(w-1)C_S^\prime+C_g^c L_1 \,, \\[0.5em]
%%%%%%%%%%%%%%%%%%%%%%%%%%
   \delta \hat h_{T}^{(\alpha_sm_c)}
   = 
   &\, (C_{T_1}-C_{T_2}+C_{T_3})L_1-C_{T_1}L_4+2(w-1)(C_{T_1}^\prime-C_{T_2}^\prime+C_{T_3}^\prime) \nonumber\\[0.3em]
   & +(C_{T_2}+C_{T_3})L_4-2C_{T_3}L_5+C_g^c L_1 \,, \\[0.5em]
%%%%%%%%%%%%%%%%%%%%%%%%%%
%%%%%%%%%%%%%%%%%%%%%%%%%%
 \delta \hat h_{V}^{(\alpha_sm_c)}
 = 
 &\, C_{V_1}(L_2-L_5)-C_{V_3}(L_4-L_5)+2(w-1)C_{V_1}^\prime+C_g^c L_2 \,, \\[0.5em]
%%%%%%%%%%%%%%%%%%%%%%%%%%
 \delta \hat h_{A_1}^{(\alpha_sm_c)}
 = 
 &\, C_{A_1}\left(L_2-\frac{w-1}{w+1}L_5\right)+2(w-1)C_{A_1}^\prime+\frac{w-1}{w+1}C_{A_3}(L_4-L_5)+C_g^c L_2 \,, \\[0.5em]
%%%%%%%%%%%%%%%%%%%%%%%%%%
 \delta \hat h_{A_2}^{(\alpha_sm_c)}
 = 
 &\, C_{A_1}(L_3+L_6)+C_{A_2}\left[L_2+(w-1) L_3+L_5-(w+1)L_6\right]\nonumber\\[0.2em]
 & -\frac{1}{w+1}C_{A_3}(L_4-3L_5)+2(w-1)C_{A_2}^\prime+C_g^c L_3 \,, \\[0.5em]
%%%%%%%%%%%%%%%%%%%%%%%%%%
 \delta \hat h_{A_3}^{(\alpha_sm_c)}
 =
 &\, C_{A_1}\left(L_2-L_3-L_5+L_6\right)+\frac{w}{w+1}C_{A_3}(L_4-3L_5)\nonumber\\[0.2em]
 & +C_{A_3}\left[L_2+(w-1)L_3+L_5-(w+1)L_6\right] \nonumber \\[0.2em]
 & +2(w-1)\left(C_{A_1}^\prime +C_{A_3}^\prime\right)+C_g^c \left(L_2-L_3\right) \,,  \\[0.5em]
%%%%%%%%%%%%%%%%%%%%%%%%%%
 \delta \hat h_{P}^{(\alpha_sm_c)}
 =
 &\, C_P\left[L_2+(w-1)L_3+L_5-(w+1)L_6\right]+2(w-1)C_P^\prime\nonumber\\[0.2em]
 & +C_g^c\left[L_2+(w-1)L_3\right] \,, \\[0.5em]
%%%%%%%%%%%%%%%%%%%%%%%%%%
 \delta \hat h_{T_1}^{(\alpha_sm_c)}
 =
 &\, C_{T_1}L_2+\frac{w-1}{2}(C_{T_2}-C_{T_3})(L_2-L_5) \nonumber\\[0.2em] 
 &+2(w-1)\left[C_{T_1}^\prime +\frac{w-1}{2}(C_{T_2}^\prime-C_{T_3}^\prime) \right]  +(w-1)C_{T_3}L_5+C_g^c L_2 \,, \\[0.5em]
%%%%%%%%%%%%%%%%%%%%%%%%%%
 \delta \hat h_{T_2}^{(\alpha_sm_c)}
 = 
 &\, C_{T_1}L_5 +\frac{w+1}{2}(C_{T_2}+C_{T_3})(L_2-L_5) \nonumber\\[0.2em]
 & +C_{T_3}(L_4-wL_5)+(w^2-1)(C_{T_2}^\prime+C_{T_3}^\prime) \,, \\[0.5em]
%%%%%%%%%%%%%%%%%%%%%%%%%%
 \delta \hat h_{T_3}^{(\alpha_sm_c)}
 =
 &\, -C_{T_1}(L_3-L_6)+C_{T_2}(L_2-L_5) \nonumber\\[0.2em] 
 & -\frac{1}{w+1}C_{T_3}(L_4-3L_5)+2(w-1)C_{T_2}^\prime-C_g^cL_3 \,,
\end{align}
for the $\alpha_s/m_c$ corrections, with 
\begin{align}
 C_{V_1} = 
 &\, \frac{1}{6z_{cb}(w-w_z)}\biggl[2(w+1)\left((3w-1)z_{cb}-z_{cb}^2-1\right)r(w)+4z_{cb}(w-w_z)\Omega(w)\nonumber\\
 &\hspace{2em} +\left(12z_{cb}(w_z-w)-(z_{cb}^2-1)\ln z_{cb}\right)\biggl],\\[0.3em]
 C_{V_2} =
 &\, \frac{-1}{6z_{cb}^2(w-w_z)^2}\biggl[ \left(-(w+1)z_{cb}^3+2w(2w+1)z_{cb}^2-(2w^2+5w-1)z_{cb}+2 \right) r(w) \nonumber\\
 &\hspace{2em} +z_{cb}\left( 2(z_{cb}-1)(w_z-w)+\left(z_{cb}^2-2(2w-1)z_{cb}+(3-2w)\right)\ln z_{cb}\right)\biggl],\\[0.3em]
 C_{V_3} = 
 &\, \frac{1}{6z_{cb}(w-w_z)^2}\biggl[ \left( -2z_{cb}^3+(2w^2+5w-1)z_{cb}^2-2w(2w+1)z_{cb}+w+1 \right) r(w)\nonumber\\
 &\hspace{2em} + \left( 2z_{cb}(z_{cb}-1)(w_z-w)+\left((3-2w)z_{cb}^2-2(2w-1)z_{cb}+1 \right)\ln z_{cb}\right) \biggl],\\[0.3em]
 C_{A_1} =
 &\, \frac{1}{6z_{cb}(w-w_z)}\biggl[2(w-1)\left((3w+1)z_{cb}-z_{cb}^2-1\right)r(w) +4z_{cb}(w-w_z)\Omega(w) \nonumber\\
 &\hspace{2em} +\left(12z_{cb}(w_z-w)-(z_{cb}^2-1)\ln z_{cb}\right)\biggl],\\[0.3em]
 C_{A_2} = 
 &\, \frac{-1}{6z_{cb}^2(w-w_z)^2}\biggl[ \left((1-w)z_{cb}^3+2w(2w-1)z_{cb}^2+(2w^2-5w-1)z_{cb}+2 \right) r(w)\nonumber\\
 &\hspace{2em} +z_{cb}\left( 2(z_{cb}+1)(w_z-w)+\left(z_{cb}^2-2(2w+1)z_{cb}+(2w+3)\right)\ln z_{cb}\right)\biggl],\\[0.3em]
 C_{A_3} = 
 &\, \frac{1}{6z_{cb}(w-w_z)^2}\biggl[ \left(2z_{cb}^3+ (2w^2-5w-1)z_{cb}^2+2w(2w-1)z_{cb}-2z_{cb}^3-w+1 \right) r(w)\nonumber\\
 &\hspace{2em} +\left( 2z_{cb}(z_{cb}+1)(w_z-w)+\left((2w+3)z_{cb}^2-2(2w+1)z_{cb}+1 \right)\ln z_{cb}\right)\biggl],\\[0.3em]
 C_S = 
 &\, \frac{1}{3z_{cb}(w-w_z)}\left[2z_{cb}(w-w_z)\Omega(w)-(w-1)(z_{cb}+1)^2r(w)+(z_{cb}^2-1)\ln z_{cb} \right],\\[0.3em]
 C_P = 
 &\, \frac{1}{3z_{cb}(w-w_z)}\left[2z_{cb}(w-w_z)\Omega(w)-(w+1)(z_{cb}-1)^2r(w)+(z_{cb}^2-1)\ln z_{cb} \right],\\[0.3em]
 C_{T_1} = 
 &\, \frac{1}{3z_{cb}(w-w_z)}\biggl[ (w-1)\left(2(2w+1)z_{cb}-z_{cb}^2-1\right)r(w) +2z_{cb}(w-w_z)\Omega(w) \nonumber\\
 &\hspace{2em} + \left( 6z_{cb}(w_z-w)-(z_{cb}^2-1)\ln z_{cb}\right) \biggl],\\[0.3em]
 C_{T_2} =
 &\, \frac{2}{3z_{cb}(w-w_z)}\biggl[ (1-w z_{cb})r(w)+z_{cb}\ln z_{cb}\biggl],\\[0.5em]
 C_{T_3} =
 &\, \frac{2}{3(w-w_z)}\biggl[ (w-z_{cb})r(w)+\ln z_{cb}\biggl], \\[0.5em] %RW: modified 8 June
 C_g^c =
 &\, -\frac{3}{2}\left[\ln\left(\frac{m_c}{\mu}\right)-\frac{13}{9}\right], \\[0.5em]
 \Omega(w) = 
 &\, \frac{w}{2\sqrt{w^2-1}} \Big[ 2 {\rm Li}_2(1-z_{cb}\,w_-)-2{\rm Li}_2(1-z_{cb}\,w_+ )+{\rm Li}_2(1-w_+^2)-{\rm Li}_2(1-w_-^2) \Big]\nonumber\\[0.3em]
 &\hspace{0em} -w r(w)\ln z_{cb} +1.
\end{align} 
Here $C'_X = \partial C_X/\partial w$, $w_\pm = w \pm \sqrt{w^2-1}$, $r(w) = \frac{\ln w_+}{\sqrt{w^2-1}}$, and $\mathrm{Li}_2(x) = \int^0_x dt \frac{\ln (1-t)}{t}$ is the dilogarithm, with $z_{cb}=m_c/m_b$ and $w_z = 1/2(z_{cb}+1/z_{cb})$. 

Note that the analytical forms of the functions $C_X$ given above are obtained in the $\overline{\mathrm{MS}}$ scheme with dimensional regularization and at the matching scale $\mu_{bc}=\sqrt{m_b m_c}$~\cite{Bernlochner:2017jka,Bernlochner:2022ywh,Neubert:1992qq}. Their results at an arbitrary scale $\mu$ can be obtained through~\cite{Bernlochner:2017jka}
\begin{align}
 C_{S,\,P}(\mu) &= C_{S,\,P}-\frac{2}{3} \big[ 2wr(w)+1 \big] \ln \left(\frac{\sqrt{m_bm_c}}{\mu}\right) , \\[0.3em]
 C_{V_1,\,A_1}(\mu) &= C_{V_1,\,A_1}-\frac{4}{3} \big[wr(w)+1 \big] \ln \left(\frac{\sqrt{m_bm_c}}{\mu}\right) , \\[0.3em]
 C_{T_1}(\mu) &= C_{T_1}-\frac{2}{3} \big[2wr(w)-3 \big] \ln \left(\frac{\sqrt{m_bm_c}}{\mu}\right) , 
\end{align}
and all the other $C_X$ are scale-independent. Specifically, we employ $\mu=\mu_b = 4.2\,\text{GeV}$ for our analyses. It should also be noted that the terms proportional to $C_g^c$ in the above formulae of $\delta \hat h_{X}^{(\alpha_sm_c)}$ are valid only up to $\mathcal{O}(\alpha_s/m_c^2)$~\cite{Bernlochner:2022ywh}. 

Finally, the $w$-dependent functions $L_i$ can be expressed in terms of the subleading IW functions as~\cite{Bernlochner:2022ywh,Falk:1992wt} 
\begin{align}
 L_1 & = -4 (w-1) \bl{\hat\chi_2(w)} +12\bl{\hat\chi_3(w)} \,, \\[0.3em]
 L_2 & = -4\bl{\hat\chi_3(w)} \,, \\[0.3em] 
 L_3 & = 4\bl{\hat\chi_2(w)} \,, \\[0.3em] 
 L_4 & = 2\bl{\eta(w)} -1 \,, \\[0.3em] 
 L_5 & = -1 \,, \\[0.3em]
 L_6 & = -2(\bl{\eta(w)} +1)/(w+1) \,. 
\end{align}

%%%%%%%%%%%%%%%%%%%%%%%%%%%%%
\section{Fit details}
\label{App:FFresult_all}
%%%%%%%%%%%%%%%%%%%%%%%%%%%%%

In this appendix, we summarize all the fit results of the form-factor parameters and the CKM matrix element $|V_{cb}|$ in different fit scenarios within the SM for the BSZ, BGL and HQET parameterizations, 
which are listed in Tables~\ref{Tab:SMfit_BSZ}, \ref{Tab:SMfit_BGL} and \ref{Tab:SMfit_HQET}, respectively. 
\bl{
The corresponding correlation matrices of the fitted form-factor parameters are not shown here to avoid potential violations of positive semidefiniteness caused by the rounding required for presentation.
The reliable results of BGL ($N_F=2$) and HQET ($3/2/1$) are packaged in the auxiliary file on arXiv version 2, where the correlation table is also available.}
%since listing them for all fit scenarios and form-factor parameterizations would be too lengthy; they will be provided upon request.

%%%%%%%%%%%%%%%%%%%%%%%%%%%%%%%%%%%%%%%%%
%% BSZ 
%%%%%%%%%%%%%%%%%%%%%%%%%%%%%%%%%%%%%%%%%
\begin{table}[htbp]
\centering
\renewcommand{\arraystretch}{1.48}
\begin{adjustbox}{width=0.98\textwidth,center,keepaspectratio}
\begin{tabular}{cccccc}
\hline\hline
BSZ ($N_F=1$) & (A, $w$) & (A, $\cos\theta_\ell$) & (A, $\cos\theta_V$)  & (A, $\chi$)  & (B) \\ \hline
$a_0^{f_+}$ & $\phantom{-}0.68 \pm 0.01$ & $\phantom{-}0.69 \pm 0.01$ & $\phantom{-}0.68 \pm 0.01$ & $\phantom{-}0.68 \pm 0.01$ & $\phantom{-}0.69 \pm 0.01$ \\
$a_1^{f_+}$ & $-2.65 \pm 0.10$ & $-2.64 \pm 0.10$ & $-2.65 \pm 0.10$ & $-2.71 \pm 0.10$ & $-2.59 \pm 0.11$ \\
$a_1^{f_0}$ & $\phantom{-}0.51 \pm 0.10$ & $\phantom{-}0.52 \pm 0.10$ & $\phantom{-}0.51 \pm 0.10$ & $\phantom{-}0.44 \pm 0.10$ & $\phantom{-}0.60 \pm 0.11$ \\
$a_0^{V}$ & $\phantom{-}0.80 \pm 0.03$ & $\phantom{-}0.72 \pm 0.02$ & $\phantom{-}0.78 \pm 0.03$ & $\phantom{-}0.82 \pm 0.03$ & $\phantom{-}0.77 \pm 0.02$ \\
$a_1^{V}$ & $-3.16 \pm 0.38$ & $-3.42 \pm 0.37$ & $-3.32 \pm 0.37$ & $-3.10 \pm 0.37$ & $-3.25 \pm 0.37$ \\
$a_1^{A_0}$ & $-3.56 \pm 0.23$ & $-3.34 \pm 0.25$ & $-3.51 \pm 0.25$ & $-4.29 \pm 0.29$ & $-3.56 \pm 0.24$ \\
$a_0^{A_1}$ & $\phantom{-}0.60 \pm 0.01$ & $\phantom{-}0.59 \pm 0.01$ & $\phantom{-}0.59 \pm 0.01$ & $\phantom{-}0.61 \pm 0.01$ & $\phantom{-}0.59 \pm 0.01$ \\
$a_1^{A_1}$ & $-0.30 \pm 0.18$ & $-0.56 \pm 0.20$ & $-0.58 \pm 0.20$ & $-0.29 \pm 0.20$ & $-0.46 \pm 0.19$ \\
$a_1^{A_{12}}$ & $\phantom{-}0.06 \pm 0.03$ & $\phantom{-}0.12 \pm 0.06$ & $\phantom{-}0.09 \pm 0.05$ & $-0.23 \pm 0.07$ & $\phantom{-}0.04 \pm 0.04$ \\ \hline 
$|V_{cb}|\times 10^3$ & $39.3 \pm 0.3$ & $39.2 \pm 0.4$ & $39.3 \pm 0.4$ & $39.9 \pm 0.4$ & $38.9 \pm 0.3$ \\
\hline\hline
BSZ ($N_F=2$) & (A, $w$) & (A, $\cos\theta_\ell$) & (A, $\cos\theta_V$)  & (A, $\chi$)  & (B) \\ \hline
$a_0^{f_+}$ & $\phantom{-}0.68 \pm 0.01$ & $\phantom{-}0.68 \pm 0.01$ & $\phantom{-}0.68 \pm 0.01$ & $\phantom{-}0.67 \pm 0.01$ & $\phantom{-}0.69 \pm 0.01$ \\
$a_1^{f_+}$ & $-1.80 \pm 0.25$ & $-1.84 \pm 0.24$ & $-1.84 \pm 0.24$ & $-1.91 \pm 0.24$ & $-1.78 \pm 0.26$ \\
$a_2^{f_+}$ & $\phantom{-}12.0 \pm \phantom{0}2.6$ & $\phantom{-}12.0 \pm \phantom{0}2.6$ & $\phantom{-}12.0 \pm \phantom{0}2.6$ & $\phantom{-}12.3 \pm \phantom{0}2.6$ & $\phantom{-}12.0 \pm \phantom{0}2.7$ \\
$a_1^{f_0}$ & $\phantom{-}0.93 \pm 0.24$ & $\phantom{-}0.87 \pm 0.24$ & $\phantom{-}0.87 \pm 0.24$ & $\phantom{-}0.72 \pm 0.24$ & $\phantom{-}0.98 \pm 0.28$ \\
$a_2^{f_0}$ & $\phantom{-}5.59 \pm 2.32$ & $\phantom{-}5.30 \pm 2.27$ & $\phantom{-}5.31 \pm 2.31$ & $\phantom{-}4.72 \pm 2.27$ & $\phantom{-}5.82 \pm 2.42$ \\
$a_0^{V}$ & $\phantom{-}0.74 \pm 0.03$ & $\phantom{-}0.67 \pm 0.02$ & $\phantom{-}0.71 \pm 0.03$ & $\phantom{-}0.78 \pm 0.03$ & $\phantom{-}0.73 \pm 0.03$ \\
$a_1^{V}$ & $-2.88 \pm 0.38$ & $-3.00 \pm 0.38$ & $-3.00 \pm 0.39$ & $-2.88 \pm 0.38$ & $-2.88 \pm 0.38$ \\
$a_2^{V}$ & $\phantom{-}23.5 \pm \phantom{0}6.2$ & $\phantom{-}28.8 \pm \phantom{0}6.0$ & $\phantom{-}25.2 \pm \phantom{0}6.2$ & $\phantom{-}18.9 \pm \phantom{0}6.2$ & $\phantom{-}23.6 \pm \phantom{0}6.2$ \\
$a_1^{A_0}$ & $-3.56 \pm 0.35$ & $-3.83 \pm 0.36$ & $-3.89 \pm 0.36$ & $-4.50 \pm 0.37$ & $-3.78 \pm 0.36$ \\
$a_2^{A_0}$ & $\phantom{-}4.86 \pm 5.72$ & $\phantom{-}5.01 \pm 5.90$ & $\phantom{-}4.94 \pm 5.84$ & $\phantom{-}10.47 \pm 5.85$ & $\phantom{-}4.30 \pm 5.79$ \\
$a_0^{A_1}$ & $\phantom{-}0.60 \pm 0.01$ & $\phantom{-}0.59 \pm 0.01$ & $\phantom{-}0.58 \pm 0.01$ & $\phantom{-}0.60 \pm 0.01$ & $\phantom{-}0.60 \pm 0.01$ \\
$a_1^{A_1}$ & $-0.53 \pm 0.27$ & $-0.85 \pm 0.28$ & $-0.88 \pm 0.27$ & $-0.46 \pm 0.28$ & $-0.58 \pm 0.27$ \\
$a_2^{A_1}$ & $-4.21 \pm 3.03$ & $-6.07 \pm 3.17$ & $-6.14 \pm 3.16$ & $-3.33 \pm 3.18$ & $-5.34 \pm 3.20$ \\
$a_1^{A_{12}}$ & $-0.34 \pm 0.07$ & $-0.55 \pm 0.10$ & $-0.55 \pm 0.09$ & $-0.87 \pm 0.11$ & $-0.48 \pm 0.09$ \\
$a_2^{A_{12}}$ & $-6.90 \pm 1.22$ & $-9.10 \pm 1.37$ & $-9.01 \pm 1.37$ & $-8.79 \pm 1.40$ & $-8.37 \pm 1.41$\\ \hline
$|V_{cb}|\times 10^3$ & $39.5 \pm 0.4$ & $39.8 \pm 0.4$ & $39.8 \pm 0.4$ & $40.6 \pm 0.5$ & $39.3 \pm 0.3$ \\ 
\hline\hline
\end{tabular}
\end{adjustbox}
\caption{Fit results of the form-factor parameters and $|V_{cb}|$ for the BSZ parameterization with $N_F=1,2$ in different fit scenarios within the SM. \label{Tab:SMfit_BSZ}}
\end{table}

%%%%%%%%%%%%%%%%%%%%%%%%%%%%%%%%%%%%%%%%%
%% BGL 
%%%%%%%%%%%%%%%%%%%%%%%%%%%%%%%%%%%%%%%%%
\begin{table}[htbp]
\centering
\renewcommand{\arraystretch}{1.5}
\begin{adjustbox}{width=0.98\textwidth,center,keepaspectratio}
\begin{tabular}{cccccc}
\hline\hline
BGL ($N_F=1$) & (A, $w$) & (A, $\cos\theta_\ell$) & (A, $\cos\theta_V$)  & (A, $\chi$) & (B) \\ \hline
$b_0^{f_+}$ & $0.0154 \pm 0.0001$ & $0.0154 \pm 0.0001$ & $0.0154 \pm 0.0001$ & $0.0153 \pm 0.0001$ & $0.0154 \pm 0.0001$ \\
$b_1^{f_+}$ & $-0.037 \pm 0.002$ & $-0.037 \pm 0.002$ & $-0.037 \pm 0.002$ & $-0.039 \pm 0.002$ & $-0.036 \pm 0.003$ \\
$b_1^{f_0}$ & $-0.197 \pm 0.011$ & $-0.199 \pm 0.011$ & $-0.200 \pm 0.011$ & $-0.210 \pm 0.011$ & $-0.194 \pm 0.013$ \\
$b_0^{f}$ & $0.0121 \pm 0.0001$ & $0.0122 \pm 0.0001$ & $0.0122 \pm 0.0001$ & $0.0122 \pm 0.0001$ & $0.0121 \pm 0.0001$ \\
$b_1^{f}$ & $-0.001 \pm 0.004$ & $-0.010 \pm 0.004$ & $-0.011 \pm 0.004$ & $-0.005 \pm 0.004$ & $-0.005 \pm 0.004$ \\
$b_1^{F_1}$ & $-0.001 \pm 0.000$ & $-0.001 \pm 0.001$ & $-0.001 \pm 0.001$ & $-0.005 \pm 0.001$ & $-0.001 \pm 0.001$ \\
$b_1^{F_2}$ & $-0.135 \pm 0.014$ & $-0.134 \pm 0.017$ & $-0.141 \pm 0.016$ & $-0.213 \pm 0.020$ & $-0.140 \pm 0.015$ \\
$b_0^{g}$ & $\phantom{-}0.028 \pm 0.001$ & $\phantom{-}0.026 \pm 0.001$ & $\phantom{-}0.027 \pm 0.001$ & $\phantom{-}0.028 \pm 0.001$ & $\phantom{-}0.027 \pm 0.001$ \\
$b_1^{g}$ & $-0.031 \pm 0.013$ & $-0.053 \pm 0.013$ & $-0.045 \pm 0.013$ & $-0.030 \pm 0.013$ & $-0.039 \pm 0.013$ \\ \hline
$|V_{cb}|\times 10^3$ & $40.0 \pm 0.4$ & $40.1 \pm 0.4$ & $40.2 \pm 0.4$ & $41.1 \pm 0.4$ & $39.7 \pm 0.3$  \\
\hline\hline
BGL ($N_F=2$) & (A, $w$) & (A, $\cos\theta_\ell$) & (A, $\cos\theta_V$)  & (A, $\chi$) & (B) \\ \hline
$b_0^{f_+}$ & $0.0154 \pm 0.0001$ & $0.0154 \pm 0.0001$ & $0.0154 \pm 0.0001$ & $0.0153 \pm 0.0001$ & $0.0154 \pm 0.0001$ \\
$b_1^{f_+}$ & $-0.039 \pm 0.003$ & $-0.040 \pm 0.003$ & $-0.040 \pm 0.003$ & $-0.042 \pm 0.003$ & $-0.039 \pm 0.003$ \\
$b_2^{f_+}$ & $\phantom{-}0.094 \pm 0.045$ & $\phantom{-}0.097 \pm 0.045$ & $\phantom{-}0.097 \pm 0.044$ & $\phantom{-}0.100 \pm 0.044$ & $\phantom{-}0.093 \pm 0.045$ \\
$b_1^{f_0}$ & $-0.228 \pm 0.016$ & $-0.231 \pm 0.016$ & $-0.232 \pm 0.016$ & $-0.239 \pm 0.015$ & $-0.225 \pm 0.016$ \\
$b_2^{f_0}$ & $\phantom{-}0.730 \pm 0.236$ & $\phantom{-}0.722 \pm 0.237$ & $\phantom{-}0.718 \pm 0.233$ & $\phantom{-}0.697 \pm 0.232$ & $\phantom{-}0.743 \pm 0.237$ \\
$b_0^{f}$ & $0.0121 \pm 0.0001$ & $0.0121 \pm 0.0001$ & $0.0121 \pm 0.0001$ & $0.0121 \pm 0.0001$ & $0.0121 \pm 0.0001$ \\
$b_1^{f}$ & $\phantom{-}0.008 \pm 0.004$ & $\phantom{-}0.006 \pm 0.005$ & $\phantom{-}0.005 \pm 0.005$ & $\phantom{-}0.009 \pm 0.005$ & $\phantom{-}0.009 \pm 0.005$ \\
$b_2^{f}$ & $-0.131 \pm 0.056$ & $-0.177 \pm 0.058$ & $-0.168 \pm 0.059$ & $-0.142 \pm 0.059$ & $-0.155 \pm 0.058$ \\
$b_1^{F_1}$ & $\phantom{-}0.004 \pm 0.001$ & $\phantom{-}0.005 \pm 0.001$ & $\phantom{-}0.005 \pm 0.001$ & $\phantom{-}0.001 \pm 0.001$ & $\phantom{-}0.005 \pm 0.001$ \\
$b_2^{F_1}$ & $-0.080 \pm 0.012$ & $-0.113 \pm 0.015$ & $-0.111 \pm 0.015$ & $-0.109 \pm 0.016$ & $-0.100 \pm 0.015$ \\
$b_1^{F_2}$ & $-0.131 \pm 0.027$ & $-0.149 \pm 0.030$ & $-0.154 \pm 0.030$ & $-0.226 \pm 0.032$ & $-0.140 \pm 0.028$ \\
$b_2^{F_2}$ & $-0.162 \pm 0.360$ & $-0.219 \pm 0.364$ & $-0.223 \pm 0.369$ & $-0.062 \pm 0.371$ & $-0.248 \pm 0.366$ \\
$b_0^{g}$ & $\phantom{-}0.027 \pm 0.001$ & $\phantom{-}0.026 \pm 0.001$ & $\phantom{-}0.027 \pm 0.001$ & $\phantom{-}0.028 \pm 0.001$ & $\phantom{-}0.027 \pm 0.001$ \\
$b_1^{g}$ & $-0.061 \pm 0.020$ & $-0.092 \pm 0.019$ & $-0.084 \pm 0.020$ & $-0.047 \pm 0.020$ & $-0.065 \pm 0.019$ \\
$b_2^{g}$ & $\phantom{-}0.335 \pm 0.188$ & $\phantom{-}0.539 \pm 0.188$ & $\phantom{-}0.500 \pm 0.189$ & $\phantom{-}0.222 \pm 0.187$ & $\phantom{-}0.357 \pm 0.186$ \\ \hline
$|V_{cb}|\times 10^3$ & $39.5 \pm 0.4$ & $39.9 \pm 0.4$ & $39.9 \pm 0.4$ & $40.6 \pm 0.4$ & $39.3 \pm 0.3$  \\
\hline\hline
\end{tabular}
\end{adjustbox}
\caption{Fit results of the form-factor parameters and $|V_{cb}|$ for the BGL parameterization with $N_F=1,2$ in different fit scenarios within the SM. \label{Tab:SMfit_BGL}}
\end{table}

%%%%%%%%%%%%%%%%%%%%%%%%%%%%%%%%%%%%%%%%%
%% HQET 
%%%%%%%%%%%%%%%%%%%%%%%%%%%%%%%%%%%%%%%%%
\begin{table}[htbp]
\centering
\renewcommand{\arraystretch}{1.04}
\begin{adjustbox}{width=0.98\textwidth,center,keepaspectratio}
\begin{tabular}{cccccc}
\hline\hline
HQET ($2/1/0$) & (A, $w$) & (A, $\cos\theta_\ell$) & (A, $\cos\theta_V$)  & (A, $\chi$) & (B) \\ \hline
$c_1^{\xi}$ & $-1.19 \pm 0.02$ & $-1.23 \pm 0.02$ & $-1.22 \pm 0.02$ & $-1.22 \pm 0.03$ & $-1.20 \pm 0.02$ \\
$c_2^{\xi}$ & $\phantom{-}1.14 \pm 0.04$ & $\phantom{-}1.19 \pm 0.04$ & $\phantom{-}1.19 \pm 0.04$ & $\phantom{-}1.19 \pm 0.04$ & $\phantom{-}1.16 \pm 0.04$ \\
$c_0^{\eta}$ & $\phantom{-}0.43 \pm 0.05$ & $\phantom{-}0.49 \pm 0.05$ & $\phantom{-}0.44 \pm 0.05$ & $\phantom{-}0.38 \pm 0.06$ & $\phantom{-}0.44 \pm 0.06$ \\
$c_1^{\eta}$ & $\phantom{-}0.02 \pm 0.03$ & $0.01 \pm 0.03$ & $0.01 \pm 0.03$ & $\phantom{-}0.01 \pm 0.03$ & $\phantom{-}0.02 \pm 0.03$ \\
$c_0^{\chi_2}$ & $-0.04 \pm 0.02$ & $-0.04 \pm 0.02$ & $-0.04 \pm 0.02$ & $-0.06 \pm 0.02$ & $-0.05 \pm 0.02$ \\
$c_1^{\chi_2}$ & $-0.04 \pm 0.02$ & $-0.04 \pm 0.02$ & $-0.04 \pm 0.02$ & $-0.04 \pm 0.02$ & $-0.03 \pm 0.02$ \\
$c_1^{\chi_3}$ & $0.01 \pm 0.01$ & $0.03 \pm 0.01$ & $0.02 \pm 0.01$ & $0.01 \pm 0.01$ & $0.00 \pm 0.01$ \\
$c_0^{\ell_1}$ & $\phantom{-}0.53 \pm 0.20$ & $\phantom{-}0.54 \pm 0.20$ & $\phantom{-}0.51 \pm 0.20$ & $\phantom{-}0.36 \pm 0.20$ & $\phantom{-}0.26 \pm 0.21$ \\
$c_0^{\ell_2}$ & $-2.26 \pm 0.17$ & $-2.23 \pm 0.18$ & $-2.21 \pm 0.18$ & $-2.23 \pm 0.18$ & $-2.13 \pm 0.20$ \\
$c_0^{\ell_3}$ & $0.30 \pm 0.63$ & $0.63 \pm 0.65$ & $0.49 \pm 0.62$ & $-1.73 \pm 0.75$ & $-0.01 \pm 0.67$ \\
$c_0^{\ell_4}$ & $-0.75 \pm 0.49$ & $-1.13 \pm 0.48$ & $-0.81 \pm 0.49$ & $-0.32 \pm 0.50$ & $-0.81 \pm 0.50$ \\
$c_0^{\ell_5}$ & $\phantom{-}0.24 \pm 0.44$ & $\phantom{-}0.87 \pm 0.40$ & $\phantom{-}0.22 \pm 0.42$ & $\phantom{-}-0.11 \pm 0.45$ & $\phantom{-}0.02 \pm 0.43$ \\
$c_0^{\ell_6}$ & $\phantom{-}0.96 \pm 0.46$ & $\phantom{-}1.11 \pm 0.47$ & $\phantom{-}0.81 \pm 0.45$ & $\phantom{-}1.48 \pm 0.49$ & $\phantom{-}0.81 \pm 0.47$ \\ \hline
$|V_{cb}|\times 10^3$ & $39.3 \pm 0.3$ & $39.4 \pm 0.4$ & $39.5 \pm 0.4$ & $40.3 \pm 0.4$ & $39.8 \pm 0.3$ \\
\hline\hline
HQET ($3/2/1$) & (A, $w$) & (A, $\cos\theta_\ell$) & (A, $\cos\theta_V$)  & (A, $\chi$) & (B) \\ \hline
$c_1^{\xi}$ & $-1.29 \pm 0.03$ & $-1.34 \pm 0.04$ & $-1.35 \pm 0.04$ & $-1.36 \pm 0.04$ & $-1.32 \pm 0.04$ \\
$c_2^{\xi}$ & $\phantom{-}2.36 \pm 0.12$ & $\phantom{-}2.49 \pm 0.13$ & $\phantom{-}2.51 \pm 0.12$ & $\phantom{-}2.52 \pm 0.12$ & $\phantom{-}2.44 \pm 0.12$ \\
$c_3^{\xi}$ & $-4.64 \pm 0.37$ & $-4.88 \pm 0.35$ & $-4.93 \pm 0.33$ & $-4.88 \pm 0.33$ & $-4.79 \pm 0.35$ \\
$c_0^{\eta}$ & $\phantom{-}0.38 \pm 0.05$ & $\phantom{-}0.44 \pm 0.05$ & $\phantom{-}0.40 \pm 0.05$ & $\phantom{-}0.34 \pm 0.05$ & $\phantom{-}0.41 \pm 0.05$ \\
$c_1^{\eta}$ & $\phantom{-}0.05 \pm 0.03$ & $\phantom{-}0.05 \pm 0.03$ & $\phantom{-}0.04 \pm 0.03$ & $\phantom{-}0.04 \pm 0.03$ & $\phantom{-}0.04 \pm 0.03$ \\
$c_2^{\eta}$ & $\phantom{-}0.00 \pm 0.07$ & $\phantom{-}-0.01 \pm 0.07$ & $\phantom{-}-0.01 \pm 0.07$ & $0.00 \pm 0.07$ & $\phantom{-}-0.01 \pm 0.07$ \\
$c_0^{\chi_2}$ & $-0.04 \pm 0.02$ & $-0.03 \pm 0.02$ & $-0.04 \pm 0.02$ & $-0.07 \pm 0.02$ & $-0.05 \pm 0.02$ \\
$c_1^{\chi_2}$ & $0.00 \pm 0.02$ & $0.00 \pm 0.02$ & $0.00 \pm 0.02$ & $0.00 \pm 0.02$ & $0.00 \pm 0.02$ \\
$c_2^{\chi_2}$ & $-0.01 \pm 0.02$ & $-0.01 \pm 0.02$ & $-0.01 \pm 0.02$ & $-0.01 \pm 0.02$ & $-0.01 \pm 0.02$ \\
$c_1^{\chi_3}$ & $0.02 \pm 0.01$ & $0.04 \pm 0.01$ & $0.04 \pm 0.01$ & $0.02 \pm 0.01$ & $0.02 \pm 0.01$ \\
$c_2^{\chi_3}$ & $\phantom{-}0.00 \pm 0.03$ & $\phantom{-}0.00 \pm 0.03$ & $\phantom{-}0.00 \pm 0.03$ & $\phantom{-}0.00 \pm 0.03$ & $\phantom{-}-0.01 \pm 0.03$ \\
$c_0^{\ell_1}$ & $\phantom{-}0.31 \pm 0.20$ & $\phantom{-}0.32 \pm 0.20$ & $\phantom{-}0.29 \pm 0.20$ & $\phantom{-}0.13 \pm 0.20$ & $\phantom{-}0.29 \pm 0.21$ \\
$c_1^{\ell_1}$ & $\phantom{-}-0.66 \pm 0.69$ & $\phantom{-}-0.87 \pm 0.69$ & $-0.73 \pm 0.68$ & $-0.52 \pm 0.69$ & $-0.68 \pm 0.68$ \\
$c_0^{\ell_2}$ & $-2.06 \pm 0.16$ & $-2.13 \pm 0.18$ & $-2.11 \pm 0.18$ & $-2.10 \pm 0.18$ & $-2.23 \pm 0.20$ \\
$c_1^{\ell_2}$ & $-2.22 \pm 0.55$ & $-2.15 \pm 0.60$ & $-1.77 \pm 0.59$ & $-1.74 \pm 0.61$ & $-1.92 \pm 0.61$ \\
$c_0^{\ell_3}$ & $\phantom{-}-0.40 \pm 0.57$ & $\phantom{-}0.63 \pm 0.57$ & $\phantom{-}0.71 \pm 0.53$ & $-0.65 \pm 0.69$ & $\phantom{-}0.44 \pm 0.67$ \\
$c_1^{\ell_3}$ & $-2.32 \pm 0.81$ & $-2.33 \pm 0.74$ & $-2.85 \pm 0.78$ & $-1.46 \pm 0.83$ & $-2.06 \pm 0.83$ \\
$c_0^{\ell_4}$ & $-0.31 \pm 0.46$ & $\phantom{-}-0.73 \pm 0.46$ & $-0.42 \pm 0.45$ & $0.05 \pm 0.49$ & $-0.54 \pm 0.47$ \\
$c_1^{\ell_4}$ & $-0.23 \pm 0.51$ & $-0.31 \pm 0.50$ & $-0.19 \pm 0.50$ & $0.05 \pm 0.52$ & $-0.29 \pm 0.51$ \\
$c_0^{\ell_5}$ & $\phantom{-}-0.26 \pm 0.43$ & $\phantom{-}0.43 \pm 0.37$ & $\phantom{-}0.05 \pm 0.39$ & $\phantom{-}-0.63 \pm 0.42$ & $\phantom{-}-0.24 \pm 0.41$ \\
$c_1^{\ell_5}$ & $-0.83 \pm 0.63$ & $-0.23 \pm 0.60$ & $-0.66 \pm 0.58$ & $-1.05 \pm 0.60$ & $-0.77 \pm 0.60$ \\
$c_0^{\ell_6}$ & $\phantom{-}0.21 \pm 0.38$ & $0.19 \pm 0.39$ & $\phantom{-}0.08 \pm 0.37$ & $\phantom{-}0.90 \pm 0.41$ & $\phantom{-}0.44 \pm 0.41$ \\
$c_1^{\ell_6}$ & $\phantom{-}1.09 \pm 0.53$ & $1.12 \pm 0.55$ & $\phantom{-}0.78 \pm 0.52$ & $\phantom{-}1.75 \pm 0.57$ & $\phantom{-}1.20 \pm 0.56$ \\ \hline
$|V_{cb}|\times 10^3$ & $39.4 \pm 0.3$ & $39.4 \pm 0.4$ & $39.6 \pm 0.4$ & $40.4 \pm 0.4$ & $39.5 \pm 0.3$  \\
\hline\hline
\end{tabular}
\end{adjustbox}
\caption{Fit results of the form-factor parameters and $|V_{cb}|$ for the HQET $(2/1/0)$ and HQET $(3/2/1)$ parameterizations in different fit scenarios within the SM. \label{Tab:SMfit_HQET}}
\end{table}

\clearpage

\bibliographystyle{JHEP}
\bibliography{ref}

\end{document}